\documentclass[a4paper,11pt]{article}
\pdfoutput=1 

\usepackage{jcappub} 
\usepackage[T1]{fontenc} 
\usepackage{bm}

\usepackage{soul}

\newcommand{\T}{\mathbf{\hat{\mathcal{T}}}}

\newcommand{\Dk}[1]{\frac{d^3#1}{(2\pi)^3}}

\newcommand{\ve}[1]{{\text{\bf #1}}} 
\newcommand{\vk}{\ve k}
\newcommand{\vp}{\ve p}
\newcommand{\vq}{\ve q}
\newcommand{\vx}{\ve x}

\newcommand{\A}{\mathcal{A}}
\newcommand{\B}{\mathcal{B}}
\newcommand{\mA}{\mathcal{A}}
\newcommand{\mB}{\mathcal{B}}
\newcommand{\mC}{\mathcal{C}}
\newcommand{\mO}{\mathcal{O}}
\newcommand{\mP}{\mathcal{P}}

\newcommand{\tm}{\text{m}}

\newcommand{\ikk}{\underset{\vk_{12}= \vk}{\int}}
\newcommand{\ikkk}{\underset{\vk_{123}= \vk}{\int}}
\newcommand{\ip}{\int_{\vp}} 

\newcommand{\dD}{\delta_\text{D}}

\newcommand{\vhn}{\hat{\ve n}}

\usepackage[dvipsnames]{xcolor}

\definecolor{Georgios}{HTML}{379683}

\usepackage{booktabs}
\usepackage{float} 
\newcommand{\ra}[1]{\renewcommand{\arraystretch}{#1}}

\title{\boldmath Redshift space power spectrum beyond Einstein-de Sitter kernels}

\author[a,b]{Alejandro Aviles,}
\emailAdd{avilescervantes@gmail.com}

\author[c]{Georgios Valogiannis,}
\emailAdd{gvalogiannis@g.harvard.edu}

\author[b]{Mario A.~Rodriguez-Meza,}
\emailAdd{marioalberto.rodriguez@inin.gob.mx}

\author[b]{Jorge L.~Cervantes-Cota,}
\emailAdd{jorge.cervantes@inin.gob.mx}

\author[d]{Baojiu Li,}
\emailAdd{baojiu.li@durham.ac.uk}

\author[e]{Rachel Bean}
\emailAdd{rbean@astro.cornell.edu}

\affiliation[a]{Consejo Nacional de Ciencia y Tecnolog\'ia, Av. Insurgentes Sur 1582,
Colonia Cr\'edito Constructor, Del. Benito Ju\'arez, 03940, Ciudad de M\'exico, M\'exico}
\affiliation[b]{Departamento de F\'isica, Instituto Nacional de Investigaciones Nucleares,
Apartado Postal 18-1027, Col. Escand\'on, Ciudad de M\'exico,11801, M\'exico.}
\affiliation[c]{Department of Physics, Harvard University, Cambridge, MA 02138, USA}
\affiliation[d]{Institute for Computational Cosmology, Department of Physics, Durham University, South Road, Durham DH1 3LE, UK}
\affiliation[e]{Department of Astronomy, Cornell University, Ithaca, NY 14853, USA}


\abstract{
We develop a framework to compute the redshift space power spectrum (PS), with kernels beyond Einstein-de Sitter (EdS), that can be applied to a wide variety of generalized cosmologies.  We build upon a formalism that was recently employed for standard cosmology in Chen, Vlah \& White (2020), and utilize an expansion of the density-weighted velocity moment generating function that explicitly separates the magnitude of the $k$-modes and their angle to the line-of-sight direction dependencies. We compute the PS for matter and biased tracers to 1-loop Perturbation Theory (PT) and show that the expansion has a correct infrared and ultraviolet behavior, free of unwanted divergences. We also add Effective Field Theory (EFT) counterterms, necessary to account for small-scale contributions to PT, and employ an IR-resummation prescription to properly model the smearing of the BAO due to large scale bulk flows within Standard-PT.  To demonstrate the applicability of our formalism, we apply it on the $\Lambda$CDM and the Hu-Sawicki $f(R)$ models, and compare our numerical results against the {\sc elephant} suite of $N$-body simulations, finding very good agreement up to $k= 0.27\, \text{Mpc}^{-1} h$ at $z=0.5$ for the first three non-vanishing Legendre multipoles of the PS. To our knowledge, the model presented in this work is the most accurate theoretical EFT-PT  for modified gravity to date, being the only one that accounts for beyond linear local biasing in redshift-space. Hence, we argue our RSD modeling is a promising tool to construct theoretical templates in order to test deviations from $\Lambda$CDM using real data obtained from the next stage of cosmological surveys such as DESI and LSST. 
}

\begin{document} 
\maketitle
\flushbottom

\begin{section}{Introduction}

In recent years galaxy and quasar surveys such as 2dF\cite{Cole:2005sx}, WiggleZ \cite{Blake:2011en}, BOSS\cite{Samushia:2013yga}, eBOSS \cite{Zhai:2016gyu} and DES \cite{Abbott:2017wau} have been measuring statistical properties of the large scale structure of the Universe, and the future is near for more refined, Stage IV probes such as DESI \cite{Aghamousa:2016zmz}, LSST \cite{Abate:2012za}, and EUCLID \cite{Scaramella:2015rra}.   
Certainly, the science of clustering has been boosted over the last two decades and has become essential to extract cosmological information. Two point statistics have been developed in different aspects of theory and numerical simulations, and  the successful comparison of both approaches in the region of common validity is an important consistency check to ensure that we are doing well and to gain physical insight into simulation results.  Once trustable simulations and mocks are at hand, one is able to compare with observations at the level of precision needed in forthcoming probes, $\sim 1\%$ or less in the two point statistics observables. 

In perturbation theory (PT) we firstly lead with dark matter density fields, and through bias they connect to tracers such as galaxies.  Then, redshift space distortions (RSD),  originated by the underlying velocity field along the line of sight to galaxies, are used to measure galaxy clustering \cite{10.1093/mnras/227.1.1}.
RSD is a solid observable to measure large scale structure, as accounted
since the pioneering works of Refs.~\cite{Peacock:2001gs,Hawkins:2002sg,Tegmark:2006az} to present.

RSD maps galaxy clustering from real to redshift space, therefore, it introduces an anisotropy to the measured power spectrum (PS) and correlation function, apart from the 
Alcock-Paczynski anisotropic effect \cite{AP_effect}, stemming 
from the coordinate conversion with an incorrect fiducial cosmology.   
The RSD mapping is 
nonlinear and intricate due to cross-correlations between density and velocity fields, but with time advances have been made on its nonlinear 
modeling, initially carried out by  \cite{Hivon:1994qb,Fisher:1995ec,Taylor:1996ne,Heavens:1998es,Magira:1999bn,Scoccimarro:2004tg,Matsubara:2008wx,Percival:2008sh,Taruya:2010mx}. The mapping is 
cumbersome also because of the Finger-of-God (FoG) feature due to the randomness of the small-scale peculiar velocity field \cite{Bernardeau:2001qr}.  There is a variety of RSD perturbation models developed in the literature, with the TNS model \cite{Taruya:2010mx} arguably being one of the most popular choices. Refinements of this model have been put forward, joined by other approaches in a long list of contributions aiming at understanding different issues: the nonlinear mappings and the role of cross correlations between density and velocity fields; the treatment of possible primordial non-Gaussian initial conditions; the regularization of perturbative expansions; effective field theory (EFT) implementations; ways to maximize RSD model performance; applications to biased tracers; and to successfully  test models with simulations, among other topics.

RSD measures the growth of structure and therefore can serve as a powerful probe to test gravity.  In fact, different clustering measurements have hinted possible deviations from the $\Lambda$CDM model, assuming WMAP \cite{Samushia:2012iq} and Planck baseline cosmologies \cite{Macaulay:2013swa,Beutler:2013yhm,Beutler:2016arn}. Though deviations are mild and tensions can,  in some cases, be attributed to differences in the determination of the clustering amplitude ($\sigma_8$) between clustering and Planck data, one is motivated to test other gravity models, beyond General Relativity (GR), that at the same time can account for the accelerated expansion of the Universe, whose theoretical origin is still unknown. This is why modified gravity (MG)  models could be interesting, among other reasons.           

Many studies in MG are based on the linear aspects of the gravitational interactions, since theory gets more complicated than GR. Hence limited models have been often explored in the literature that captures only part of the complexity of MG.  In fact when considering perturbations the right-hand-side  (RHS) of the Poisson equation is often subject to approximations, such as considering that the effective density perturbation is separable in a function of the wavenumber and scale factor or even that it does not depend on $k$, and that certainly limits the range of possible gravity theories. In a more general scheme, MG induces, even at linear level, scale dependencies and the nonlinear theory becomes cumbersome. In the last years, however, efforts have been made to develop MG cosmological structure formation PT in full generality at one loop in standard perturbation theory (SPT) \cite{Koyama:2009me} and in Lagrangian perturbation theory (LPT) \cite{Aviles:2017aor}. Nonlinear PT is limited to weakly nonlinear scales, but it is nevertheless important because of the baryon acoustic oscillations (BAO) and RSD effects that manifest themselves there. Non-linearities permit us to understand what the important couplings are, among the different Fourier modes that are defined by the kernels \cite{Taruya:2016jdt}, and this in turn helps us to understand the role of screening mechanisms in structure formation \cite{Aviles:2018qotF}.  These studies have been important to confirm simulation results \cite{Li:2011vk,Cataneo:2018cic} and to construct mocks that employ linear and quasilinear physics analytically \cite{Valogiannis:2016ane,Winther:2017jof}, and in this way, to achieve the above-mentioned level of precision that new generation of stage IV experiments demand to test gravity \cite{Alam:2020jdv}.

RSD in MG has been developed, firstly measuring possible deviations from the growth index and growth rate 
\cite{Linder:2007nu,Wang:2007ht,Guzzo:2008ac,Yamamoto:2008gr,Simpson:2009zj,Stril:2009ey,Bean:2010zq}, then estimating and testing linear perturbation theory 
for different MG models \cite{Song:2008qt,Guzik:2009cm,Song:2010fg,Asaba:2013xql}.  RSD nonlinear perturbation theory in MG was built in the context of 
SPT \cite{Taruya:2013quf,Taruya:2014faa} and then in LPT using the Gaussian streaming model \cite{Bose:2017dtl,Valogiannis:2019nfz}.  In addition to hybrid approaches combining techniques \citep{Bose:2019yjp}, RSD effects have been also computed through 
simulations \cite{Oyaizu:2008sr,Oyaizu:2008tb,Schmidt:2008tn,Schmidt:2009sg,Zhao:2010qy,Li:2009sy,Li:2010mqa,Li:2010re,Li:2012by,Jennings:2012pt,Arnalte-Mur:2016alq,Hernandez-Aguayo:2018oxg} and compared to observations \cite{Song:2015oza}, finding still modest constraints on the parameter $f_{R0} (< 8 \times 10^{-4}$) of the Hu-Sawicki (HS) MG model using the 
DR11 CMASS data, among other constraints \cite{Dossett:2014oia,Johnson:2015aaa}; {see however Ref.~\cite{He:2018oai}}. Further developments have been made to compute the RSD effect for Vainshtein and Chameleon screening models in the PS \cite{Bose:2017dtl} and correlation function \cite{Bose:2017dtl} by numerically computing the SPT kernels using Taruya's method \cite{Taruya:2016jdt}.  Although current clustering data is not sufficiently precise to discriminate between different gravity models, the analysis of different systematics is an important task to be better studied. For instance,  comparing MG models to observations demands first pipeline validation of theory and simulations, otherwise an issue arises by which  GR nuisance parameters and nonlinear RSD modeling can result in biasing MG physical parameters, since there is an overlap of their effects \cite{Taruya:2013my,Barreira:2016ovx,Bose:2017myh}.  A consistent pipeline treatment and  more precise data from upcoming surveys will make possible to disentangle these effects.  

Recently a general scheme for RSD has been put 
forward \cite{Vlah:2018ygt} in which perturbative expansions are expressed in terms of the density-weighted velocity moment generating function, that is based on the pioneering work of \cite{Scoccimarro:2004tg}, which serves to compare in a unified manner the different expansions of RSD modeling and to formulate a Fourier streaming model. This formalism was later applied and tested against simulations in \cite{Chen:2020fxs} in view of the precision required for the next generation of clustering surveys. We will follow this approach aiming at developing the general theory of RSD for generalized  cosmologies to 1-loop approximation. We will find that our expressions are more elaborate because generalized models possess additional scales, and hence, more complex $k$-dependencies. We find generalized kernels, beyond Einstein-de Sitter (EdS), that can be applied to $\Lambda$CDM, MG models, dark energy with scale-dependent Poisson equation or even massive neutrinos. 

In this work we extend the moment expansion approach of \cite{Chen:2020fxs} to be applied to generalized cosmologies, on which their deviations to $\Lambda$CDM are encapsulated in the SPT kernels and linear growth function. To do this, we Taylor expand the density-weighted velocity moment generating function accounting for all terms that arise in the 1-loop redshift space PS. In order to model small scales out of the reach of SPT and the effects of the damping of the density and velocity fields along the line-of-sight direction, we utilize the EFT formalism of \cite{Ivanov:2019pdj,Chen:2020fxs}. Further, to properly damp the BAO features we use the IR-resummation technique as implemented in \cite{Ivanov:2018gjr,Chudaykin:2020aoj}. Our tracers are constructed using an extended version of the well-known formalism of \cite{McDonald:2009dh,Saito:2014qha}, that accounts for the additional scales introduced in models beyond GR. Within this work we exemplify our result using the HS $n=1$ MG model, for which we present a brief summary in appendix \ref{app:fRHS}, focusing on the F6, F5, and F4 realizations. We compare our results for the monopole, quadrupole, and hexadecapole of the spectra against the \verb|ELEPHANT| MG N-body simulations \citep{Cautun:2017tkc}, and find good agreement with our theory up to scales about $k = 0.27\,\text{Mpc}^{-1}h$ at redshifts $z=0.5$ and $z=1$, for both tracers and matter.   
To our knowledge, this is the first PT work dealing with the redshift space PS for MG/generalized cosmologies for tracers beyond linear bias.

The rest of this work is organized as follows: In section \ref{sect:basic_model} we explain the scope of alternative models that are included in our theory and the basics of PT power spectra. In section \ref{sect:BiasExp} we present the bias expansion of \cite{McDonald:2009dh} extended for generalized cosmologies, where we also discuss about third order bias and renormalization; in addition, we present here one-dimensional spectra for tracers. In section \ref{RSD_PT_GC} we develop the RSD model, using the moment expansion approach for generalized kernels. We discuss large and small scale behaviors, counterterms from EFT, and use an IR-resummation scheme to properly compute the BAO damping to high-$k$. In section \ref{num_results} we compute redshift-space power spectra and compare our results to simulations. Complementary information is sent to appendices, where we present expansions of the $I^\tm_{n}(k)$ functions in Appendix \ref{app:ImnFunctions}; formulae for specific MG chosen to present our results, the HS model, in Appendix \ref{app:fRHS}, but we emphasize that other MG or alternative models can be straightforwardly employed. The SPT kernels for generalized cosmologies are  in Appendix \ref{app:Kernels}. 

\end{section}

\begin{section}{Gravity models, fluid equations, and generalized kernels} \label{sect:basic_model}
The scope of gravity theories or matter/energy models of the present work are those governed by the Poisson equation that can be written, in Fourier space, as 
\begin{equation} \label{PoissonEq}
 -\frac{k^2}{a^2} \Phi(\vk) = A(k,t) \delta(\vk)  + S(\vk), 
\end{equation}
in which the functions $A(k,t)$ and $S(\vk)$ determine specific models. Theories such as MG, $k$-dependent dark energy models  or massive neutrinos in $\Lambda$CDM can be written in this way. To be specific we will exemplify our results,  throughout this work, using the $\Lambda$CDM model and HS $n=1$ F6, F5 and F4 $f(R)$ models; for the latter we present a brief summary in the context of LSS in appendix \ref{app:fRHS}.
One identifies $A(k,t)$ with the most commonly used $\mu(k,t)$ function through
\begin{equation} \label{Ak_Eq}
 A(k,t) = A_0 \mu(k,t), \qquad A_0=4 \pi G \bar{\rho},
\end{equation}
such that these theories can be understood within an effective modification of Newton's constant to $G_{\rm eff} = \mu(k,t) G$. This function has served to  parametrize linear effects of different MG and dark energy models, but it can be obtained directly from a specific theory.

The function $S$ can also have different origins:  In the case of MG models, $S$ comes from the nonlinearities in the  Klein-Gordon-like equations and it is responsible for screening mechanisms that drive theories to  GR at small scales. In MG theories posed in the Jordan frame it appears sourcing the Poisson  equation, as in our eq.~(\ref{PoissonEq}); instead, in theories defined in the Einstein Frame,  such as the symmetrons, $S$ appears as a fifth force sourcing geodesic equation, and in this way it will source the Euler equation through the effective gravitational potential \cite{Brax:2013fna,Aviles:2018qotF}. 
On the other hand, for massive neutrinos in the $\Lambda$CDM model it has a special 
form to account for the nonlinear neutrino density  field \cite{Aviles:2020cax}. Dark energy models can also be accommodated using particular $A(k,t)$ and $S(k)$ functions.

To find $n$-order perturbative solutions, we find $A(k,t)$ analytically (in HS it is given by eq.~\eqref{AktHS}), and assume that $S(\vk)$ can be expanded in a Taylor series in Fourier space as
\begin{equation} \label{s_of_k}
 S(\vk) = 
 \frac{1}{2} \ikk \mathcal{S}^{(2)}(\vk_1,\vk_2) \delta(\vk_1)\delta(\vk_2) + \frac{1}{6} \ikkk \mathcal{S}^{(3)}(\vk_1,\vk_2,\vk_3) \delta(\vk_1)\delta(\vk_2)\delta(\vk_3) + \cdots , 
\end{equation}
whereby the specific  $S^{(n)}(k)$ are determined by a particular theory. Note that if $S$ has a linear piece $S = S^{(1)}\delta + \mO(\delta^2)$, it can be absorbed by $A(k)$. 

\bigskip

We now turn to the fluid equations, considering a CDM fluid element with peculiar velocity
\begin{equation}
 v^i(t) = \frac{d x^i(t)}{d\tau} = a \dot{x}^i(t),
\end{equation}
where $\vx$ is its comoving coordinate, $t$ the cosmic time, and $d\tau = \frac{1}{a}dt$ the conformal time, 
such that the total velocity is $v_T^i=a H x^i + v^i$.  
The fluid equations in the absence of velocity dispersion are
\begin{align}
\partial_t\delta(\vx,t)  + \frac{1}{a}\partial_i \big[(1+\delta)v^i \big] &= 0, \label{FEcontpre}\\  
\partial_t v^i(\vx,t)+ \frac{1}{a} v^j\partial_j v^i  + H v^i  + \frac{1}{a} \partial_i \Phi & =0 \label{FEEulerpre},  
\end{align}
with $\Phi$ the gravitational potential. 
We use the (dimensionless) velocity divergence
\begin{equation} \label{thetadef}
 \theta(\vx,t) = -\frac{\partial_i v^i}{a H f_0 },
\end{equation}
where $f_0(t)$ is an arbitrary function of time that will be fixed to be the logarithmic growth rate at a convenient scale. 
We assume the transverse piece of the velocity is negligible at large scales, hence it is a longitudinal field fully specified by $\theta$, hence the above equation 
can be inverted and $v^i(\vx,t) = - a H f_0 \partial_i  \nabla^{-2} \theta(\vx,t)$. 

In Fourier space the continuity and Euler equations can be written as\footnote{We use the shorthand notations
\begin{equation} \label{int_conv1}
 \ip =  \int \Dk{p},
\end{equation}
and
\begin{equation} \label{int_conv2}
\underset{\vk_{1\cdots n}= \vk}{\int}  = \int \frac{d^3k_1\cdots d^3k_n}{(2\pi)^{3(n-1)}} \dD(\vk-\vk_{1\cdots n}),
\end{equation}
with $\vk_{1\cdots n}= \vk_1 + \cdots + \vk_n$.}
\begin{align}
\frac{1}{H}  \frac{\partial\delta(\vk)}{\partial t} - f_0 \theta(\vk) &=  f_0  \ikk \alpha(\vk_1,\vk_2) \theta(\vk_1) \delta(\vk_2), \label{FScontEq}\\
 \frac{1}{H}  \frac{\partial f_0 \theta(\vk)}{\partial t} + \left( 2+ \frac{\dot{H}}{H^2}\right) f_0 \theta(\vk)
&- \frac{A(k)}{H^{2}} \delta(\vk)  -\frac{S(\vk)}{H^2} =   f_0^2 \ikk \beta(\vk_1,\vk_2) \theta(\vk_1) \theta(\vk_2), \label{FSEulerEq}
\end{align}
with
\begin{equation}
 \alpha(\vk_1,\vk_2) = 1+\frac{\vk_{1}\cdot\vk_2}{k_1^2},  \qquad  \beta(\vk_1,\vk_2) = \frac{k_{12}^2(\vk_1\cdot\vk_2)}{2 k_1^2 k_2^2}. 
\end{equation}

One finds perturbative solutions at the different orders according to eqs.~(\ref{FScontEq}, \ref{FSEulerEq}) and the nonlinear expansion (\ref{s_of_k}).  To linear order one gets
\begin{align}
 \delta^{(1)}(\vk,t) &= D_+(\vk,t)\delta^{(1)}(\vk,t_0), \label{delta1}\\
 \theta^{(1)}(\vk,t) &=  \frac{f(\vk,t)}{f_0} \delta^{(1)}(\vk,t), \label{theta1}
\end{align}
with $D_+$ the fastest growing solution to the equation
\begin{equation}\label{1stOrderEq}
\big( \T - A(k) \big)  D_+(k,t) \equiv \left( \frac{d^2 \,}{d t^2} + 2H\frac{d\,}{d t}- A(k,t) \right) D_+(k,t) = 0,
\end{equation}
where the LHS defines the operator $\T$ \cite{Matsubara:2015ipa}, and
\begin{equation}
 f(k,t) = \frac{d \log D_+(k,t)}{d \log a(t)}
\end{equation}
is the scale and time dependent growth rate. We choose $f_0(t) \equiv f(k=0,t)$, such that at large scales one recovers the linear order solution $\theta=\delta$, valid in $\Lambda$CDM; which is natural since at very large scales many MG theories reduce to GR, at least in the quasi-static approximation.

In SPT, the $n$-th order velocity and density fields are written as weighted convolutions of $n$ linear density fields,
\begin{align} 
 \delta^{(n)} (\vk,t) &= \underset{\vk_{1\cdots n}= \vk}{\int} F_n(\vk_1,\cdots,\vk_n;t) \delta_L(\vk_1,t) \cdots \delta_L(\vk_n,t), \label{deltanK} \\
 \theta^{(n)} (\vk,t) &= \underset{\vk_{1\cdots n}= \vk}{\int} G_n(\vk_1,\cdots,\vk_n;t) \delta_L(\vk_1,t) \cdots \delta_L(\vk_n,t)  \label{thetanK}
\end{align}
with SPT kernels $F_n$ and $G_n$ and we have written explicitly their temporal dependence. 
For linear order, the kernels can be read from eqs.~\eqref{delta1} and \eqref{theta1}, giving
\begin{align}
 F_1(\vk) =1, \qquad
 G_1(\vk) = \frac{f(k)}{f_0}.
\end{align}
Higher order kernels can be found by solving eqs.~(\ref{FScontEq}) and (\ref{FSEulerEq}) iteratively. However, this approach is lengthy, especially for third order kernels, and we found it more efficient to obtain them by means of mappings from LPT known kernels. This is done in appendix \ref{app:Kernels} (see also  \cite{Aviles:2018saf}), where we obtain
\begin{align}
F_2(\vk_1,\vk_2) &= \frac{1}{2} + \frac{3}{14}\A + \left( \frac{1}{2} - \frac{3}{14}\B  \right)   \frac{(\vk_1\cdot\vk_2)^2}{k_1^2 k_2^2}
        + \frac{\vk_1\cdot\vk_2}{2 k_1k_2} \left(\frac{k_2}{k_1} + \frac{k_1}{k_2} \right), \label{F2_kernel}\\
G_2(\vk_1,\vk_2) &= \frac{3\A(f_1+f_2) + 3 \dot{\A}/H }{14 f_0} +
\left(\frac{f_1+f_2}{2 f_0} - \frac{3\B(f_1+f_2) + 3 \dot{\B}/H }{14 f_0}\right) \frac{(\vk_1\cdot\vk_2)^2}{k_1^2 k_2^2} \nonumber\\
&\quad + \frac{\vk_1\cdot\vk_2}{2 k_1k_2} \left( \frac{f_2}{f_0}\frac{k_2}{k_1} + \frac{f_1}{f_0}\frac{k_1}{k_2} \right), \label{G2_kernel}
\end{align}
where $f_{1,2} = f(\vk_{1,2})$.  The functions $\mA$ and $\mB$ are scale and time dependent:
\begin{equation} \label{AandBdef}
 \mA(\vk_1,\vk_2,t) = \frac{7 D^{(2)}_{\mA}(\vk_1,\vk_2,t)}{3 D_{+}(k_1,t)D_{+}(k_2,t)}, 
 \qquad \mB(\vk_1,\vk_2,t) = \frac{7 D^{(2)}_{\mB}(\vk_1,\vk_2,t)}{3 D_{+}(k_1,t)D_{+}(k_2,t)},
\end{equation}
with second order growth functions $D^{(2)}_{\mA,\mB}$ are solutions of the linear second order differential equations \cite{Aviles:2017aor}:
\begin{align}
\big(\T - A(k)\big)D^{(2)}_{\mA} &= \Bigg[A(k) + (A(k)-A(k_1))\frac{\vk_1\cdot\vk_2}{k_2^2} + (A(k)-A(k_2))\frac{\vk_1\cdot\vk_2}{k_1^2} \nonumber\\
             &     \qquad   +  \mathcal{S}^{(2)}(\vk_1,\vk_2) \Bigg]  D_{+}(k_1)D_{+}(k_2), \label{DAeveq} \\
\big(\T - A(k)\big)D^{(2)}_{\mB} &= \Big[A(k_1) + A(k_2) - A(k) \Big]  D_{+}(k_1)D_{+}(k_2), \label{DBeveq}
\end{align}
with appropriate initial conditions to project out the homogeneous first order growth functions, i.e., the kernel of the linear differential operator $(\T - A(k))$. Hereafter, when it does not lead to confusion, we omit to write the time dependence of these functions. 

Our kernel expressions show where and how the different growth solutions enter to distinguish kernels in different cosmologies. Solutions so far can be specified to the known EdS and $\Lambda$CDM cases. In fact for $\Lambda$CDM, $A(k,t)=A_0 = \frac{3}{2}\Omega_m H^2$, and one obtains 
\begin{equation}
D^{(2)\text{$\Lambda$CDM}}_{\mA,\mB}(t) = \frac{3}{7}D_+^2(t) + \frac{4}{7}\left(\T - \frac{3}{2}\Omega_m H^2 \right)^{-1} \left[\frac{3}{2}\Omega_m H^2\left( 1 - \frac{f^2}{\Omega_m }\right) \right],
\end{equation}
such that $\mA^{\Lambda{\rm CDM}}=\mB^{\Lambda{\rm CDM}}$ are only time dependent and close to unity.
For EdS,  $f=\Omega_m=1$, and  $\mA^{\rm EdS} =\mB^{\rm EdS} = 1$, recovering the EdS standard kernels. 

The expressions for $F_3$ and $G_3$ are large and not displayed here, these are given in Appendix \ref{app:Kernels}, eqs.~(\ref{F3_kernel}) and (\ref{G3_kernel}). Having found the kernels up to third order, we can now construct spectra and cross-spectra of the velocity and density fields. At 1-loop in PT one can use the general expressions
\begin{equation}\label{Pab}
P_{ab}(k) = P^L_{ab}(k) +P_{ab}^{22}(k) + P_{ab}^{13}(k),  
\end{equation}
where $a$ and $b$ refer to $\delta$ or $\theta$ fields, and linear power spectra $P^L_{ab}(k)$  
 \begin{equation} \label{Plinear}
 P^L_{\delta\delta}(k) \equiv P_L(k) , \quad
 P^L_{\delta\theta}(k) =  \frac{f(k)}{f_0} P_L(k), \quad
 P^L_{\theta\theta}(k) =  \left(\frac{f(k)}{f_0} \right)^2 P_L(k),
 \end{equation}
and leading nonlinear contributions 
\begin{align}
P_{\delta\delta}^{22}(k) &=   2 \ip \big[F_2(\vp,\vk-\vp)\big]^2 P_L(p) P_L(|\vk -\vp|), \\
P_{\delta\theta}^{22}(k) &=   2 \ip F_2(\vp,\vk-\vp)G_2(\vp,\vk-\vp) P_L(p) P_L(|\vk -\vp|), \\
P_{\theta\theta}^{22}(k) &=   2 \ip \big[G_2(\vp,\vk-\vp)\big]^2 P_L(p) P_L(|\vk -\vp|), \\
P_{\delta\delta}^{13}(k) &=   6  P_L(k) \ip F_3(\vk,-\vp, \vp) P_L(p),\\
P_{\delta\theta}^{13}(k) &=   3  P_L(k) \ip \big[F_3(\vk,-\vp, \vp)G_1(\vk) + G_3(\vk,-\vp, \vp) \big] P_L(p),\\
P_{\theta\theta}^{13}(k) &=   6  P_L(k) \ip G_3(\vk,-\vp, \vp)G_1(\vk) P_L(p). 
\end{align}

The above expressions are valid for any cosmological model, with the same expressions for power spectra and kernels, the difference being the corresponding growth solutions, and hence the kernels, for each cosmological model. 
In figure \ref{fig:PddPdtPtt} we present our results at 1-loop for the $\Lambda$CDM and the F6 HS models, showing their ratios to linear power spectra for cases $ab= \delta \delta, \delta \theta, \theta \theta$; similar results are presented in \cite{Bose:2017dtl}, 
in which the numerical method of \cite{Taruya:2016jdt} was used to compute the kernels. 
We note the differences in the gravity models are more pronounced for spectra containing velocity fields, particularly $P_{\theta\theta}$. This is a because even in linear theory velocity fields receive a boost given by the $\frac{f(k)}{f_0}$ factors and the 
 relation $\theta^{(1)}=\delta^{(1)}$ does not hold, as it does in $\Lambda$CDM. 
 Instead, linear fields are related by eq.~(\ref{theta1}), 
and as a consequence, the differences among models are more pronounced in redshift space than in real space statistics.

\begin{figure}[tbp]
\centering 
\includegraphics[width=.45\textwidth]{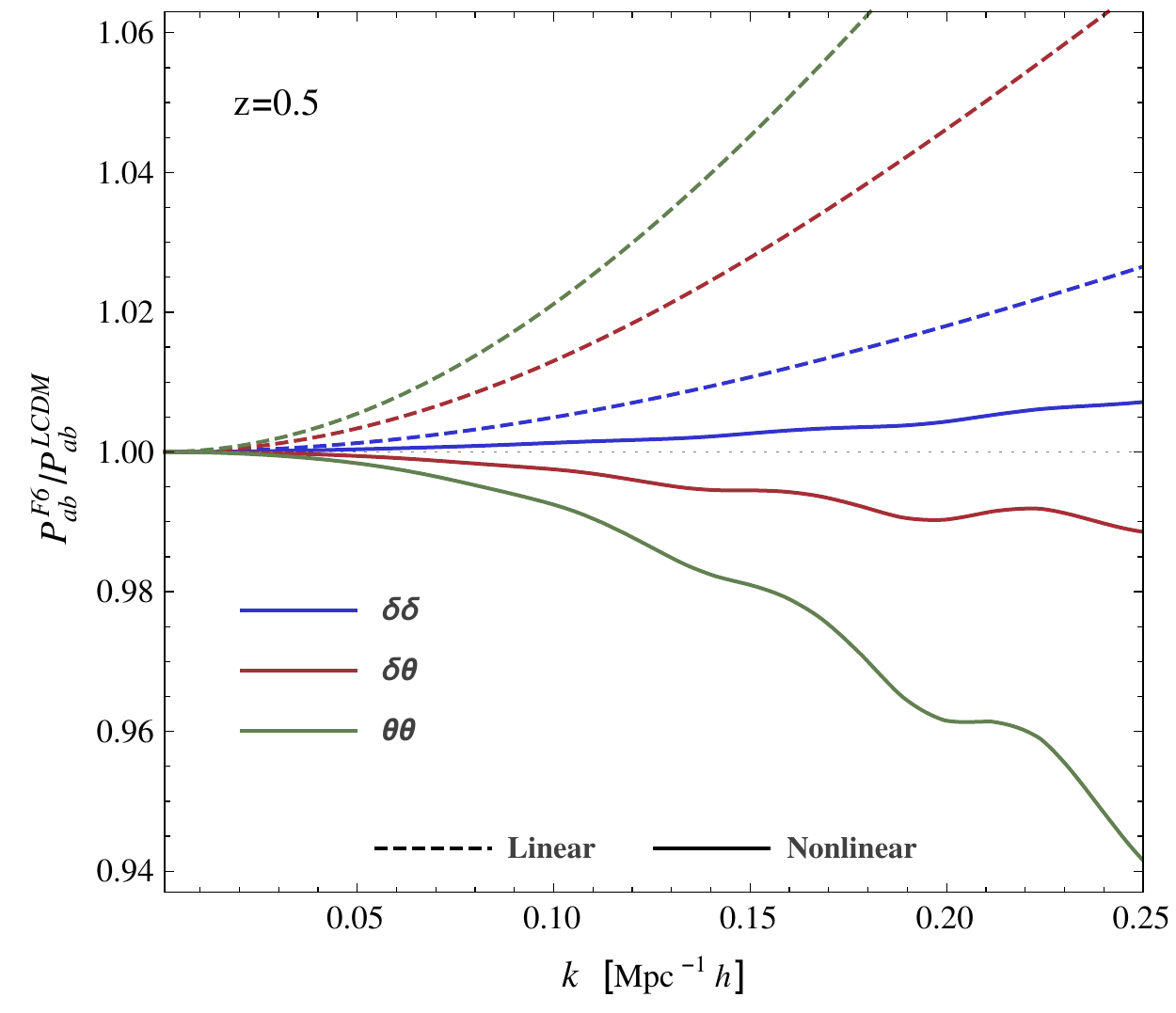}
\includegraphics[width=.45\textwidth]{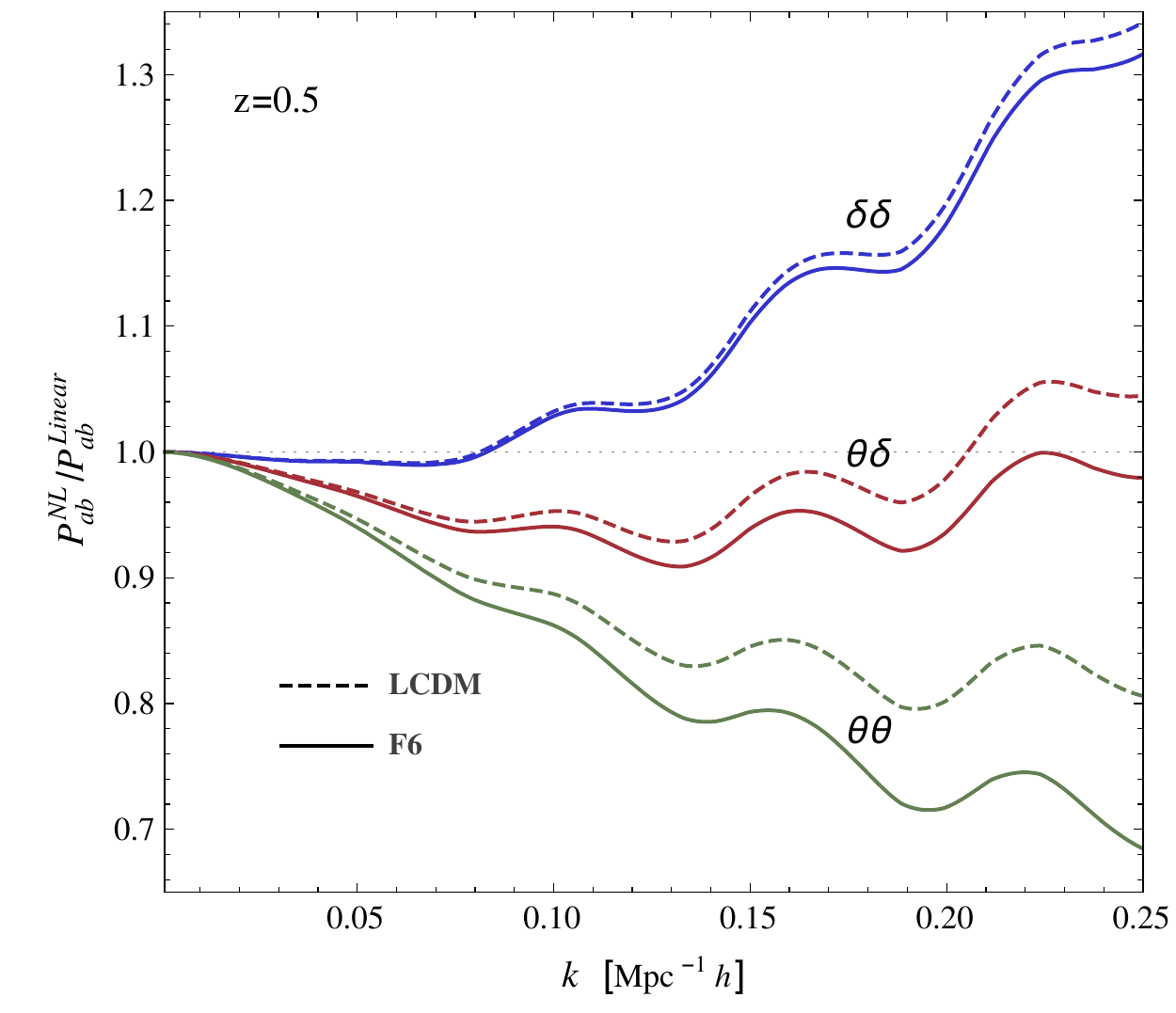}
\caption{\label{fig:PddPdtPtt} $P_{ab}(k)$ linear and 1-loop power spectra for cases $ab= \delta \delta, \delta \theta, \theta \theta$ at $z=0.5$ computed using eq.~\eqref{Pab}. {\it Left panel:} ratios of F6 to $\Lambda$CDM linear (dashed lines) and non-linear (solid lines) spectra. {\it Right panel:} ratios of non-linear to linear power spectra for $\Lambda$CDM (dashed lines) and F6 (solid lines).}
\end{figure}

\end{section}

\begin{section}{Bias expansion}\label{sect:BiasExp}
 
This section closely follows the work of McDonald and Roy \cite{McDonald:2009dh}, see also \cite{McDonald:2006mx,Saito:2014qha}, slightly adapted to account for the effects of cosmologies beyond $\Lambda$CDM.
It is well known that for theories with extra degrees of freedom even linear bias becomes scale-dependent, for example in MG \cite{Hui:2007zh} or in the presence of massive neutrinos, particularly when biasing the total matter field \cite{Villaescusa-Navarro:2013pva,Castorina:2013wga,LoVerde:2014pxa,Vagnozzi:2018pwo,Banerjee:2019omr}. 
Our first assumption is the existence of higher-curvature bias operators $\nabla^2\delta$,  $\nabla^4\delta$, ...,  that effectively encapsulate the effects 
of a function $A(k)$ that is scale-dependent; see for example section 8 of \cite{Desjacques:2016bnm} and \cite{Aviles:2018saf}.
We expand the tracers' density in terms of a set of operators, including the leading curvature operators, labeled with ``$m$'' to make reference to matter fields, as
\begin{align}\label{biasexp}
\delta(\vx) &= c_\delta \delta_m + c_{\nabla^2\delta} \nabla^2 \delta_m + \frac{1}{2}c_{\delta^2} \delta_m^2 + \frac{1}{2}c_{s^2}s^2     \nonumber\\
   &\quad  + \frac{1}{6}c_{\delta^3} \delta_m^3 +\frac{1}{2}c_{\delta s^2} \delta s^2  + c_\psi \psi + c_{st}s t  + \frac{1}{2} c_{s^3} s^3, \\
\theta(\vx) &=     \theta_m +  c_{\nabla^2\theta} \nabla^2 \theta_m.     
\end{align}
We note however that the above bias expansion is not complete since the linear growth function cannot be factorized in time and scale dependent pieces. But by expanding $A(k)$ in powers of $k^2$ we can partially tame the new scale introduced in beyond $\Lambda$CDM models with curvature operators \cite{Desjacques:2016bnm}. In the MG models studied here, the parameter expansion is the inverse squared of the mass of the associated scalar field (that can be identified from eq.~\eqref{AktHS} as $m^2= M_1/3$), so we expect that our modeling better fits for length scales larger than the inverse of this mass \cite{Desjacques:2016bnm,Aviles:2018saf}.

In eq.~\eqref{biasexp}, we have used the standard definitions $s^2 = s_{ij}s_{ij}, \, st = s_{ij}t_{ij},  \, s^3=s_{ij}s_{jk}s_{ki}$, 
\begin{align}
s_{ij}(\vk) &= \left( \frac{k_i k_j}{k^2} - \frac{1}{3}\delta_{ij} \right) \delta_m(\vk),
\quad t_{ij}(\vk) = \left( \frac{k_i k_j}{k^2} - \frac{1}{3} \delta_{ij} \right) \eta(\vk),
\end{align}
and
\begin{align}
\eta(\vk) &=\theta_m(\vk) - \frac{f(k)}{f_0}  \delta_m(\vk), \label{etadef}
\end{align}
such that in virtue of eq.~\eqref{theta1}, $\eta$ vanishes at linear order in PT. This also means that $t_{ij}$ is second order and hence $st$ is third order. 
We further define
\begin{equation}
 \psi(\vk) = \eta(\vk) +\frac{f(k)}{f_0}\left(- \frac{2}{7} s^2(\vk) + \frac{4}{21} \delta_m^2(\vk) \right).
\end{equation}
For EdS kernels, this operator is third order in PT. But even using the $\Lambda$CDM kernels it is still second order, as we will see below.

We further have considered an operator $\nabla^2 \theta_m$, to be consistent with the inclusion of spatial derivatives of linear overdensities; see section 2.7 in  \cite{Desjacques:2018pfv}.  
In the following, we will only keep up to $\nabla^2 \theta_m$ and $\nabla^2 \delta_m$ derivatives and consider them only at the linear level. We do this for simplicity since the addition of higher than second order derivatives complicates 
largely the algebra and give small contributions; also, although curvature operators are linear, their contributions are of the similar magnitude as those coming from nonlinear operators. Moreover, we will not rely on curvature large bias coefficients 
for fitting purposes, as for modeling the FoG. Such large contributions, are plausible to come from EFT corrections, which will be introduced in section \ref{counterterms}, and are somewhat degenerated with $\nabla^2$ biasing, 
but with a different origin. 
Hence, we include $\nabla^2 \theta_m$ and $\nabla^2 \delta_m$ biasing operators mainly for consistency and renormalization purposes, and will treat their bias 
coefficients as small. Otherwise we can always write the tracer PS as an arbitrary polynomial in $k^2$ times the matter PS and fit to simulations as much as we want. 

\begin{figure}[tbp]
\centering 
\includegraphics[width=.49\textwidth]{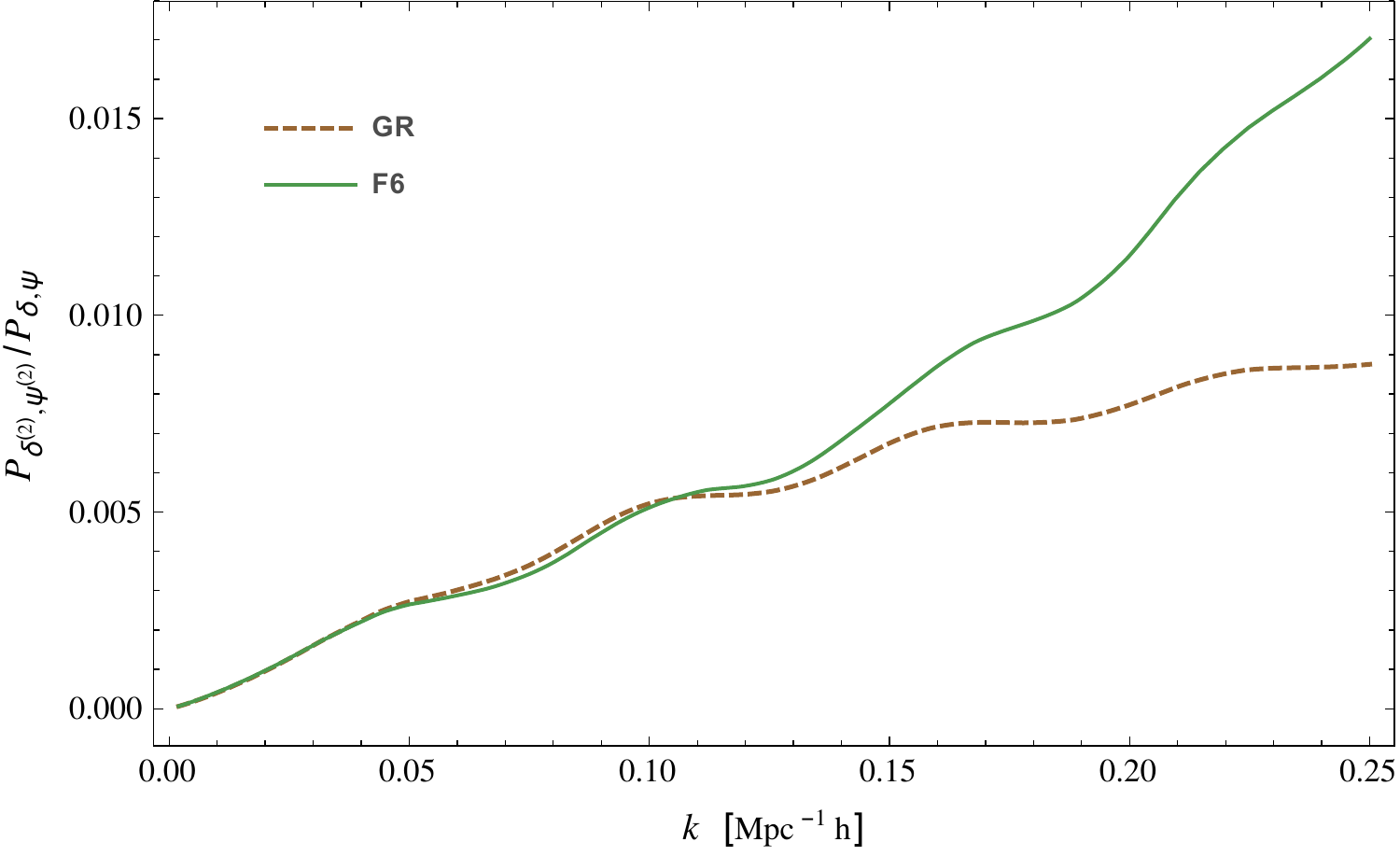}
\caption{\label{fig:Pdpsi} Ratio of $P_{\delta^{(2)}\psi^{(2)}}(k)$ to  $P_{\delta\psi}(k)=P_{\delta^{(2)}\psi^{(2)}}(k)+ P_{\delta^{(1)}\psi^{(3)}}(k)$ for $\Lambda$CDM and F6 models at redshift $z=0.5$. 
This figure shows that the cross contribution of $\psi^{(2)}$ is very small
at the scales of interest in PT, about $0.5\, \%$ at $k\sim 0.1 \, h/\text{Mpc}$ for both gravity models.}
\end{figure}

The bias expansion in eq.~(\ref{biasexp}) can be written as $\delta(\vx) = \sum c_\mathcal{O} \mathcal{O}(\vx)$,
with each operator written in Fourier space at first, second, and third order in PT as 
\begin{align}
 \mathcal{O}^{(1)}(\vk) &=       K_\mathcal{O}^{(1)}(\vk)               \delta_m^{(1)}(\vk), \\
 \mathcal{O}^{(2)}(\vk) &= \ikk  K_\mathcal{O}^{(2)}(\vk_1,\vk_2)       \delta_m^{(1)}(\vk_1)\delta_m^{(1)}(\vk_2), \\
 \mathcal{O}^{(3)}(\vk) &= \ikkk K_\mathcal{O}^{(3)}(\vk_1,\vk_2,\vk_3) \delta_m^{(1)}(\vk_1)\delta_m^{(1)}(\vk_2) \delta_m^{(1)}(\vk_3).
\end{align}

We now proceed to compute explicitly the $K_\mathcal{O}$ kernels for each bias operator order by order: 
\begin{itemize}
\item The first order contributions are trivially 
\begin{align}
\delta^{(1)}(\vk) &= (c_\delta - c_{\nabla^2\delta} k^2)\delta_m^{(1)}(\vk), \,\, {\rm and} \\ 
\theta^{(1)}(\vk) &= (1 - c_{\nabla^2\theta} k^2)\theta_m^{(1)}(\vk). 
\end{align}

\item To second order,
\begin{align}
 K_{\delta}^{(2)}(\vk_1,\vk_2)     &= F_2(\vk_1,\vk_2),  \\
 K_{\delta^2}^{(2)}(\vk_1,\vk_2)   &= 1,  \label{kerneld22} \\
 K_{s^2}^{(2)}(\vk_1,\vk_2)        &= S_2(\vk_1,\vk_2), \label{kernels22} \\
 K_{\psi}^{(2)}(\vk_1,\vk_2)       &= G_2(\vk_1,\vk_2) - G_1(\vk_{12})\left[F_2(\vk_1,\vk_2) + \frac{2}{7}S_2(\vk_1,\vk_2) - \frac{4}{21}\right],  \label{kernelpsi2}
\end{align}
with
\begin{equation}
 S_2(\vk_1,\vk_2) = \frac{(\vk_1\cdot\vk_2)^2}{k_1^2 k_2^2} -\frac{1}{3}.
\end{equation}
Note that $G_1^\text{EdS}=1$, and $\frac{2}{7} S_2(\vk,\vp) -\frac{4}{21}  =  G_2^\text{EdS}(\vk,\vp)-F_2^\text{EdS}(\vk,\vp)$,
and hence for EdS kernels the operator $\psi$ vanishes at second order. However, if $\Lambda$CDM kernels are used, we have instead 
\begin{equation}\label{K2psiLCDM}
K^{(2),\text{$\Lambda$CDM}}_\psi(\vk_1,\vk_2)= \frac{3}{14}\left(\mA^\text{$\Lambda$CDM} -1 -\frac{\dot{\mA}^\text{$\Lambda$CDM}}{f_0H} \right)\left( 1-\frac{(\vk_1 \cdot \vk_2)^2}{k_1^2 k_2^2} \right), 
\end{equation}
so the operator $\psi$ vanishes only at first order in PT
as we commented above. Despite $\psi$ being second order, we will treat it as third order because its influence is indeed small, as can be seen in figure \ref{fig:Pdpsi}, where we show the cross-power spectrum 
$P_{\delta^{(2)}\psi^{(2)}}(k) = \langle \delta^{(2)}(\vk)\psi^{(2)}(\vk') \rangle'$ divided by the full spectrum $P_{\delta\psi}(k) = \langle \delta(\vk)\psi(\vk') \rangle'$, which also receives a contribution from $\psi$ at third order, $\langle \delta^{(1)}(\vk)\psi^{(3)}(\vk') \rangle'$. 
We do this for $\Lambda$CDM and F6 noticing that the error introduced by neglecting the $P_{\delta^{(2)}\psi^{(2)}}(k)$ piece is about the $1\, \%$ or smaller for $k \lesssim 0.2 \, h \text{Mpc}^{-1}$ for both models. This suggests to make $\psi^{(2)}=0$, as we do in the following. 
In $\Lambda$CDM, regardless of $\psi$ being second order, one can absorb the residual given by eq.~\eqref{K2psiLCDM} in the bias parameters $c_{s^2}$ and $c_{\delta^2}$ in virtue of eqs.~\eqref{kerneld22} and \eqref{kernels22}. Alternatively, $\psi$ can be easily redefined to be third order, as in \cite{Donath:2020abv}.

\item Third order kernels are
\begin{align}
 K_{\delta}^{(3)}(\vk_1,\vk_2,\vk_3)     &= F_3(\vk_1,\vk_2,\vk_3),  \\
 K_{\delta^2}^{(3)}(\vk_1,\vk_2,\vk_3)   &= 2  F_2(\vk_2,\vk_3),  \\
 K_{\delta^3}^{(3)}(\vk_1,\vk_2,\vk_3)   &= 1,  \\
 K_{s^2}^{(3)}(\vk_1,\vk_2,\vk_3)        &= 2  S_2(\vk_1,\vk_2+\vk_3) F_2(\vk_2,\vk_3), \\
 K_{\delta s^2}^{(3)}(\vk_1,\vk_2,\vk_3)        &=  S_2(\vk_1,\vk_2), \\
 K_{s^3}^{(3)}(\vk_1,\vk_2,\vk_3)        &=   S_3(\vk_1,\vk_2,\vk_3),\\
 K_{st}^{(3)}(\vk_1,\vk_2,\vk_3)         &=   S_2(\vk_1,\vk_2+\vk_3) \left( G_2(\vk_2,\vk_3) - \frac{f(|\vk_{23}|)}{f_0} F_2(\vk_2,\vk_3) \right),\\
 K_{\psi}^{(3)}(\vk_1,\vk_2,\vk_3)       &=   G_3(\vk_1,\vk_2,\vk_3) - G_1(\vk_{123}) F_3(\vk_1,\vk_2,\vk_3)   \nonumber\\
                                         &\quad  - 2  G_1(\vk_{123}) F_2(\vk_2,\vk_3)\left( \frac{2}{7} S_2(\vk_1,\vk_2+\vk_3) -\frac{4}{21} \right),
\end{align}
with
\begin{equation}
 S_3(\vk_1,\vk_2,\vk_3) = \Bigg( \frac{k_1^ik_1^j}{k_1^2} - \frac{1}{3}\delta_{ij} \Bigg)\Bigg( \frac{k_2^jk_2^k}{k_2^2} - \frac{1}{3}\delta_{jk} \Bigg)\Bigg( \frac{k_3^kk_3^i}{k_3^2} - \frac{1}{3}\delta_{ki} \Bigg).
\end{equation}

Note also that $\mA(\vp,-\vp)=\mB(\vp,-\vp)$, as implied from eqs.~\eqref{DAeveq} and \eqref{DBeveq}, and by means that the source vanishes at very large scales $S(\vk = 0)=0$. 
Therefore, $F_2(\vp,-\vp) = 0$;  more generally, we have  $F_2(\vp,\vq\rightarrow - \vp)  \propto |\vp+\vq|^2$, which is a consequence of momentum conservation.

\end{itemize}

\begin{subsection}{Third order bias and renormalization}

It is standard to define the function \cite{McDonald:2009dh,Saito:2014qha}
\begin{equation} \label{sigma23EdS}
 \sigma^{2}_{3}(k) = \frac{105}{16} \ip P_L(p) \left[ S_2(\vp,\vk-\vp)\left(\frac{2}{7}S_2(-\vp,\vk)  -\frac{4}{21} \right) + \frac{8}{63} \right],
\end{equation}
that serves to collect some of the biasing terms constructed out from operators at third order in PT. Here, we outline the procedure to do that. First consider the linear matter overdensity field $\delta_m^{(1)}$ correlated with operators $st$, $s^{2}$ and $\psi$, yielding the spectra $P_{\delta^{(1)}_m,st^{(3)}} = \langle \delta^{(1)}_m(\vk)st^{(3)}(\vk') \rangle'$, $P_{\delta^{(1)}_m,s^{2(3)}} = \langle \delta^{(1)}_m(\vk)s^{2(3)}(\vk') \rangle'$, and $P_{\delta^{(1)}_m,\psi^{(3)}} = \langle \delta^{(1)}_m(\vk)\psi^{(3)}(\vk') \rangle'$, contributing  to the matter-tracer cross-power spectrum as 
\begin{align}
 P(k) &\ni \frac{1}{2}c_{s^2} P_{\delta^{(1)}_m,s^{2(3)}}(k) + c_{st} P_{\delta^{(1)}_m,st^{(3)}}(k) + c_\psi P_{\delta^{(1)}_m,\psi^{(3)}}(k) \nonumber\\
 &= c_s^2 P_L(k) \ip P_L(p) \big[ K_{s^2}^{(3)}(\vp,-\vp,\vk) -K_{s^2}^{(3)}(\vp,-\vp,0) \big] \nonumber\\
 &+ 2 c_{st} P_L(k) \ip P_L(p) \big[K_{st}^{(3)}(\vp,-\vp,\vk) - K_{st}^{(3)}(\vp,-\vp,0)\big] \nonumber\\
 &+ 2 c_{\psi} P_L(k) \ip P_L(p) \Bigg\{\frac{3}{2} \big[ G_3(\vk,-\vp,\vp) - \frac{f(k)}{f_0}  F_3(\vk,-\vp,\vp) \big] \nonumber\\ 
 &\quad   \qquad - 2 \frac{f(k)}{f_0} F_2(-\vp,\vk)\left( \frac{2}{7} S(\vp,\vk-\vp) -\frac{4}{21} \right) \Bigg\}.
\end{align}
A direct computation does not give the terms  $K_{s^2}^{(3)}(\vp,-\vp,0)$ and $K_{st}^{(3)}(\vp,-\vp,0)$. These should be included in order to make the large scales insensitive to the small scales, otherwise
when these integrals are regularized, for example  by a sharp cutoff at $p=\Lambda\gg 1$, they go as $\Lambda^{3+n}$ for a scale invariant PS $P_L(k\gg1) \propto k^{n}$, 
becoming UV divergent for a typical PS.  More explicitly, these terms are 
\begin{align}
\ip P_L(p) K_{s^2}^{(3)}(\vp,-\vp,0) &=  \ip P_L(p) \left[ \frac{8}{9} + \frac{2}{21}\big(3\mA(0,\vp)-\mB(0,\vp)\big) \right]  = \frac{68}{63} \mC_{s^2} \sigma^2  \\
\ip P_L(p) K_{st}^{(3)}(\vp,-\vp,0)  &=  \ip P_L(p)  \Big\{ \frac{1}{9} -\frac{3 f(p)}{9 f_0} + \frac{f(p) \big(3\mA(0,\vp)-\mB(0,\vp)\big)}{21 f_0} \nonumber\\
                                          &\quad \qquad +  \frac{3 \dot{\mA}(0,\vp)-\dot{\mB}(0,\vp)}{21 f_0 H}  \Big\} = -\frac{8}{63} \mC_{st} \sigma^2.
\end{align}
For large values of the inner integration moment $\vp$, both
$\mA(0,\vp)$ and $\mB(0,\vp)$ converge (this is for the case in which function $\mu(k)$ is bounded), hence $\mC_{s^2}$ and $\mC_{st}$ are constants.
For $\Lambda$CDM kernels, $\mC_{s^2}=\frac{63}{68}\left[ \frac{8}{9} + \frac{4}{21}\mA^\text{$\Lambda$CDM} \right]$  
and $\mC_{st} = \frac{63}{8}\left[ \frac{2}{9} -\frac{2}{21}\big( \mA^\text{$\Lambda$CDM} + \dot{\mA}^\text{$\Lambda$CDM}/f_0H \big) \right]$, and both reducing to 1 for EdS kernels. These $\vk\rightarrow 0$ terms contribute
to the PS as $\propto \sigma^2 P_L(k)$, so they can be absorbed by the linear bias parameter by redefining it as $c_\delta \rightarrow c_\delta  - \frac{68}{63} \mC_{s^2} \sigma^2 + \frac{8}{63} \mC_{st} \sigma^2$.

We introduce the functions 
\begin{align}
  \sigma_{3, s^2}^2(k) &=  -\frac{21}{16} \ip P_L(p) \big[ K_{s^2}^{(3)}(\vp,-\vp,\vk) -K_{s^2}^{(3)}(\vp,-\vp,0) \big], \\
  \sigma_{3, st}^2(k) &=  \frac{105}{16} \ip P_L(p) \big[ K_{st}^{(3)}(\vp,-\vp,\vk) -K_{st}^{(3)}(\vp,-\vp,0) \big], \\
  \sigma_{3, \psi}^2(k) &=  \frac{2205}{256} \ip P_L(p) \Bigg\{\frac{3}{2} \big[ G_3(\vk,-\vp,\vp) - \frac{f(k)}{f_0}  F_3(\vk,-\vp,\vp) \big] \nonumber\\ 
   &\quad   \qquad - 2 \frac{f(k)}{f_0} F_2(-\vp,\vk)\left( \frac{2}{7} S_2(\vp,\vk-\vp) -\frac{4}{21} \right) \Bigg\}. 
\end{align}
For EdS kernels, these three functions are identical to $\sigma^2_3(k)$, given above in eq.~(\ref{sigma23EdS}), and hence all the biasing terms involving operators $s^2$, $st$ and $\psi$ at third order can be grouped into a single one, contributing to the PS as $b_{3nl} \sigma^2_3(k) P_L(k)$.\footnote{There are other three spectra of the form $\langle \delta^{(1)}_m \mO^{(3)}\rangle$ involving an operator at third order: $\langle \delta(\vk) s^{3}(\vk')\rangle' = 0$, so $s^3$ does not contribute to 1-loop PS; $\langle \delta(\vk) \delta^{3}(\vk') \rangle' \propto \sigma^2 P_L$, so it is UV sensitive but absorbed by linear bias parameter; and $\langle \delta(\vk) \delta^{2(3)}(\vk') \rangle'$, which is treated separately and contributes to the term $P_{b_1b_2}$ in eq.~\eqref{PddTNL} below.} In $\Lambda$CDM, the $k$-dependence of these functions is the same as well, proportional to $\sigma^2_3(k)$, differing only by multiplicative constants, which are irrelevant because they can be absorbed by their own bias parameters. In generalized cosmologies, instead, not only the normalization is different, but also the $k$-dependence. At very large scales the $\sigma^2_{3, \{ s^2,st,\psi \}}(k)$ functions have the same $k$-dependence, but they differ significantly at smaller, but yet linear scales. In figure \ref{fig:sigma32} we plot these functions for the F6 model at $z=0.5$ normalized to have the same large scales values and divided by the EdS $\sigma^{2}_{3}(k)$, showing significant differences between them even at linear scales.  The most straightforward route we can follow is to consider the three contributions separately with three different bias parameters $c_{s^2}$, $c_{st}$, and $c_\psi$. However, we note that in general we can approximate
\begin{equation} \label{sigma23approx}
 \sigma^2_{3, \{ s^2,st,\psi \}}(k) \approx  \sigma^2_3(k) - \alpha_{1,\{ s^2,st,\psi \}} k^2 - \alpha_{2,\{ s^2,st,\psi \}  } k^4 + \cdots
\end{equation}
as much as we desire over a given finite interval, and absorb $\alpha_{1,\{ s^2,st,\psi \}} k^2$, $\alpha_{2,\{ s^2,st,\psi \}} k^4$, $\dots$ into the higher-order curvature biases $c_{\nabla^2\delta}$, $c_{\nabla^4\delta}$, $\dots$.  Then, we write
\begin{align} \label{bnlps}
 \frac{1}{2}c_{s^2} P_{\delta^{(1)}_m,s^{2(3)}}(k) + c_{st} P_{\delta^{(1)}_m,st^{(3)}}(k) + c_\psi P_{\delta^{(1)}_m,\psi^{(3)}}(k) = b_{3nl} \sigma^2_3(k) P_L(k) 
\end{align}
with 
\begin{equation} \label{bnl}
 b_{3nl} =  \frac{32}{105} c_{st} -\frac{16}{21} c_{s^2} + \frac{512}{2205} c_\psi, 
\end{equation}
together with the redefinitions
\begin{align}
c_\delta &\longrightarrow  c_\delta  - \frac{68}{63} \mC_{s^2} \sigma^2 + \frac{16}{63} \mC_{st} \sigma^2, \label{cdbnl}\\
c_{\nabla^{2m}\delta} &\longrightarrow  c_{\nabla^{2m}\delta} -      \frac{32}{105} c_{st}\alpha_{m,st} +\frac{16}{21} c_{s^2}\alpha_{m,s^2} - \frac{512}{2205} c_\psi \alpha_{m,\psi}. \label{cnbnl}
\end{align}
In figure \ref{fig:sigma32} we also show the curves for these functions once corrected by a polynomial in $k^2$ that we take linear, i.e., we plot (dashed lines) $\sigma^2_{3, \{ s^2,st,\psi \}}(k) + \alpha_{1,\{ s^2,st,\psi \}} k^2$ divided by  
$\sigma^2_3(k)$, showing an agreement of less than $2$-$3\%$ over the interval $k\in \{0,0.2\}$, covering the mild non-linear scales reached by PT. Clearly as higher degree polynomials are considered, a better agreement we can
get; however, this requires to add the corresponding higher-order bias operators into the theory. Here, we opted to include only the first curvature bias $\nabla^2 \delta_m$.

\begin{figure}[tbp]
\centering 
\includegraphics[width=.69\textwidth]{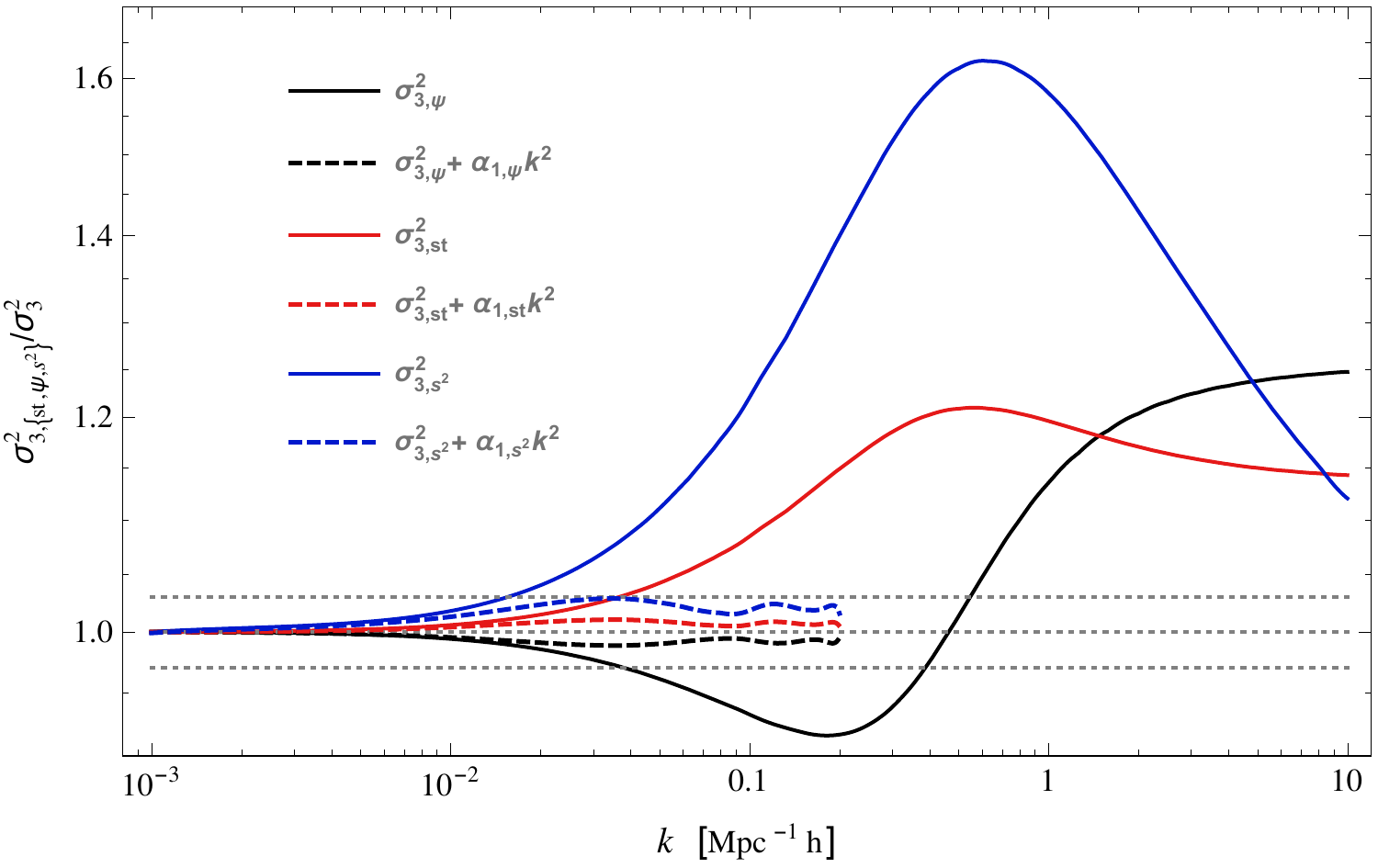}
\caption{\label{fig:sigma32} $\sigma^2_{3}$ functions. Solid lines show the ratio of biasing functions $\sigma^2_{3, \{ s^2,st,\psi \}}(k)$ in the F6 model with respect to $\sigma^{2 \,\ \rm EdS}_{3}$ given by eq.~\eqref{sigma23EdS}. Dashed lines show the same as solid's, but now correcting the functions with a quadratic term $k^2$ that becomes absorbed by higher curvature bias.}
\end{figure}

The computation of power spectra involving a linear matter velocity field is almost identical since $\langle \theta_m^{(1)}(\vk)\mO^{(3)}(\vk') \rangle=\frac{f(k)}{f_0}\langle \delta_m^{(1)}(\vk)\mO^{(3)}(\vk') \rangle$. Hence, we only have to keep track of 
the $f(k)/f_0$ factors: instead of eq.~(\ref{bnlps}), we obtain 
\begin{align} \label{bnlpst}
 \frac{1}{2}c_{s^2} P_{\theta^{(1)}_m,s^{2(3)}}(k) + c_{st} P_{\theta^{(1)}_m,st^{(3)}}(k) + c_\psi P_{\theta^{(1)}_m,\psi^{(3)}}(k) = b_{3nl} \sigma^2_3(k) \frac{f(k)}{f_0}P_L(k).
\end{align}
and the polynomials in $k^2$ should be absorbed by higher-order velocity derivative bias parameters
\begin{align}
c_{\nabla^{2}\theta} &\longrightarrow  c_{\nabla^{2}\theta} -      \frac{32}{105} c_{st}\alpha_{1,st} +\frac{16}{21} c_{s^2}\alpha_{1,s^2} - \frac{512}{2205} c_\psi \alpha_{1,\psi}. \label{cntbnl}
\end{align}

We emphasize that collecting some of the biasing terms into $\sigma^2_3(k)$ is not necessary on theoretical grounds, but is adopted here
in order to keep as less bias parameters as possible. 

The above treatment completes the biasing formalism adopted in this work, valid for $\Lambda$CDM and alternative cosmological models.

\end{subsection}

\begin{subsection}{Tracers one-dimensional spectra}\label{PS_1dim}
We now compute the PS for biased tracers using the power spectra expressions at the end of section \ref{sect:basic_model}, but for tracers developed in the previous subsection.  The spectra $P_{\delta\delta}$, $P_{\delta\theta}$, and $P_{\theta\theta}$ are computed using the standard tools developed in \cite{McDonald:2009dh}, see also \cite{Saito:2014qha}, obtaining 

\begin{align}
 P_{\delta\delta}(k) &= (b_1 -  b_{\nabla^2\delta}k^2)^2 P_L(k) + b_1^2 P^\text{loop}_{m,\delta\delta}(k)  + 2 b_1 b_2 P_{b_1b_2}(k) + 2 b_1 b_{s^2} P_{b_1b_{s^2}}(k) \nonumber\\
                     &\quad + b_2^2 P_{b_2^2}(k) + 2 b_2 b_{s^2} P_{b_2 b_{s^2}}(k) + b_{s^2}^2 P_{b_{s^2}^2}(k) + 2 b_1 b_{3nl} \sigma^2_3(k) P_L(k),  \label{PddTNL}\\
 P_{\delta\theta}(k) &=  (b_1 - b_{\nabla^2\delta} k^2 )(1 -  b_{\nabla^2\theta} k^2)\frac{f(k)}{f_0}P_L(k) + b_1 P^\text{loop}_{m,\delta\theta}(k) + b_2 P_{b_2,\theta}(k) \nonumber\\
                     &\quad    + b_{s^2} P_{b_{s^2},\theta}(k) + b_{3nl} \sigma^2_3(k) \frac{f(k)}{f_0} P_L(k), \label{PdtTNL}\\
 P_{\theta\theta}(k) &=  (1 - b_{\nabla^2\theta}k^2 )^2\left(\frac{f(k)}{f_0}\right)^2 P_L(k) + P^\text{loop}_{m,\theta\theta}(k),  \label{PttTNL}                 
\end{align}
with
\begin{align}
 P_{b_1b_2}(k)     &= \ikk F_2(\vk_1,\vk_2) P_L(\vk_1)P_L(\vk_2), \\
 P_{b_1b_{s^2}}(k) &= \ikk F_2(\vk_1,\vk_2) S_2(\vk_1,\vk_2) P_L(\vk_1)P_L(\vk_2),  \\
 P_{b_2^2}(k)      &= \frac{1}{2} \ikk P_L(\vk_1) \big[P_L(\vk_2) - P_L(\vk_1) \big], \\
 P_{b_2 b_{s^2}}(k)&= \frac{1}{2} \ikk P_L(\vk_1) \left[P_L(\vk_2)S_2(\vk_1,\vk_2) - \frac{2}{3}P_L(\vk_1) \right], \\
 P_{b_{s^2}^2}(k)&= \frac{1}{2} \ikk P_L(\vk_1) \left[P_L(\vk_2)[S_2(\vk_1,\vk_2)]^2 - \frac{4}{9}P_L(\vk_1) \right],
\end{align}
and
%
\begin{align}
P_{b_2,\theta}(k) &=\ikk G_2(\vk_1,\vk_2) P_L(\vk_1)P_L(\vk_2),  \\
P_{b_{s^2},\theta}(k) &=\ikk G_2(\vk_1,\vk_2) S_2(\vk_1,\vk_2) P_L(\vk_1)P_L(\vk_2), 
\end{align}
with renormalized bias parameters  
\begin{align}
 b_1 = c_\delta + \left[ \frac{31}{24}c_{\delta^2} \mC_{\delta^2} + \frac{1}{2}c_{\delta^3}\mC_{\delta^3} + \frac{1}{3} c_{\delta s^2}\mC_{\delta s^2}  + \frac{68}{63} c_{s^{2}} \mC_{s^2}  - \frac{16}{63}c_{st} \mC_{st} \right] \sigma^2,
\end{align}
with $\mC_{\mO}$ constants of order unity and they are required in cosmologies beyond EdS, as we have seen for $\mC_{s^2}$ and $\mC_{st}$. At 1-loop in the PS the rest of biasing parameters remain equal: $b_2=c_{\delta^2}$ and $b_{s^2}=c_{s^2}$.

\end{subsection}

\end{section}

\begin{section}{Perturbation theory in redshift-space for generalized cosmologies}\label{RSD_PT_GC}

As we observe objects in the sky we map them through their angular position $\hat{\ve n}$ and radial 
position as inferred from their redshift. The latter is given by the Hubble flow and their peculiar velocity $\ve v$. Hence, an object located at a comoving distance $\vx$ is observed to be at an apparent position $\ve s$, such that the map between 
real and redshift space positions is given by the non-relativistic, longitudinal Doppler effect,
\begin{equation}\label{RSDmap}
 \ve s = \vx + \ve u,
\end{equation}
with line-of-sight ``velocity'' $\ve u$ defined as
\begin{equation}\label{defu}
 \ve u \equiv  \vhn \frac{\ve v \cdot \vhn}{aH}. 
\end{equation}

We use the plane-parallel approximation, on which $\vhn$ is a constant vector pointing in the direction of the objects sample, instead of being equal to the position unit vector $\hat{\vx}$. We also assume that the velocity is longitudinal, with divergence field $\theta$, defined in eq.~(\ref{thetadef}), for which
\begin{equation}\label{tton}
\ve u(\vk) =  i f_0 \vhn \frac{\vk \cdot \vhn}{k^2} \theta(\vk), 
\end{equation}
that we will mostly use.

Clearly, the map to redshift coordinates conserves the number of tracers, $\big[1+\delta_s(\ve s)\big]d^3s = \big[1+\delta(\vx)\big]d^3x$, yielding
\begin{equation}
 (2\pi)^3\delta_\text{D}(\vk) + \delta_s(\vk) = \int d^3x \big(1+\delta(\vx)\big) e^{-i \vk \cdot(\vx+ \ve u(\vx))},  
\end{equation}
and the redshift-space PS becomes  \cite{Scoccimarro:2004tg,Vlah:2018ygt}
\begin{equation}\label{RSDPS}
  (2\pi)^3\delta_\text{D}(\vk) + P_s(\vk) = \int d^3x e^{-i\vk\cdot \vx} \Big[ 1+\mathcal{M}(\ve J= \vk,\vx) \Big], 
\end{equation}
with velocity moments generating function  
\begin{equation}\label{VDgenF}
1+\mathcal{M}(\ve J,\vx) =  \left\langle \big(1+\delta(\vx_1)\big)\big(1+\delta(\vx_2)\big)  e^{-i \ve J \cdot \Delta \ve u}  \right\rangle,
\end{equation}
where $\Delta \ve u = \ve u(\vx_2)-\ve u(\vx_1)$ and $\ve x = \vx_2 - \vx_1$. Function $\mathcal{M}$ (or its Fourier transform) plays a central role in RSD. Different expansion procedures of eq.~\eqref{VDgenF} yield different approaches to RSD modeling, grouped in \cite{Vlah:2018ygt} as: direct Lagrangian, moment expansion, streaming model, and smoothing kernel. We will follow here the moment expansion approach, in which the exponential in the generating function is expanded and the moments are evaluated. Thereafter we will consider EFT contributions which yield the damping along the line-of-sight direction produced by non-coherent motions of particles at small scales. The connection to smoothing kernels will be discussed in section \ref{smooth_K}.

\begin{subsection}{Velocity moments of the generating function}

The $\tm$-th density weighted velocity field moment of the generating function is an $\tm$-rank tensor defined as  \cite{Scoccimarro:2004tg,Vlah:2018ygt}
\begin{align}
 \Xi^{\tm}_{ i_1 \cdots i_\tm}(\vx) &\equiv i^\tm \frac{\partial^\tm}{\partial J_{i_1}\cdots \partial J_{i_\tm}} \big[ 1+\mathcal{M}(\ve J,\vx) \big] \Big|_{\ve J = 0}
 = \langle \big(1+\delta_1\big)\big(1+\delta_2\big)\Delta u_{i_1}\cdots\Delta u_{i_\tm} \rangle,
\end{align}
with $\delta_1 =\delta(\vx_1)$ and $\delta_2 =\delta(\vx_2)$. 
The PS in the moment expansion approach becomes
\begin{align} \label{PSmomexp}
   (2\pi)^3 \dD(\vk) + P_s(\vk)
      &= \sum_{\tm=0}^\infty \frac{(-i)^\tm}{\tm!} k_{i_1}\dots k_{i_\tm}  \tilde{\Xi}_{i_{1}\cdots i_{\tm}}^{\tm}(\vk),
\end{align}
where the $\tilde{\Xi}^{\tm}_{i_{1}\cdots i_\tm}(\vk)$  are the Fourier moments of the generating function 
---the Fourier transforms of their configuration space counterparts, $\Xi^{\tm}_{i_{1}\cdots i_{n}}(\vx)$. The hope is that by cutting the sum in eq.~(\ref{PSmomexp}) at a finite, low moment ($\tm$) yields a good approximation to the PS. Indeed, we see that terms linear in the PS appear only for moments $\tm=0,\,1,$ and $2$ (higher moments involve correlators of at least three fields), while terms $\mathcal{O}(P_L^2)$ show up to $\tm=4$. From the fifth moment upwards all terms are at least $\mathcal{O}(P_L^4)$.  Since in this work we want to compute the moments that appear in the 1-loop PS, it is sufficient to cut the sum in eq.~(\ref{PSmomexp}) at $\tm=4$, and search for moments up to $\tilde{\Xi}^{4}_{ijkl}(\vk)$. This is what we do in the next subsection. We choose to take a slow route to do it in order to isolate the moments depending on their dependence on velocity and density fields. This approach will ease to understand further approximations and to compare to different approaches in the literature. 

\end{subsection}

\begin{subsection}{Computation of moments}\label{momenta_comp}
We define the $\tm$-th scalar velocity moment of the redshift-space PS as
\begin{equation}\label{Pmdef}
 P^\tm(k,\mu) \equiv \frac{(-i)^\tm}{\tm!} k_{i_1}\cdots k_{i_\tm} \tilde{\Xi}^{\tm}_{i_1\cdots i_\tm}(\vk) = \sum_{n=0}^\tm \mu^{2n} f_0^\tm I^\tm_{n}(k),
\end{equation}
such that the total PS is  
\begin{equation}\label{PsInPm}
P_s(k,\mu)=\sum_{\tm=0}^\infty   P^\tm(k,\mu),    
\end{equation}
up to a Dirac delta function localized at $\vk=0$. Whereas the first equality (definition) in eq.~(\ref{Pmdef})  depends on $\vk$, implicitly on $k$ and $\mu\equiv \hat{\vk} \cdot \vhn$, the second equality is an ansatz with explicit angular dependence as even powers of $\mu$. This ansatz has been shown to work, at least for moments $\tm=0,\dots,4$, in Ref.~\cite{Jalilvand:2019brk}. The functions $I^\tm_{n}(k)$  will be key to our approach since they encode in a compact way the different expansions entering in the power spectra.

Following, we compute the required moments in which we leave some of the long computations to Appendix \ref{app:ImnFunctions}, where we write the $I^\tm_{n}(k)$ as 2-dimensional integrals. 
\begin{itemize}
    \item The moment $\tm=0$ is simply the 2-point real space correlation function
\begin{equation}
\Xi^{0}(\ve x) = \langle \big(1+\delta_1\big)\big(1+\delta_2\big) \rangle = 1 + \xi(\vx).
\end{equation}
In Fourier space $\tilde{\Xi}^{0}(\vk) = (2\pi)^3 \dD(\vk) +  P_{\delta\delta}(k)$, or
\begin{align} \label{Pm0}
 P^{\tm=0}(k,\mu) =  P_{\delta\delta}(k), 
\end{align}
plus a Dirac delta function localized at $\vk=0$, which in the following we will omit. 
Hence $I_0^0(k)=P_{\delta\delta}(k)$ is the real space, full nonlinear PS.

\item The $\tm=1$ moment is
\begin{align}
\Xi_{i}^{1}(\ve x) &= \langle \big(1+\delta_1\big)\big(1+\delta_2\big)\Delta u_i \rangle = \langle \Delta u_i  (\delta_1+\delta_2)  \rangle +\langle  \Delta u_i \delta_1 \delta_2 \rangle \nonumber\\
&\equiv \Xi_{i}^{1,ud}(\ve x) + \Xi_{i}^{1,udd}(\ve x)
\end{align} 
where the moment with the label ``$ud$'' refers to the correlator constructed by the product of one velocity ($u$) field and one density ($d$) field, and ``$udd$'' refers to the correlator containing one velocity and two density fields. 

To show how the computations are performed, and for the only time in this work, we work in the detail one of these correlators
\begin{align}
&\langle \delta(\vx_1)\Delta u_i  \rangle = \int \Dk{k_1}\Dk{k_2} e^{i\vk_1\cdot\vx_1}\big(e^{i\vk_2\cdot\vx_2}-e^{i\vk_2\cdot\vx_1} \big) \left(i f_0 \frac{\vk_2 \cdot \vhn}{k_2^2} \hat{n}_i \right)
\langle \delta(\vk_1)\theta(\vk_2)\rangle \nonumber\\
&\quad = \int \Dk{k_1}\Dk{k_2} e^{i\vk_1\cdot\vx_1}\big(e^{i\vk_2\cdot\vx_2}-e^{i\vk_2\cdot\vx_1} \big) \left(i f_0 \frac{\vk_2 \cdot \vhn}{k_2^2} \hat{n}_i \right)
 (2\pi)^3\dD(\vk_1+\vk_2) P_{\delta\theta}(k_1)\nonumber\\
&\quad = -i f_0 \hat{n}_i  \int \Dk{k_1} \big(e^{-i\vk_1\cdot\vx}-1 \big)  \frac{\vk_1 \cdot \vhn}{k_1^2} P_{\delta\theta}(k_1) 
  = i f_0 \hat{n}_i  \int_{\vp} e^{i\vp\cdot\vx} \frac{\vp \cdot \vhn}{p^2}  P_{\delta\theta}(p),
\end{align}
where we used eq.~(\ref{tton}) to write
\begin{equation}
 \Delta u_i \equiv u_i(\vx_2)-u_i(\vx_1) = \int_{\vp} \big(e^{i\vp \cdot\vx_2}-e^{i\vp \cdot\vx_1} \big) \left( i f_0 \frac{\vp \cdot \vhn}{p^2} \hat{n}_i \right) \theta(\vp).
\end{equation}

Hence, in Fourier space,
\begin{equation} \label{xi1ud}
 \tilde{\Xi}_{i}^{1,ud}(\vk) = 2i f_0 \hat{n}_i \frac{\mu}{k}P_{\delta\theta}(k).
\end{equation}
Analogously, we compute for the other first order moment,
\begin{equation} \label{xi1udd}
 \tilde{\Xi}_{i}^{1,udd}(\vk) = i\hat{n}_i f_0 \int_{\vp}  \frac{\vp\cdot \vhn}{p^2} \big[B_{\theta\delta\delta}(\vp,-\vk,\vk-\vp)- B_{\theta\delta\delta}(\vp,-\vk-\vp,\vk) \big], 
\end{equation}
where the cross bispectrum is 
\begin{equation} \label{cbtdd}
(2\pi)^3 \dD(\vk_1+\vk_2+\vk_3) B_{\theta\delta\delta}(\vk_1,\vk_2,\vk_3) = \langle \theta(\vk_1)  \delta(\vk_2) \delta(\vk_3)\rangle.
\end{equation}
Summing the two contributions (\ref{xi1ud}) and (\ref{xi1udd}) and contracting with $-ik_i$ we obtain $P^{\tm=1}(\vk) = -ik_i \tilde{\Xi}^{1}_i(\vk)$,
\begin{align}\label{rXi1}
P^{\tm=1}(k,\mu) &= 2 \mu^2 f_0 P_{\delta\theta}(k)  
+2 k \mu f_0  \int_{\vp}  \frac{\vp\cdot \vhn}{p^2} B_{\theta\delta\delta}(\vp,-\vk,\vk-\vp) \nonumber\\
 &= \mu^2 f_0 I^{1}_{1}(k),
\end{align}
with $ I^{1}_{1}(k) =  I^{1,ud}_{1}(k) +  I^{1,udd}_{1}(k)$, where $I^{1,ud}_{1}(k) = 2 P_{\delta\theta}(k)$ and $I^{1,udd}_{1}(k)$ given in eq.~(\ref{I1udd1ATracers}).

\item The second moment, $\tm=2$, is
\begin{align}
\Xi_{ij}^{2}(\ve x) &= \langle \big(1+\delta_1\big)\big(1+\delta_2\big)\Delta u_i \Delta u_j\rangle  =
\langle \Delta u_i\Delta u_j  \rangle+\langle \Delta u_i\Delta u_j (\delta_1 +\delta_2 ) \rangle  +\langle \Delta u_i\Delta u_j   \delta_1 \delta_2 \rangle \nonumber\\
    & =  \Xi_{ij}^{2,uu}(\vx)+\Xi_{ij}^{2,uud}(\vx)+\Xi_{ij}^{2,uudd}(\vx).
\end{align}
Working them out based on their fields dependence, we obtain
\begin{align}\label{xi2uu}
 \tilde{\Xi}^{2,uu}_{ij}(\vk) =  
 - 2 f_0^2  \hat{n}_i \hat{n}_j  \frac{\mu^2}{k^2} P_{\theta\theta}(k), 
\end{align}
\begin{align}\label{xi2uud}
&\tilde{\Xi}^{2,uud}_{ij}(\vk) = -2 f_0 \hat{n}_i\hat{n}_j \int_{\vp} \frac{\vp\cdot\vhn}{p^2}\frac{(\vk-\vp)\cdot\vhn}{|\vk-\vp|^2}   B_{\theta\theta\delta} (\vp,\vk-\vp,-\vk) \nonumber\\
&+2 f_0 \hat{n}_i\hat{n}_j \int_{\vp} \frac{\vp\cdot\vhn}{p^2}\frac{\vk\cdot\vhn}{k^2} \big[  B_{\theta\theta\delta} (\vp,\vk,-\vk-\vp)-  B_{\theta\theta\delta} (\vp,-\vk,\vk-\vp)\big],
\end{align}
and, to 1-loop corrections,
\begin{align}\label{xi2uudd}
\tilde{\Xi}^{2,uudd}_{ij}(\vk) 
&=- 2 f_0^2 \hat{n}_i\hat{n}_j \int_{\vp} \frac{(\vp\cdot\vhn)^2}{p^4} P_{\theta\theta}(p)\Big[ P_{\delta\delta}(|\vk - \vp|)-P_{\delta\delta}(k) \Big]\nonumber\\
&\quad - 2 f_0^2 \hat{n}_i\hat{n}_j \int_{\vp} \frac{\vp\cdot\vhn}{p^2}\frac{(\vk-\vp)\cdot\vhn}{|\vk-\vp|^2}  
  P_{\delta\theta}(p) P_{\delta\theta}(|\vk-\vp|).
\end{align}
Summing up the three contributions, contracting with $-\frac{1}{2}k_ik_j$, and rearranging terms we have 
\begin{align}\label{rXi2}
P^{\tm=2}(\vk) &=    f_0^2  \mu^4 P_{\theta\theta}(k) \nonumber\\
&+ k\mu f_0 \int \Dk{p} \frac{\vp\cdot\vhn}{p^2}\Bigg[f_0\frac{[(\vk-\vp)\cdot\vhn]^2}{|\vk-\vp|^2}   B_{\theta\delta\theta} (\vp,-\vk,\vk-\vp) \nonumber\\ 
&\qquad\qquad -f_0\frac{[(\vk+\vp)\cdot\vhn]^2}{|\vk+\vp|^2}   B_{\theta\delta\theta} (\vp,\vk,-\vk-\vp) \Bigg] \nonumber\\ 
&+ k\mu f_0 \int \Dk{p} \frac{\vp\cdot\vhn}{p^2} \Bigg[f_0 \frac{(\vk\cdot\vhn)^2}{k^2} B_{\theta\theta\delta} (\vp,-\vk,\vk-\vp) \nonumber\\ 
&\qquad\qquad - f_0 \frac{(\vk\cdot\vhn)^2}{k^2} B_{\theta\theta\delta} (\vp,\vk,-\vk-\vp)\Bigg] \nonumber\\
&+(k\mu f_0)^2 \int \Dk{p} \frac{(\vp\cdot\vhn)^2}{p^4} P_{\theta\theta}(p)\Big[ P_{\delta\delta}(|\vk - \vp|)-P_{\delta\delta}(k) \Big]\nonumber\\
&+ (k\mu f_0)^2 \int \Dk{p} \frac{\vp\cdot\vhn}{p^2}\frac{(\vk-\vp)\cdot\vhn}{|\vk-\vp|^2}  
  P_{\delta\theta}(p) P_{\delta\theta}(|\vk-\vp|),
\end{align}
which can be written  as
\begin{align}
 P^{\tm=2}(k,\mu) &= f_0^2 \big[ \mu^2  I^{2}_1(k) + \mu^4  I^{2}_2(k) \big],
\end{align}
with
\begin{align}
 I^2_1(k) &= I^{2,uu}_1(k) + I^{2,uud}_1(k) + I^{2,uudd}_1(k), \nonumber\\
 I^2_2(k) &= I^{2,uu}_2(k) + I^{2,uud}_2(k) + I^{2,uudd}_2(k),
\end{align}
with $I^{2,uu}_1(k)=0$, $I^{2,uu}_2(k) = P_{\theta\theta}(k)$, and the rest of the $I^2_{1,2}(k)$ functions are given by eqs. (\ref{I2uud12ATracers}) and (\ref{I2nuudd}).


\item The third velocity moment, $\tm=3$, is
\begin{align}
\Xi_{ijk}^{3}(\ve x) &= \langle \big(1+\delta_1\big)\big(1+\delta_2\big)\Delta u_i \Delta u_j\Delta u_k \rangle 
= \langle \Delta u_i\Delta u_j \Delta u_k  \rangle +\langle (\delta_1 + \delta_2) \Delta u_i\Delta u_j \Delta u_k  \rangle \nonumber\\
     &=\Xi_{ijk}^{3,uuu}(\vx) +\Xi_{ijk}^{3,uuud}(\vx),
\end{align}
where we have not written the term $\langle \delta_1 \delta_2 \Delta u_i\Delta u_j \Delta u_k  \rangle$ since it is order $\mathcal{O}(P_L^3)$, and not considered here: contrary to the previous moments, whose expressions are valid to arbitrary PT order, in the following we keep only terms up to $\mathcal{O}(P_L^2)$. The ``$uuu$'' correlator yields
\begin{align}\label{Xi3uuu}
 &\tilde{\Xi}_{ijk}^{3,uuu}(\vk) = \int d^3x e^{-i\vk \cdot\vx}  \langle \Delta u_i\Delta u_j\Delta u_k  \rangle 
 =3i f_0^3 \hat{n}_i  \hat{n}_j  \hat{n}_k \int_{\vp} \frac{\vp\cdot\vhn}{p^2} \frac{\vk\cdot\vhn}{k^2} \nonumber\\ 
 &\qquad \times \Bigg[ \frac{(\vk+\vp)\cdot\vhn}{|\vk+\vp|^2}  B_{\theta\theta\theta}(\vp,\vk,-\vk-\vp) -\frac{(\vk-\vp)\cdot\vhn}{|\vk-\vp|^2}  B_{\theta\theta\theta}(\vp,-\vk,\vk-\vp) \Bigg] , 
\end{align}
and the ``$uuud$'' term,
\begin{align}\label{Xi3uuud}
 \tilde{\Xi}_{ijk}^{3,uuud}(\vk) &
 = 12 i f_0^3  \hat{n}_i \hat{n}_j \hat{n}_k \int_{\vp} 
            \frac{(\vp\cdot\vhn)^2}{p^4}  \frac{\vk\cdot\vhn}{k^2} P_{\theta\theta}(p)P_{\delta\theta}(k) \nonumber\\  
&\quad -12 i f_0^3  \hat{n}_i \hat{n}_j \hat{n}_k \int_{\vp} 
            \frac{(\vp\cdot\vhn)^2}{p^4}  \frac{(\vk-\vp)\cdot\vhn}{|\vk-\vp|^2} P_{\theta\theta}(p)P_{\delta\theta}(|\vk-\vp|). 
\end{align}
Summing up the two contributions, rearranging some terms and contracting with $\frac{i}{6}k_ik_jk_k$ we have 
\begin{align}\label{rXi3}
P^{\tm=3}(\vk) &=-2  k^2 \mu^4 f_0^3  \sigma^2_v  P_{\delta\theta}(\vk) \nonumber\\  
&\quad  + 2 k \mu^3 f_0^3 \int_{\vp} \frac{\vp\cdot\vhn}{p^2}   
               \frac{((\vk-\vp)\cdot\vhn)^2}{|\vk-\vp|^2}  B_{\theta\theta\theta}(\vp,-\vk,\vk-\vp) \nonumber\\ 
 &\quad + 2 k^2\mu^2 f_0^3   \int_{\vp}  \frac{(\vp\cdot\vhn)^2}{p^4} \frac{\big[(\vk-\vp)\cdot\vhn\big]^2}{|\vk-\vp|^2}  
                 P_{\theta\theta}(p) P_{\delta\theta}(|\vk-\vp|)  \nonumber\\
&\quad  + 2 k^2\mu^2 f_0^3   \int_{\vp} \frac{\vp\cdot\vhn}{p^2}\frac{\big[(\vk-\vp)\cdot\vhn\big]^3}{|\vk-\vp|^4} P_{\delta\theta}(p)  
                  P_{\theta\theta}(|\vk-\vp|),
\end{align}
with the velocity variance
\begin{equation}
 \sigma^2_v = \frac{1}{6\pi^2} \int_0^\infty dp P^L_{\theta\theta}(p). 
\end{equation}
This can be written as
\begin{align}
 P^{\tm=3}(k,\mu) &= f_0^3 \big[ \mu^2  I^{3}_1(k) + \mu^4  I^{3}_2(k) + \mu^6  I^{3}_3(k) \big],
\end{align}
with $ I^{3}_1(k) = 0$, $ I^{3}_2(k) =  I^{3,uuu}_2(k) +  I^{3,uuud}_2(k)$, $ I^{3}_3(k) =  I^{3,uuu}_3(k) +  I^{3,uuud}_3(k)$, where these $I^\tm_n$ functions are giving by eqs.~(\ref{I3uud23ATracers}) and (\ref{I3nuuud}).

\begin{figure}[tbp] 
\vspace{-2cm}
\centering 
\includegraphics[width=1\textwidth]{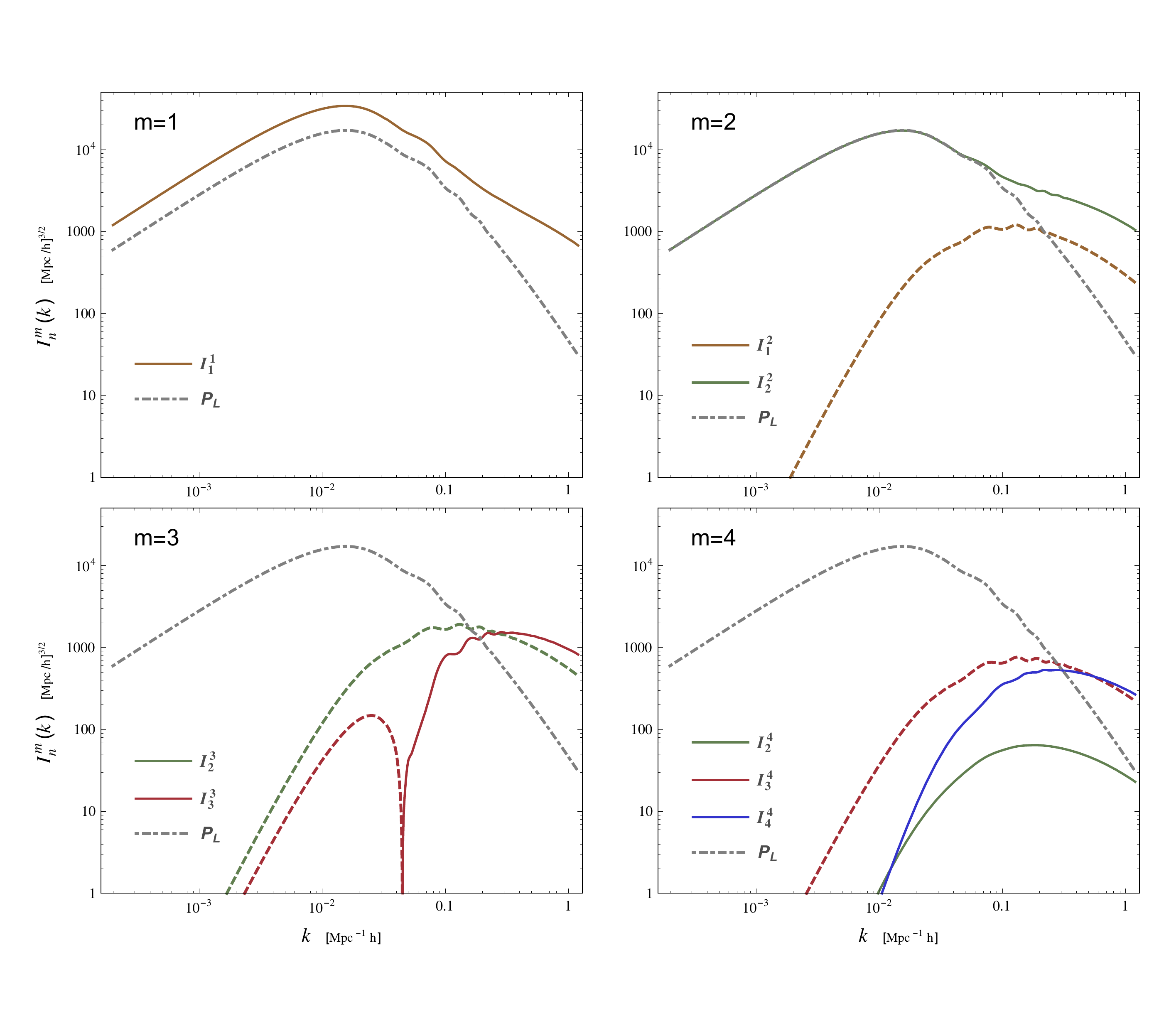}
\vspace{-1.5cm}
\caption{\label{fig:moments} Functions $I^\tm_n(k)$ for dark matter that contribute to the velocity moments through eq.~(\ref{Pmdef}). We are using F6 MG model at redshift $z=0.5$. Solid lines are for moments, and dashed lines for their negatives. Each panel shows the different $\tm$ pieces in which brown lines are for $n=1$, green for $n=2$, red for $n=3$, and blue for $n=4$. For comparison in all panels we
also plot the linear matter power spectrum $P_L(k)$ which is the linear piece of $I_0^0(k)$.}
\end{figure}

\item Up to 1-loop, the fourth moment is 
$\Xi_{ijkl}^{\tm=4}(\vx)=\langle \Delta u_i \Delta u_j \Delta u_k \Delta u_l \rangle$, yielding
\begin{align}\label{xi4uuuu}
\tilde{\Xi}^{4,uuuu}_{ijkl}(\vk) 
&= -24 \hat{n}_i \hat{n}_j \hat{n}_k \hat{n}_l f_0^4 \int_{\vp} 
    \frac{(\vp\cdot\vhn )^2}{p^4 }\frac{(\vk\cdot\vhn )^2}{k^4} P_{\theta\theta}(p)P_{\theta\theta}(k)    \nonumber\\
&\quad   +12 \hat{n}_i \hat{n}_j \hat{n}_k \hat{n}_l f_0^4 \int_{\vp}  
    \frac{(\vp\cdot\vhn )^2}{p^4 }\frac{((\vk-\vp)\cdot\vhn )^2}{|\vk-\vp|^4} P_{\theta\theta}(p)P_{\theta\theta}(|\vk-\vp|).
\end{align}
After some manipulations we have\footnote{By contracting eq.~(\ref{xi4uuuu}) with $\frac{1}{4!} k_ik_jk_kk_l$, the second term carries a factor $(k\mu)^4$. Thereafter, we decompose two of these four powers of $k\mu$ as 
$(k\mu)^2 = \big((\vk-\vp)\cdot\vhn \big)^2 +2((\vk-\vp) \cdot \vhn)(\vp\cdot\vhn) + (\vp\cdot\vhn)^2$.} 
\begin{align}\label{rXi4}
P^{\tm=4}(\vk)&= - k^2 \mu^6 f_0^4 \sigma^2_v P_{\theta\theta}(k) \nonumber\\
& + \mu^2 k^2 f_0^4  \int_{\vp} 
     \frac{(\vp\cdot\vhn )^3}{p^4 }\frac{((\vk-\vp)\cdot\vhn )^3}{|\vk-\vp|^4} P_{\theta\theta}(p) P_{\theta\theta}(|\vk-\vp|)\nonumber\\
& +\mu^2 k^2 f_0^4  \int_{\vp} 
     \frac{(\vp\cdot\vhn )^2}{p^4 }\frac{((\vk-\vp)\cdot\vhn )^4}{|\vk-\vp|^4} P_{\theta\theta}(p) P_{\theta\theta}(|\vk-\vp|).
\end{align}
This can be written as
\begin{align}
 P^{\tm=4}(k,\mu) &= f_0^3 \big[ \mu^2  I^{4}_1(k) + \mu^4  I^{4}_2(k) + \mu^6  I^{4}_3(k)  + \mu^8  I^{4}_4(k)\big]
\end{align}
with $ I^{4}_1(k) = 0$, $ I^{4}_2(k) =  I^{4,uuuu}_2(k)$, $ I^{4}_3(k) =  I^{4,uuuu}_4(k)$ and  $ I^{4}_4(k) =  I^{4,uuuu}_4(k)$,  where these functions are giving by eq. (\ref{I4nuuuu}). 
\end{itemize}

To get a sense of the significance of the different contributions $I^{\tm}_n(k)$, in figure \ref{fig:moments} we plot the non-vanishing of these functions with $\tm=1,\,2,\,3,\,4$,  for dark matter particles ($b=1$, and all other bias parameters equal to zero); the function $I^0_0(k)$ is simply given by the density-density PS. We employ the F6 model and evaluated results at $z = 0.5$. The larger contributions are given by $I^{1}_{1}$ and $I^{2}_{2}$, since these are the only $I$-functions that contain terms linear in the PS, while the rest of the functions yield pure 1-loop contributions. For comparison, on each panel we also show the linear density-density PS $P_L(k)$.

\end{subsection}

%
%

\begin{subsection}{Moment expansion approach} 

We have split the velocity moments as 
\begin{align}
 \Xi^{0} &=  \Xi^{0,dd},  \nonumber\\
 \Xi^{1}_{i} &=  \Xi^{1,ud}_{i} + \Xi^{1,udd}_{i}, \nonumber \\
 \Xi^{2}_{ij} &=  \Xi^{2,uu}_{ij} + \Xi^{2,uud}_{ij} + \Xi^{2,uudd}_{ij}, \nonumber\\
 \Xi^{3}_{ijk} &=  \Xi^{3,uuu}_{ijk} + \Xi^{3,uuud}_{ijk}, \nonumber\\
 \Xi^{4}_{ijkl} &=  \Xi^{4,uuuu}_{ijkl}.  \label{OrgMom}
\end{align}
Now, it should become clear how this splitting is useful to compare models in the literature.
To linear order, only the pieces $ \tilde{\Xi}^{0,dd}(\vk)$,  $\tilde{\Xi}^{1,ud}_{i}(\vk)$,  and $\tilde{\Xi}^{2,uu}_{ij}(\vk)$  contribute yielding \cite{10.1093/mnras/227.1.1,Scoccimarro:2004tg}: 
\begin{align}
P_s^K(k,\mu) &=  P^L_{\delta\delta}(k) + 2 f_0 \mu^2  P^L_{\delta\theta}(k) + f_0^2  \mu^4 P^L_{\theta\theta}(k) , \nonumber\\
&=  b_1^2 \left(1 + \mu^2 \frac{f(k)}{b_1}\right)^2 P_L(k) ,
\end{align}
where in the second equality we have omitted $\nabla^2\delta_m$ biasing to show how the Kaiser formula is recovered, although the Kaiser boost becomes scale dependent due to the growth rate $f(k)$.
We generalize, for future use, to the nonlinear Kaiser 
\begin{align}
P_s^{K,\text{NL}}(k,\mu) &=  \tilde{\Xi}^{0,dd}(\vk) -  i k_i \tilde{\Xi}^{1,ud}_{i}(\vk) - \frac{1}{2}k_i k_j \tilde{\Xi}^{2,uu}_{ij}(\vk) \nonumber\\ 
  &= P_{\delta\delta}(k) + 2 f_0 \mu^2  P_{\delta\theta}(k) + f_0^2  \mu^4 P_{\theta\theta}(k),
\end{align}
which adds the non-linear corrections to the power cross-spectra $P_{\delta\delta,\delta\theta,\theta\theta}$, but neglects some of the next to leading perturbative order, 1-loop contributions.

Consider now all the terms that contain bispectra contributions. These are given in eqs.~\eqref{xi1udd}, \eqref{xi2uud}, and \eqref{Xi3uuu}, and add up as\footnote{$A(k,\mu)$ is of course a different function than $A(k, t)$ of section \ref{sect:basic_model}.} 
\begin{align}\label{defA_mu_k}
&A(k,\mu) \equiv -i k_i \tilde{\Xi}_i^{1,udd}(\vk) -\frac{1}{2}k_ik_j\tilde{\Xi}^{2,uud}_{ij}(\vk) +\frac{i}{6}k_ik_jk_k\tilde{\Xi}_{ijk}^{3,uuu}(\vk)  \nonumber\\
&= 2 k \mu f_0 \ip  \frac{\vp\cdot \vhn}{p^2} B_{\theta\delta\delta}(\vp,-\vk,\vk-\vp) \nonumber\\
&\quad +k\mu f_0 \ip \frac{\vp\cdot\vhn}{p^2}\Bigg[f_0\frac{[(\vk-\vp)\cdot\vhn]^2}{|\vk-\vp|^2}   B_{\theta\delta\theta} (\vp,-\vk,\vk-\vp) 
       -f_0\frac{[(\vk+\vp)\cdot\vhn]^2}{|\vk+\vp|^2}   B_{\theta\delta\theta} (\vp,\vk,-\vk-\vp) \Bigg] \nonumber\\ 
&\quad +k\mu f_0 \ip \frac{\vp\cdot\vhn}{p^2} \Bigg[f_0 \frac{(\vk\cdot\vhn)^2}{k^2}   B_{\theta\theta\delta} (\vp,-\vk,\vk-\vp)- f_0 \frac{(\vk\cdot\vhn)^2}{k^2} B_{\theta\theta\delta} (\vp,\vk,-\vk-\vp)\Bigg] \nonumber\\
&\quad +2 k\mu f_0 \ip \frac{\vp\cdot\vhn}{p^2}  
   f_0^2\frac{(\vk\cdot\vhn)^2}{k^2}\frac{((\vk-\vp)\cdot\vhn)^2}{|\vk-\vp|^2}  B_{\theta\theta\theta}(\vp,-\vk,\vk-\vp).
\end{align}
Let us define \cite{Taruya:2010mx}
\begin{align}\label{defBsigma}
B_\sigma(\vk_1,\vk_2,\vk_3) &\equiv  B_{\theta\delta\delta}(\vk_1,\vk_2,\vk_3) +f_0 \frac{(\vk_3\cdot\vhn)^2}{k_3^2} B_{\theta\delta\theta}(\vk_1,\vk_2,\vk_3)  \nonumber\\
&\quad +f_0 \frac{(\vk_2\cdot\vhn)^2}{k_2^2}   B_{\theta\theta\delta} (\vk_1,\vk_2,\vk_3)
+ f_0^2\frac{(\vk_2\cdot\vhn)^2}{k_2^2} \frac{(\vk_3\cdot\vhn)^2}{k_3^2}  B_{\theta\theta\theta}(\vk_1,\vk_2,\vk_3) \nonumber\\
&= \left\langle \theta(\vk_1)\left[\delta(\vk_2) + f_0 \frac{(\vk_2\cdot\vhn)^2}{k_2^2}\theta(\vk_2) \right]\left[\delta(\vk_3) + f_0 \frac{(\vk_3\cdot\vhn)^2}{k_3^2}\theta(\vk_3) \right]\right\rangle'.
\end{align}
Then, noting the symmetries $B_\sigma(\vk_1,\vk_2,\vk_3)=B_\sigma(\vk_1,\vk_3,\vk_2)=B_\sigma(-\vk_1,-\vk_2,-\vk_3)$, one can write
\begin{align}
A(k,\mu)= 2 k \mu f_0 \ip  \frac{\vp\cdot \vhn}{p^2} B_\sigma(\vp,-\vk,\vk-\vp) \, , 
\end{align}
which is one of the corrections to the non-linear Kaiser model introduced for the TNS model in \cite{Taruya:2010mx}.

The rest of the velocity moments are collected as 
\begin{equation} \label{D_k_mu_def}
  D(k,\mu) \equiv -\frac{1}{2}k_ik_j\tilde{\Xi}^{2,uudd}_{ij}(\vk) +\frac{i}{3!}k_ik_jk_k\tilde{\Xi}_{ijk}^{3,uuud}(\vk)  +\frac{1}{4!}k_ik_jk_k k_l \tilde{\Xi}_{ijkl}^{4,uuuu}(\vk), 
\end{equation}
hence this function is constructed out of four velocity or density fields, which at 1-loop only contribute as linear fields. Hence the only  involved kernel different than EdS is $G_1(k)$.  
We make the splitting 
\begin{align} \label{D_k_mu}
 D(k,\mu) = B(k,\mu) + C(k,\mu)  -(k\mu f_0 \sigma_v)^2 P^K_s(k,\mu),
\end{align}
with \cite{Taruya:2010mx}
\begin{align}
 B(k,\mu) &= (k\mu f_0)^2  \ip F(\vp)F(\vk-\vp) , \\
  F(\vp) &= \frac{\vp\cdot\vhn}{p^2}\Bigg[ P_{\delta\theta}(p) + f_0 \frac{(\vp\cdot\vhn)^2}{p^2} P_{\theta\theta}(p) \Bigg], 
\end{align}
and 
\begin{equation}
 C(k,\mu) = (k\mu f_0)^2  \ip  \frac{(\vp\cdot\vhn)^2}{p^4}P_{\theta\theta}(p) P^K_s(|\vk-\vp|, \mu_{\vk-\vp}),
\end{equation}
with $ \mu_{\vk-\vp} $ the angle between $\vk-\vp$ and the line of sight $\vhn$. 

Finally, the redshift-space PS in the moment expansion (ME) approach to 1-loop in SPT is
\begin{align} \label{PSME}
 P_s^\text{ME}(k,\mu) &=  \sum_{\tm=0}^4 \frac{(-i)^\tm}{\tm!} k_{i_1} \cdots k_{i_\tm} \Xi^\tm_{i_1 \cdots i_\tm}(\vk) 
 =  P_s^{K,\text{NL}}(k,\mu) +  A(k,\mu) + D(k,\mu).
\end{align}
Note that eq.~(\ref{PSME}) is known from \cite{Taruya:2010mx} (eq.~(23) in that  paper), but  we have used the ME formalism of refs.~\cite{Scoccimarro:2004tg,Vlah:2018ygt} to derive it, and we have generalized it for tracers in arbitrary cosmologies. However, this is not the TNS RSD model expression, since the authors preferred to consider another expression with a phenomenological FoG damping term (eq.~(18) in their paper), instead of the exact expression. In section \ref{counterterms} we will use EFT-counterterms to model the missing effects of non-linear mappings between redshift and real space coordinates and FoG features as has been done recently in several works \cite{Perko:2016puo,Ivanov:2019pdj,Chen:2020fxs,Chudaykin:2020aoj,Philcox:2020srd}.

\end{subsection}

\begin{subsection}{Large and small scales behavior}\label{large-small-behav}

We now want to check the IR and UV behavior of the functions composing the ME PS,  eq.~\eqref{PSME}. Taking their large-scale limit, up to order $\mO(k^2)$ we obtain
\begin{align}
B(k\rightarrow 0,\mu) &= -C(k\rightarrow 0,\mu) =-(k\mu f_0)^2  \ip \mu^2_\vp \big( 1+ f_0 \mu^2_\vp \big)^2 \frac{P_L^2(p)}{p^2} \propto k^2,
\end{align}
such that UV cancellations provide $B(k\rightarrow 0,\mu)+ C(k\rightarrow 0,\mu)\propto k^4$. 
Hence at large scales the dominant term in function $D$ is $ -(k\mu f_0 \sigma_v)^2 P^K_s(k,\mu)$ and it behaves as 
\begin{equation}
D(k\rightarrow 0,\mu) \propto k^2 P_L(k) \propto k^{2+n_s},
\end{equation}
with $n_s$ the primordial spectral index.
Hence, $B$ grows faster than $k^2 P_L$ at low-$k$, which would violate momentum conservation,  
yielding that small scales largely affect the loop contributions at large scales. $C$ cancels this pathological behaviour, 
and $B+C$ goes as $k^4$. It is the term $-\sigma^2_v k^2 P_{\theta\theta}(k)$ the one that brings the correct behaviour, $D \rightarrow k^2 P_L $ at large scales.

\begin{figure}[tbp]
\vspace{-2cm}
\centering 
\includegraphics[width=0.95\textwidth]{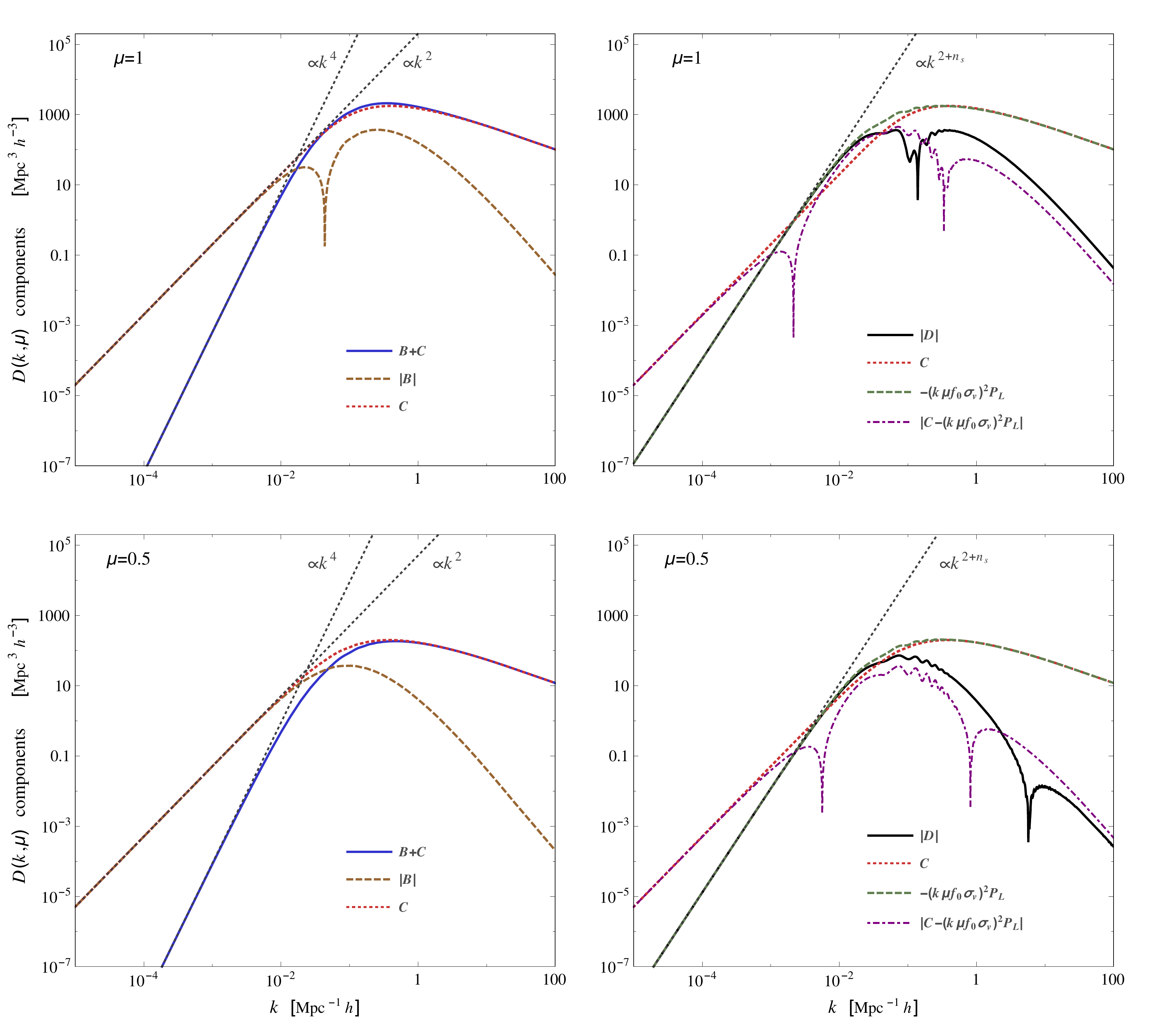}
\caption{Components of $D(k,\mu)$ according to the splitting of eq.~\eqref{D_k_mu} for the model F6 at redshift $z=0.5$ and for line-of-sight angle directions $\mu=1$ (top panels) and $\mu=0.5$ (bottom panels). On the left panels we show that $B$ and $C$ functions behave as $k^2$ at large scales, but the sum $B + C$ goes as $k^4$, while at small scales   $B + C \sim C$. On the right panels we show the effects of adding the component $\propto -k^2 P_L(k)$, which at large scales dominates rendering $D(k\rightarrow 0,\mu) \propto k^2 P_L(k) \propto k^{2+n_s}$, while at small  scales it cancels the dominant contribution to $C$ and makes $D\propto B$. \label{fig:PCcomponents} }
\end{figure}

On the other limit, at high-$k$, we obtain an IR divergence in $C(k,\mu)$ for power spectra with spectral index $n\leq -1$, which is canceled out by the term 
$-\sigma^2_v k^2 P_{\theta\theta}(k)$. Indeed, we can expand for $p \leq k$, 
assuming a scale invariant PS $P\propto k^n$ which is a good approximation for high $k$, so $k^m d^m P_L(k) / d k^m \propto P_L(k)$, and  we have
\begin{equation} \label{PsKexp}
   P_s^K(|\vk-\vp|, \mu_{\vk-\vp}) =  P_s^K(k, \mu)+ P_L(k) \sum_{m=1}^\infty \alpha_m \left(\frac{p}{k}\right)^m,
\end{equation}
where the coefficients $\alpha_m$ are constructed out of contractions of up to $m$ angular directions $\hat{p}_i$:
\begin{equation}
\alpha_m = \sum_{i=0}^{m} \tilde{\alpha}_s^{j_1 \cdots j_{i} } \hat{p}_{j_{1}} \cdots \hat{p}_{j_{i}},    
\end{equation}
such that if $m$ is odd (even) each term of the above sum contains only an odd (even) number of vectors $\hat{p}_i$.
%
Hence, 
\begin{align} \label{CsLimitki}
C(k,\mu) &- (k\mu f_0 \sigma_v)^2 P^K_s(k,\mu)=  \nonumber\\
   &\quad (k\mu f_0)^2  \int \Dk{p}  \frac{(\vp\cdot\vhn)^2}{p^4}P_{\theta\theta}(p) \Big[ P^K_s(|\vk-\vp|, \mu_{\vk-\vp}) -P^K_s(k, \mu) \Big] \nonumber\\ 
&\ni (k\mu f_0)^2 P_L(k) \underset{p \ll k}{\int} \Dk{p} \frac{(\vp \cdot\vhn)^2}{p^4} P_{\theta\theta}(p) \left[ \alpha_2 \frac{p^2}{k^2} + \mO\left( \frac{p^4}{k^4}\right)\right],   
\end{align}
where in the last line of the above equation we used that the first term in the expansion of  eq.~\eqref{PsKexp} factorizes the angular dependence as $\alpha_1 = \tilde{\alpha}_i \hat{p}^i$, such that when performing the angular integration in eq.~\eqref{CsLimitki} we find $\int d\Omega_\vp \hat{p}_i\hat{p}_j\hat{p}_k=0 $, hence this term vanishes and the first correction is of order  $p^2/k^2$; for the same reason all odd powers of $(p/k)$ are not present, and the second correction is $\mO\big((p/k)^4\big)$. Hence the combination $C(k,\mu) - (k\mu f_0 \sigma_v)^2 P^K_s(k,\mu)$ is IR divergent only for $n\leq -3$, and  safe for typical power spectra. This cancellation is equivalent to the real space 1-loop SPT PS 
for which the function $P_{22}(k)$ has an IR divergence for  $n\leq -1$, that is cured by a term $-\sigma^2_\Psi k^2 P_L(k) \in P_{13}(k)$, with $\sigma^2_\Psi$ the variance of Lagrangian displacements. %

Summarizing, we notice that the full $D(k,\mu)$ is free of UV and IR divergence for typical power spectra. However, the three pieces that compose it in our splitting, eq.~(\ref{D_k_mu}), have to be present, such that  if any of them would be missing there would exist either an IR or an UV divergence. To make this point more clear, in figure \ref{fig:PCcomponents} we plot the different components of the $D(k,\mu)$ function for angles $\mu=0.5$ (bottom panels) and $1$ (top panels). At large scales $D(k,\mu) \propto k^{2+n_s}$, although both $B(k,\mu)$ and $C(k,\mu)$ behave as $k^2$. Meanwhile, at small scales function $B(k,\mu)$ behaves equally to $C(k,\mu)-(k \mu f_0 \sigma_v)^2 P_L(k)$, both scaling approximately as  $k^{1/2} P_L(k)$.

\end{subsection}

\begin{subsection}{EFT counterterms}\label{counterterms}

So far, we have followed a standard PT approach to construct the velocity moments and the complete redshift space PS in the SPT-ME approach. However, loop integrals are of the form
$I(\vk)=\ip K(\vk,\vp)$ and are computed over all internal momentum space, although  $K(\vk,\vp)$ does not hold at all scales, particularly for high internal momentum. Though these kernels are typically suppressed for regions $p \gg k$, such that small scales do not affect considerably the $I(k)$ functions at moderate, quasilinear scales, they pose a fundamentally wrong UV behaviour ---in particular $P_{13}$. The EFT for large scale structure \cite{Baumann:2010tm,Hertzberg:2012qn} formalism cuts-off the loop integrals, by directly smoothing the overdensity fields by an arbitrary scale, and introduces counterterms necessary to remove the cut-off dependence on the final expressions. This is dramatically more important in correlators of fields that do not vanish at zero separation, as $\sigma_v^2$. The objective of EFT is to cure the spurious high-$k$ effects on statistics due to non modeled small scale physics, out of the reach of PT. Further, dark matter evolution is dictated by the Boltzmann equation, and its simplified description with momentum conservation and Euler equation breaks down by nonlinear collapse which makes different streams to converge, leading to velocity dispersion, the formation of matter caustics and, ultimately, to shell crossing. Hence, the very concept of CDM as a coherent fluid at all scales with no velocity dispersion is theoretically  inconsistent because of gravitational collapse, breaking down at shell-crossing at best, and very rapidly all the Boltzmann hierarchy is necessary to describe the dynamics; this a key concern of EFT. In real space, for the PS, the leading order EFT correction counterterm ($ct$) is given by $P_{ct}(k) = -c^2_s(t) k^2 P_L(k)$, with $c_s$ the effective speed of sound of dark matter arising from fluid equations of a non-perfect fluid. 

In redshift space, the situation is more complex, because the counterterms not only model small scales, non-perturbative physics, but also the non-linear mathematical map between real space and redshift space densities \cite{Perko:2016puo,Ivanov:2019pdj,Chen:2020fxs}. In this case, each moment $P^\tm$ carries its own counterterms of the form $\sum \tilde{\alpha}_n \mu^{2n}k^2 P_L(k)$, leading to, see e.g. \cite{Ivanov:2019pdj,Chen:2020fxs}, 
\begin{equation}
P_{ct}(k,\mu) = (\alpha_0 + \alpha_1 \mu^2 +\cdots ) k^2 P_L(k).   
\end{equation}
Finally, along the line-of-sight direction, 2-point statistics are dominated by FoG as a nonlinear coupling between the velocity and density fields, with a characteristic scale given by the velocity dispersion $\sigma_v$, for which typically $\sigma_v^{-1} \sim k_\text{NL}$, motivating to go beyond the leading order through the ansatz \cite{Ivanov:2019pdj}
\begin{equation}
 P_{ct}^\text{NLO}(k,\mu) = \tilde{c}   \big(\mu k f_0 \sigma_v \big)^4   P^K_s (k,\mu),  
\end{equation}
which can be also understood as stemming from a Taylor series expansion of a phenomenological damping factor $\mathcal{D}_\text{FoG}\big[(k \mu f_0 \sigma_v)^2\big]$ at second order on $(k^2/\sigma_v^{-2})$. But, as discussed above, $\sigma_v$ is not well modeled by PT, and historically replaced by a free parameter $\sigma_\text{FoG}$. Here, it becomes modeled by the $\tilde{c}$ time-dependent counterterm. The multipoles become
\begin{equation}
 P_{ct,\ell}^\text{NLO}(k) =     \tilde{c} \mathcal{D}_{ct,\ell}^\text{NLO}(k) f_0^4 k^4 P_L(k),
\end{equation}
with
\begin{align}
   \mathcal{D}_{ct,\ell=0}^\text{NLO}(k) &=  \frac{b_1^2  }{5}+\frac{2}{7} b_1 f(k)  +\frac{f^2(k)  }{9},  \\
    \mathcal{D}_{ct,\ell=2}^\text{NLO}(k) &= \frac{4 b_1^2  }{7}+\frac{20}{21} b_1 f(k)  +\frac{40 f^2(k)  }{99}, \\
    \mathcal{D}_{ct,\ell=4}^\text{NLO}(k) &= \frac{8 b_1^2  }{35}+\frac{48}{77} b_1 f(k)  +\frac{48 f^2(k) }{143}.
\end{align}
It is interesting to note that, regardless of the bias and growth rate, the next-to-leading order FoG counterterm contribution to the quadrupole is the largest, about  $\mathcal{D}_{ct,\ell=2}^\text{NLO} \sim 3\mathcal{D}_{ct,\ell=0}^\text{NLO}$.

We arrive to our final expression for the EFT-Moment expansion approach modeling of the redshift space PS,
 \begin{align} \label{PSEFT}
 P_s^\text{EFT}(k,\mu) &=   P_s^\text{ME}(k,\mu)  + (\alpha_0 + \alpha_1 \mu^2 + \cdots) k^2 P_L(k) + \tilde{c}   \big(\mu k f_0 \big)^4   P^K_s (k,\mu) + P_\text{shot},
\end{align} 
with $P_\text{shot}$ modeling the stochastic terms, uncorrelated with long wave-length fluctuations.  That is, we assume that stochasticity is well localized, with a small range of coherence, such that its spectra can be characterized by a constant (both in $k$ and $\mu$) shot noise $P_\text{shot}$. In contrast, in e.g. refs.~\cite{DAmico:2019fhj,Nishimichi:2020tvu,Chen:2020fxs,Schmittfull:2020trd} the stochastic noise has a component proportional to $(k \mu)^2$, leading to two free parameters to model it. Here, we are adopting the most simple prescription where the stochastic contribution is given by a constant shot noise, affecting only the monopole of the PS; see  e.g.~\cite{Ivanov:2019pdj} for a recent use with real data\footnote{In contrast, in \cite{DAmico:2019fhj,Colas:2019ret} a more cumbersome stochastic expression, including $k^2$ and $(\mu k)^2$ contributions, is used to fit DR12 BOSS data.}.  We will find that this approach yields accurate results when comparing to simulated data.

\end{subsection}

\begin{subsection}{IR-resummation}

\begin{figure}[tbp]
\centering 
\includegraphics[width=0.68\textwidth]{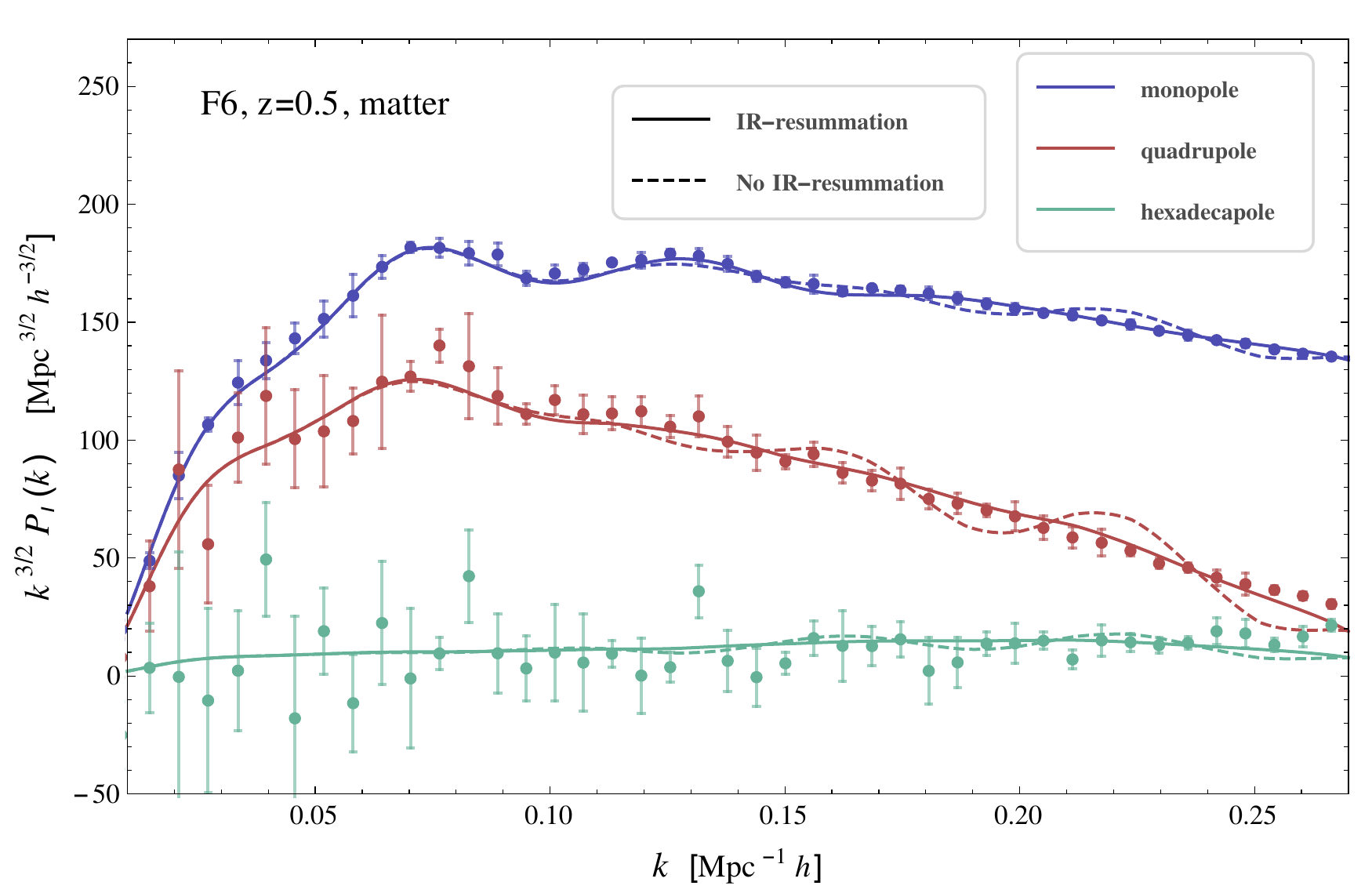}
\caption{Matter redshift-space power spectrum multipoles with (solid lines) and without (dashed lines) the IR-resummations for MG model F6 at redshift $z=0.5$. The effect of IR-resummation is clearly seen and necessary to degrade the BAO  oscillations at high-$k$. The error bars denote the RMS error over the five realizations in the  simulations. \label{fig:pkmatterF6z05} 
} 
\end{figure}

Despite the success of SPT-EFT in modeling the broadband PS, the theory yet gives poor results in modeling the BAO  since long-wavelength displacement fields, though being essentially linear, stream largely contributing to damp features in the PS in a manner that is non-perturbative under an SPT scheme \cite{Eisenstein:2006nk,Crocce:2007dt,Carlson_2012,PhysRevD.86.103528}.
Then, in order to model the spread and degradation of the BAO oscillations due to large scale bulk flows, we employ IR-resummations  \cite{Senatore:2014via,Baldauf:2015xfa,Peloso:2016qdr,Vlah:2015sea,Vlah:2015zda,Blas:2016sfa,Ding:2017gad,Senatore:2017pbn,Lewandowski:2018ywf} as implemented in \cite{Ivanov:2018gjr,Ivanov:2019pdj,Chudaykin:2020aoj}. 
The main idea is to split the real space linear PS in one piece containing the oscillations (wiggles), $P_\text{w}$, and a smooth piece without them, $P_\text{nw}$, such that $P_L(k)=P_\text{w}(k) + P_\text{nw}(k)$, to do this we use the method based on fast sine transforms of \cite{Hamann:2010pw} (proposed in appendix (A.1) of that paper). Note that this approach is somewhat arbitrary, since there is not a single way to tell what is a PS without BAO, so there is a handcrafted element in the process.\footnote{In this sense, the LPT approach is more natural, because the Lagrangian displacements of bulk flows are fully resummed from the beginning (see e.g. \cite{Chen:2020fxs}). Recently, this scheme has been extended to resum also the velocity fields \cite{Chen_2020_NoEntry}, such that a posteriori IR-resummations are no longer required.}

To leading order one gets the Kaiser, IR-resummed PS \cite{Ivanov:2018gjr}
\begin{equation}
 P_s^{K,\text{IR}}(k,\mu) = \big(b_1 + f(k) \mu^2 \big)^2 \left[ e^{-k^2 \Sigma^2_\text{tot}(k,\mu)} P_{L}(k) + \big(1-e^{-k^2 \Sigma^2_\text{tot}(k,\mu)} \big) P_\text{nw}(k) \right]   
\end{equation}
with damping, angle-dependent, factor 
\begin{equation}
\Sigma^2_\text{tot}(k,\mu) = \big[1+f(k) \mu^2 \big( 2 + f(k) \big) \big]\Sigma^2 + f^2(k) \mu^2 (\mu^2-1) \delta\Sigma^2,    
\end{equation}
with $\Sigma^2$ the real space isotropic damping and $\delta\Sigma^2$ a subdominant contribution,
\begin{align}
\Sigma^2 &= \frac{1}{6 \pi^2}\int_0^{k_s} dp \,P_\text{nw}(p) \left[ 1 - j_0\left(\frac{p}{k_\text{BAO}}\right) + 2 j_2 \left(\frac{p}{k_\text{BAO}}\right)\right],\\
\delta\Sigma^2 &= \frac{1}{2 \pi^2}\int_0^{k_s} dp \,P_\text{nw}(p)  j_2 \left(\frac{p}{k_\text{BAO}}\right),
\end{align}
where $k_s$ is a scale separating long and short modes and $k_\text{BAO}\simeq (105 \, \text{Mpc} \, h^{-1})^{-1}$ corresponds to the BAO scale. Functions $j_n$ are the spherical Bessel functions of order $n$. The choice of $k_s$ is also arbitrary, but the results depend very weakly on it, as long as $k_s\gtrsim 0.05 \, \text{Mpc}^{-1}  h$ \cite{Chudaykin:2020aoj}; in this work we use $k_s = 0.2 \, \text{Mpc}^{-1}  h$.

To 1-loop the IR-resummed PS becomes \cite{Ivanov:2018gjr}
\begin{align}\label{PsIR}
P_s^\text{IR}(k,\mu) &= 
 e^{-k^2 \Sigma^2_\text{tot}(k,\mu)} P_s^\text{EFT}(k,\mu) +  \big(1-e^{-k^2 \Sigma^2_\text{tot}(k,\mu)} \big) P_\text{s,nw}^\text{EFT}(k,\mu) \nonumber\\
 &\quad +  e^{-k^2 \Sigma^2_\text{tot}(k,\mu)} P_\text{w}(k) k^2 \Sigma^2_\text{tot}(k,\mu), 
\end{align}
where $P_s^\text{EFT}(k,\mu)$ is the 1-loop PS computed using eq.~\eqref{PSEFT}, and  $P_\text{s,nw}^\text{EFT}(k,\mu)$ is also computed with eq.~\eqref{PSEFT} but using as input the non-wiggle linear PS $P_\text{nw}$.  Equation \eqref{PsIR} is our final expression to be compared with simulated halo statistics in section \ref{num_results}.

In figure \ref{fig:pkmatterF6z05} we show the matter redshift-space PS multipoles ($\ell=0,2,4$) with and without the IR-resummations for MG model F6 at redshift $z=0.5$. We notice that the BAO degradation is not well modeled by eq.~\eqref{PSEFT} alone, as large oscillations are still present at high-$k$ values (dashed lines) but it needs the IR-resummation presented in this section to follow more appropriately the matter particle data, as shown in solid lines (the details of the simulations are presented in the next section). 

We finally notice that we have used the same resummation scheme as for $\Lambda$CDM, with the generalized cosmologies effects entering through the scale-dependent growth rate $f(k)$ and the linear PS. This pragmatic approach is natural for generalized cosmologies that reduce to $\Lambda$CDM at large scales, since the bulk flows that are treated non-perturbatively are almost indistinguishable among the different models.\footnote{MG theories like DGP \cite{Dvali:2000hr} and cubic Galileons \cite{Nicolis:2008in} do not reduce to $\Lambda$CDM at large scales. However, the linear growth rate $f$ is scale independent in these cases, suggesting this IR-resummation approximation is valid for such theories as well.}

\end{subsection}

\end{section}

\begin{section}{Numerical Results}\label{num_results}

Having laid out our perturbation theory framework in the previous sections, here we proceed to compare our theoretical predictions for the PS from eq.~\eqref{PsIR}, against the ones obtained by state-of-the-art $N$-body simulations. 

Before discussing the results, we first begin with a brief overview of the $N$-body simulations we compare with, which are the Extended LEnsing PHysics using ANalaytic ray Tracing \verb|ELEPHANT| simulations \citep{Cautun:2017tkc,Alam:2020jdv}, that were performed with a modified version of the \verb|RAMSES| code,   
the \verb|ECOSMOG| module \citep{1475-7516-2012-01-051,Bose:2016wms}. The $\Lambda$CDM runs correspond to the following set of parameters $\{\Omega_{m},\Omega_{\Lambda},h,n_s,\sigma_8, \Omega_b\}=\{0.281,0.719,0.697,0.971,0.848, 0.046\}$, while 3 instances of the HS $n=1$ $f(R)$ model were simulated, corresponding to three variations of $|\bar{f}_{R_0}|=\{10^{-6},10^{-5}, 10^{-4}\}$. Furthermore, each scenario has been run using 5 different initial random seeds, that we average over. For the purposes of simplicity, we refer to these scenarios as GR, F6, F5, and F4, respectively, and we will also focus on snapshots at $z=0.5$ and $z=1$. The simulations span a cubic volume of $V_{box}=(1024 \,\text{Mpc} \,h^{-1})^3$, with $1024^3$ dark matter particles, while gravitationally bound haloes were identified using the publicly available code \verb|ROCKSTAR| \citep{2013ApJ...762..109B}. Further details about the simulations can be found at \citep{Cautun:2017tkc,Alam:2020jdv}.

\begin{figure}[tbp]
\centering 
\includegraphics[width=1.0\textwidth]{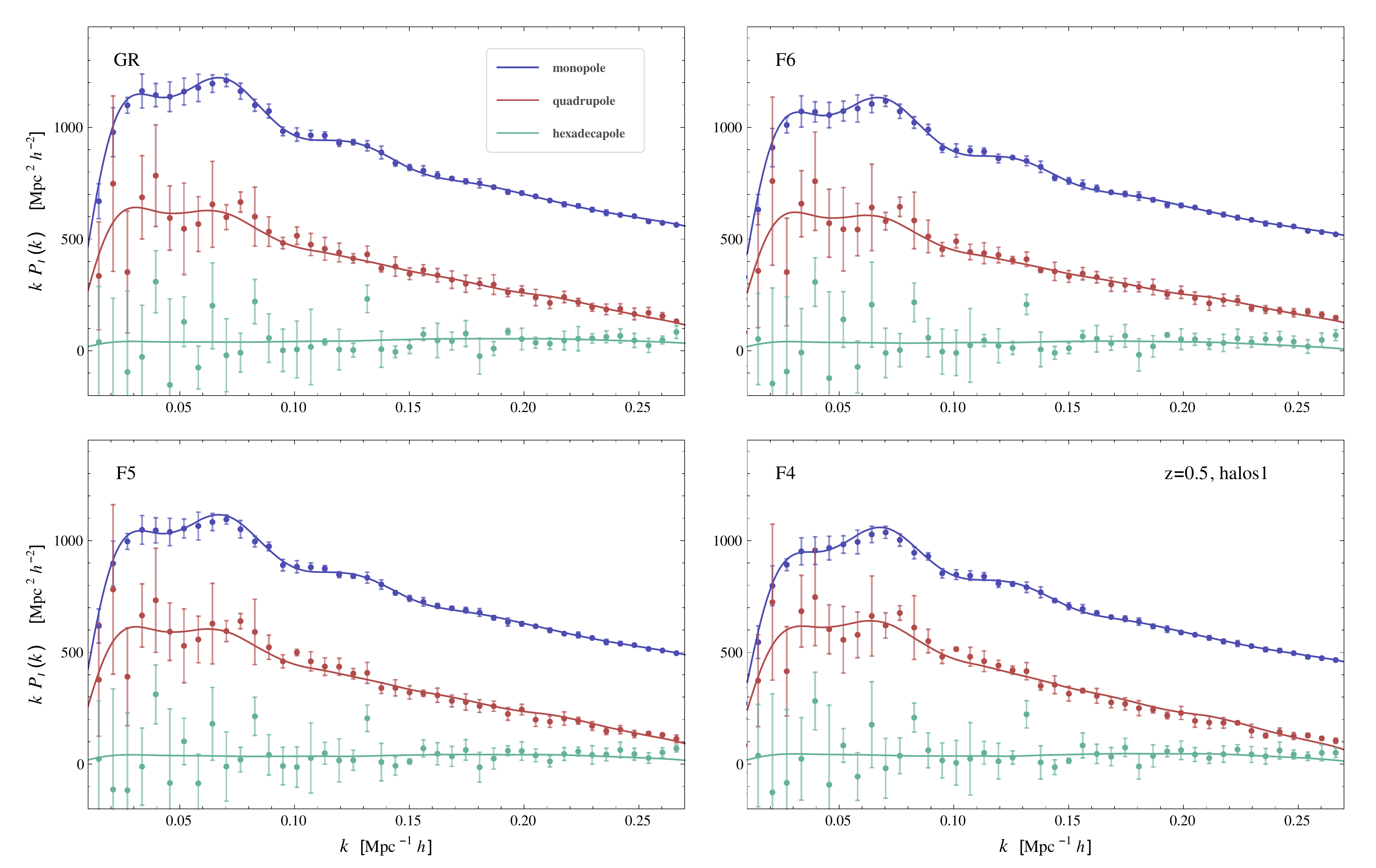}
\caption{Redshift-space power spectrum monopole, quadrupole, and hexadecapole for halo catalogue~1  ($10^{12}< M_{h} <4.5 \times 10^{12} M_{\odot} h^{-1}$) at redshift $z=0.5$.  \label{fig:halos1z05} }
\end{figure}

\begin{figure}[tbp]
\centering 
\includegraphics[width=1.0\textwidth]{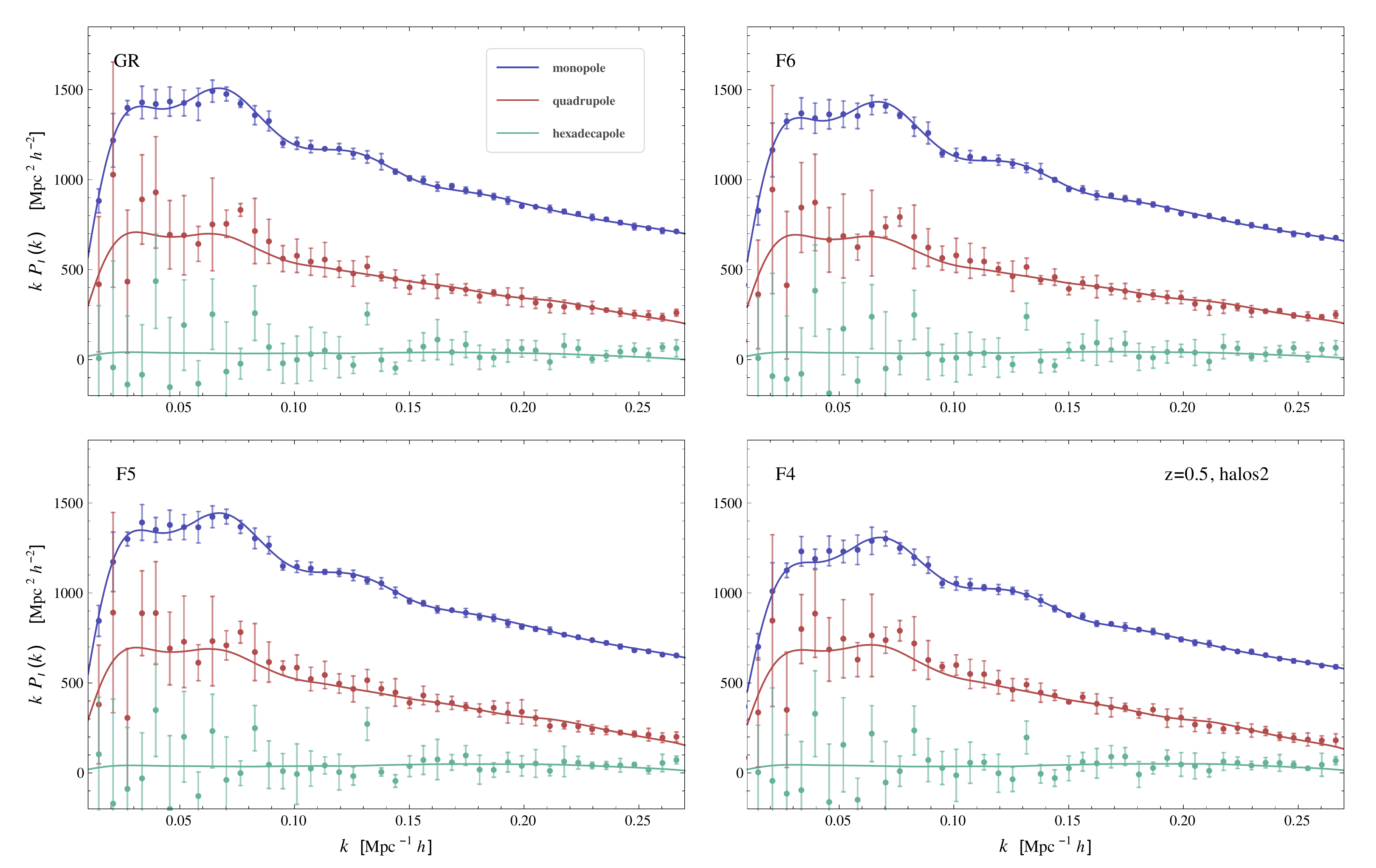}
\caption{Redshift-space power spectrum multipoles for halo catalogue 2 ($4.5\times 10^{12}<M_h < 1\times 10^{13}$ $M_{\odot} h^{-1}$) at redshift $z=0.5$. \label{fig:halos2z05} }
\end{figure}

To generate RSD outputs from the simulated dark matter and halo catalogues, we displace the particle (and halo) positions with their peculiar velocities, in accordance with the mapping ($\ref{RSDmap}$), choosing the line of sight $\vhn$ along the $\hat{\bold{z}}$ axis. We utilize the publicly available code \verb|NbodyKit|\footnote{\url{https://nbodykit.readthedocs.io/en/latest/index.html}} in order to extract the first 3 nonvanishing multipoles of the matter and halo PS, $P_0(k),P_2(k),P_4(k)$, evaluated at 512 equally spaced $k$ bins in the range  $0.00614<k< 3.135\,\text{Mpc}^{-1} h$. In the halo case, we extract two separate sub-samples, spanning halo mass ranges of $12<\log_{10} M_{h} <12.65$ and $12.65 <\log_{10}M_{h}<13$ in units of $M_{\odot} h^{-1}$, that we will refer to as halo catalogue 1 and 2, respectively, from now on. Upon the extraction of the clustering statistics from the halo catalogues, we include both main and satellite haloes in our calculation. We further note that one could alternatively only work with the main gravitationally bound structures identified by the halo finder, as was e.g. done in \citep{Valogiannis:2019nfz}, in which case our model would still be applicable, and the agreement with the simulations would likely be better. Given however that the former scenario is more realistic, we did choose to include the sub-halos and still found good agreement with our model's predictions, as we will see shortly.

The final missing piece in order to compare against the simulations pertains to the determination of the model free parameters, consisting of a total of 4 bias parameters $\{b_1, b_2, b_{s^2}, b_{3nl} \} $ and 4 additional EFT parameters $\{c_0, c_2, c_4, \tilde{c} \}$, with $c_\ell$ the multipoles of the factor $(\alpha_0 + \alpha_2 \mu^2 + \cdots)$ in eq.~\eqref{PSEFT}. We keep the shot-noise Poissonian, such that $P_\text{shot}=1/\bar{n}$, with $\bar{n}$ the mean number density of halos. Although one has the liberty to let this parameter free, or even adding a dependence on $\mu^2k^2$ \cite{Nishimichi:2020tvu,Chen:2020fxs,Schmittfull:2020trd}, we find that the Poissonian noise gives accurate results for our simulated data. We will further reduce the number of independent parameters from 8 to 6, since $b_{s^2}$ and $b_{3nl}$ can be expressed in terms of $b_1$, from EdS co-evolution \citep{Saito:2014qha}, as
\begin{align}
b_{s^2} &= - \frac{4}{7}(b_1-1),\\
b_{3nl} &=  \frac{32}{315}(b_1-1).
\end{align}
These expressions assume that the initial, Lagrangian bias is only local and that the velocity field is unbiased. The above relations were obtained using EdS evolution; however, we find that they yield acceptable results also in the MG models considered here, so we use them to reduce the number of free parameters of the theory. 
In the case of simpler local bias models, one can analytically predict the bias parameters in MG, as was done e.g. in \citep{Aviles:2018saf,Valogiannis:2019xed}. This is not the case for the higher order bias and EFT parameters considered in this work, which cannot be modeled from first principles. As a result, we determine these values by fitting our model's predictions \eqref{PsIR} to the simulated multipoles, through Markov Chain Monte Carlo runs\footnote{We use the generic sampler \href{https://github.com/rodriguezmeza/MathematicaMCMC-1.0.0.git}{https://github.com/rodriguezmeza/MathematicaMCMC-1.0.0.git} \cite{2019MNRAS.488.5127F}.}  over the range $0< k < 0.25\,\text{Mpc}^{-1} h$, and report the best-fit values given by the maximum-likelihood estimators, and 1$\sigma$ confidence intervals for each sample in Table~\ref{tabla:BiasParams}.\footnote{We should note, at this point, that during the fitting process we have assumed a diagonal covariance matrix.  Even though we know that in principle the errors in different bins are correlated and expect that the inclusion of the non-diagonal terms would tighten the constraints. We nevertheless obtain very good fits, as it was also found in, e.g., \citep{Reid:2011ar} that made similar assumptions.} Finally, we note that the linear PS for the base $\Lambda$CDM cosmology is obtained by \verb|CAMB|\footnote{\href{https://camb.info/}{https://camb.info/}} \cite{Lewis:1999bs}, and the linear PS of MG models are computed as
\begin{equation}
P^\text{MG}_L(k,z) =\left( \frac{D_+(k,z)}{D_+(k\rightarrow 0,z)}\right)^2P^\text{$\Lambda$CDM}_L(k,z),    
\end{equation}
which is an excellent approximation as long as MG effects are negligible at early times for the scales of interest, which is certainly the case for $|f_{R0}|\leq 10^{-4}$.

Before we proceed to discuss the results for haloes, we begin by considering the relatively simpler case of dark matter, which is shown in figure \ref{fig:pkmatterF6z05}. There, we compare our theoretical predictions for the monopole, quadrupole and hexadecapole of the matter PS, against the ones from the $N$-body simulations, for the F6 model at $z=0.5$. The prediction for the monopole remains consistent (within 1$\sigma$ error bars) with the simulation down to $k\sim 0.26 \,\text{Mpc}^{-1} h$, with the higher order multipoles achieving similar levels of accuracy down to wave-modes of $k\sim 0.24 \,\text{Mpc}^{-1} h$. Even though the larger error bars in the latter case (which are expected given the limited simulation volume and sample variance) do not allow us to perform strict comparisons around the BAO scale, the model predictions are nevertheless well consistent with the simulated trend. We finally add that we checked and found very similar agreement, as in the F6 case, for the rest of the gravity models, that we do not show here. In \cite{Bose:2017dtl} similar comparisons were performed to HS models using the TNS model, but without the use of EFT parameters and IR-resummation, and hence its level of accuracy is more limited than in this work. 

\begin{figure}[tbp]
\centering 
\includegraphics[width=1.0\textwidth]{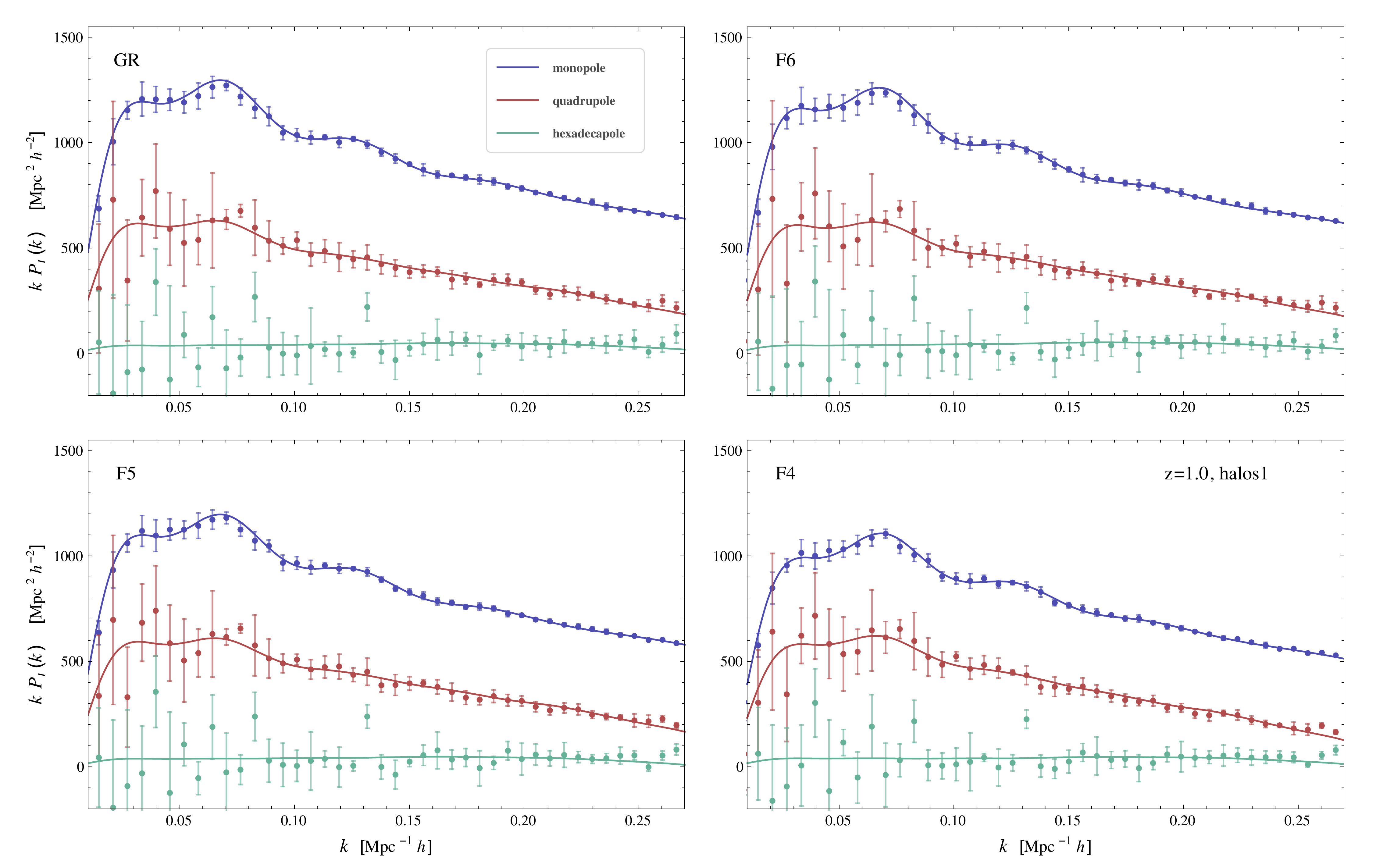}
\caption{Redshift-space power spectrum monopole, quadrupole, and hexadecapole for halo catalogue~1  ($10^{12}< M_{h} < 4.5\times 10^{12} M_{\odot} h^{-1}$) at redshift  $z=1$.  \label{fig:halos1z1} }
\end{figure}

\begin{figure}[tbp]
\centering 
\includegraphics[width=1.0\textwidth]{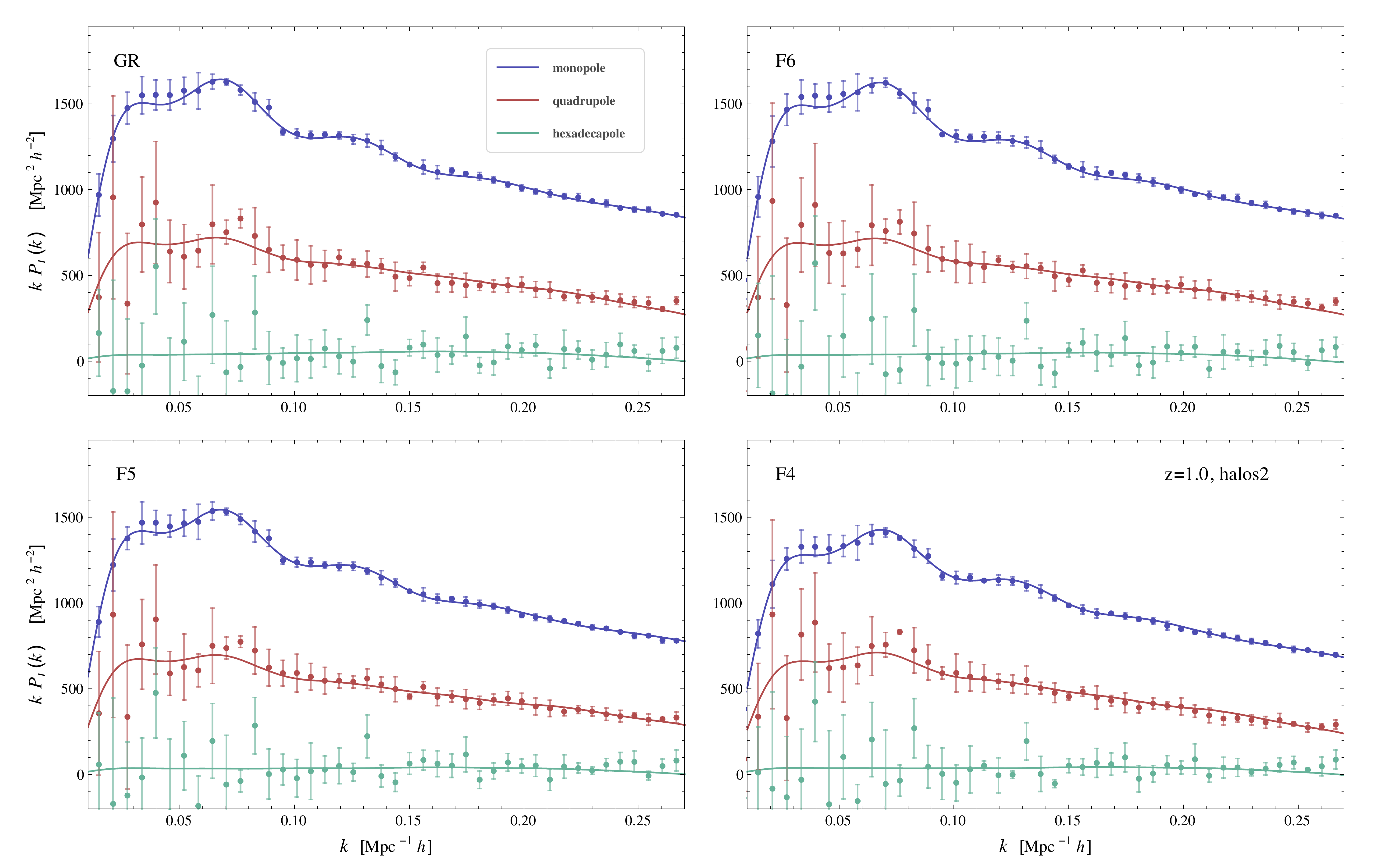}
\caption{Redshift-space power spectrum multipoles for halo catalogue 2 ($4.5\times 10^{12}<M_h < 1\times 10^{13}$  $M_{\odot} h^{-1}$) at redshift $z=1$.   \label{fig:halos2z1} }
\end{figure}

In figures \ref{fig:halos1z05} and \ref{fig:halos2z05}, we compare our model predictions~\eqref{PsIR} for the first three multipoles of the halo PS for all gravity models, against the ones obtained from the simulated halo catalogues $1$ and $2$, respectively, at $z=0.5$. The error bars show the RMS errors for the five realizations on each model. Starting with the halo 1 sample, we recover very similar levels of accuracy for the F6, F5, and F4 gravity models as in the known GR case, demonstrating a robust behavior across the parameter space of the Hu-Sawicki model. As in the dark matter case of figure \ref{fig:pkmatterF6z05}, the monopole predictions are the most accurate down to higher $k$ modes, up to $k=0.27 \,\text{Mpc}^{-1} h$, but the model also performs very well for the higher multipoles, demonstrating consistency with the corresponding simulated curves well within the non-linear regime. The more pronounced error bars, due to cosmic variance, in the higher multipoles are once again expected, especially given the relatively smaller number of halos compared to the dark matter particle case. We observe very similar trends in the halo catalogue $2$ results of figure \ref{fig:halos2z05}, and we also find this to be the case with the corresponding comparisons at redshift $z=1$, that are displayed in figures \ref{fig:halos1z1} and \ref{fig:halos2z1}. As a result, these comparisons demonstrate that our model is overall very successful at recovering the anisotropic redshift-space halo PS across a broad range of halo masses, cosmological redshifts and degrees of deviation from the base GR case, for the $f(R)$ HS MG scenario. 

In figure \ref{fig:CLs} we show contour plots for the case of model F5 at $z=0.5$ for the halo catalogue 1 (the rest of the models present similar behavior), showing some degeneracies, particularly for the $b_1$-$c_{0}$ and $b_1$-$c_{2}$ spaces. These are indeed expected since at small scales the EFT parameters, being positive, tend to lower the multipoles, while bigger $b_1$ values augment the overall power of the full spectrum. There is also a clear positive correlation between the $c_0$ and $c_2$ parameters although these affect only the monopole and the quadrupole, respectively; which can be understood as inherited from their degeneracies with the local linear bias $b_1$. 

Finally, we remark that our approach is close to the work of \cite{Chen:2020fxs}, where the authors find a good accuracy of their perturbative treatment up to $k=0.3\,\text{Mpc}^{-1} h$ for redshift $z=0.8$, that is similar to our findings.

\begin{center}
\begin{table*}
\small
\ra{1.3}
\begin{center}
\begin{tabular}{@{}l r r r r r r r r r r r @ {}}\toprule
   \multicolumn{7}{c}{Best Fit Bias and EFT Parameters}\\
   \hline 
\multicolumn{1}{c}{Model} &
\multicolumn{1}{c}{$b_{1}$ }&
\multicolumn{1}{c}{$b_{2}$ } &
\multicolumn{1}{c}{$c_{0}$} &
\multicolumn{1}{c}{$c_{2}$}& 
\multicolumn{1}{c}{$c_{4}$}& 
\multicolumn{1}{c}{$\tilde{c}$} \\

\cmidrule[0.4pt](r{0.25em}){1-1}
\cmidrule[0.4pt](lr{0.25em}){2-2}
\cmidrule[0.4pt](lr{0.25em}){3-3}
\cmidrule[0.4pt](lr{0.25em}){4-4}
\cmidrule[0.4pt](lr{0.25em}){5-5}
\cmidrule[0.4pt](lr{0.25em}){6-6}
\cmidrule[0.4pt](lr{0.25em}){7-7}
$\,$ GR, $z=0.5$, \texttt{halos1} & 1.475$\pm$0.004 & 0.49$\pm$0.01 & -28.1$\pm$0.5 & -35.3$\pm$0.5 & -3.2$\pm$0.1 & -0.48$\pm$0.01 \\ 
$\,$ GR, $z=0.5$, \texttt{halos2} & 1.655$\pm$0.003 & -1.00$\pm$0.02 & -4.3$\pm$0.1 & -45.5$\pm$0.5 & -6.3$\pm$0.2 & -0.50$\pm$0.01 \\  
$\,$ GR, $z=1.0$, \texttt{halos1} & 1.911$\pm$0.003 & 0.34$\pm$0.01 & -16.8$\pm$0.4 & -34.3$\pm$0.6 & -2.8$\pm$0.1 & -0.69$\pm$0.02 \\  
$\,$ GR, $z=1.0$, \texttt{halos2} & 2.181$\pm$0.002 & -0.04$\pm$0.01 & -7.6$\pm$0.2 & -28.5$\pm$0.7 & -1.4$\pm$0.04 & -0.92$\pm$0.02 \\ 
$\,$ F6, $z=0.5$, \texttt{halos1} & 1.415$\pm$0.003 & 0.91$\pm$0.02 & -31.0$\pm$0.5 & -26.6$\pm$0.4 & -2.2$\pm$0.1 & -0.41$\pm$0.01 \\  
$\,$ F6, $z=0.5$, \texttt{halos2} & 1.616$\pm$0.003 & 0.26$\pm$0.01 & -27.4$\pm$0.4 & -32.2$\pm$0.4 & -4.5$\pm$0.1 & -0.46$\pm$0.01 \\  
$\,$ F6, $z=1.0$, \texttt{halos1} & 1.882$\pm$0.003 & 0.59$\pm$0.01 & -18.8$\pm$0.4 & -28.6$\pm$0.5 & -0.7$\pm$0.02 & -0.70$\pm$0.02 \\  
$\,$ F6, $z=1.0$, \texttt{halos2} & 2.173$\pm$0.003 & 0.77$\pm$0.02 & -21.9$\pm$0.5 & -23.0$\pm$0.5 & -2.3$\pm$0.1 & -0.82$\pm$0.02 \\  
$\,$ F5, $z=0.5$, \texttt{halos1} & 1.392$\pm$0.003 & 0.49$\pm$0.01 & -28.7$\pm$0.4 & -30.2$\pm$0.4 & -2.5$\pm$0.1 & -0.45$\pm$0.01 \\ 
$\,$ F5, $z=0.5$, \texttt{halos2} & 1.619$\pm$0.004 & 0.64$\pm$0.02 & -40.0$\pm$0.6 & -33.8$\pm$0.5 & -3.1$\pm$0.1 & -0.48$\pm$0.01 \\  
$\,$ F5, $z=1.0$, \texttt{halos1} & 1.818$\pm$0.002 & -0.31$\pm$0.01 & -9.2$\pm$0.2 & -31.3$\pm$0.6 & -0.7$\pm$0.02 & -0.69$\pm$0.02 \\ 
$\,$ F5, $z=1.0$, \texttt{halos2} & 2.112$\pm$0.004 & 1.00$\pm$0.03 & -33.4$\pm$0.6 & -27.6$\pm$0.6 & -4.9$\pm$0.1 & -0.57$\pm$0.01 \\  
$\,$ F4, $z=0.5$, \texttt{halos1} & 1.270$\pm$0.002 & -1.00$\pm$0.02 & -8.0$\pm$0.2 & -40.1$\pm$0.4 & -2.9$\pm$0.1 & -0.40$\pm$0.01 \\ 
$\,$ F4, $z=0.5$, \texttt{halos2} & 1.441$\pm$0.002 & -1.00$\pm$0.02 & -10.0$\pm$0.2 & -44.5$\pm$0.5 & -4.2$\pm$0.1 & -0.38$\pm$0.01 \\  
$\,$ F4, $z=1.0$, \texttt{halos1} & 1.690$\pm$0.004 & 0.25$\pm$0.01 & -23.5$\pm$0.5 & -36.9$\pm$0.5 & -2.8$\pm$0.1 & -0.52$\pm$0.01 \\  
$\,$ F4, $z=1.0$, \texttt{halos2} & 1.965$\pm$0.004 & 0.96$\pm$0.02 & -40.2$\pm$0.7 & -34.1$\pm$0.7 & -5.5$\pm$0.1 & -0.52$\pm$0.01 \\ 
\bottomrule
\end{tabular}
\caption{Bias and EFT parameters fitted from simulations and used in figures \ref{fig:halos1z05} ($z=0.5$,   \texttt{halos1}), \ref{fig:halos2z05} ($z=0.5$,   \texttt{halos2}), \ref{fig:halos1z1} ($z=1$,   \texttt{halos1}), and \ref{fig:halos2z05} ($z=1$,   \texttt{halos2}).
The units of parameters $c_{0}$, $c_{2}$, and $c_{4}$ are $\text{Mpc}^{2}\,h^{-2}$.
}
\label{tabla:BiasParams}
\end{center}
\end{table*}
\end{center}

\begin{figure}[tbp]
\centering 
\includegraphics[width=.85\textwidth]{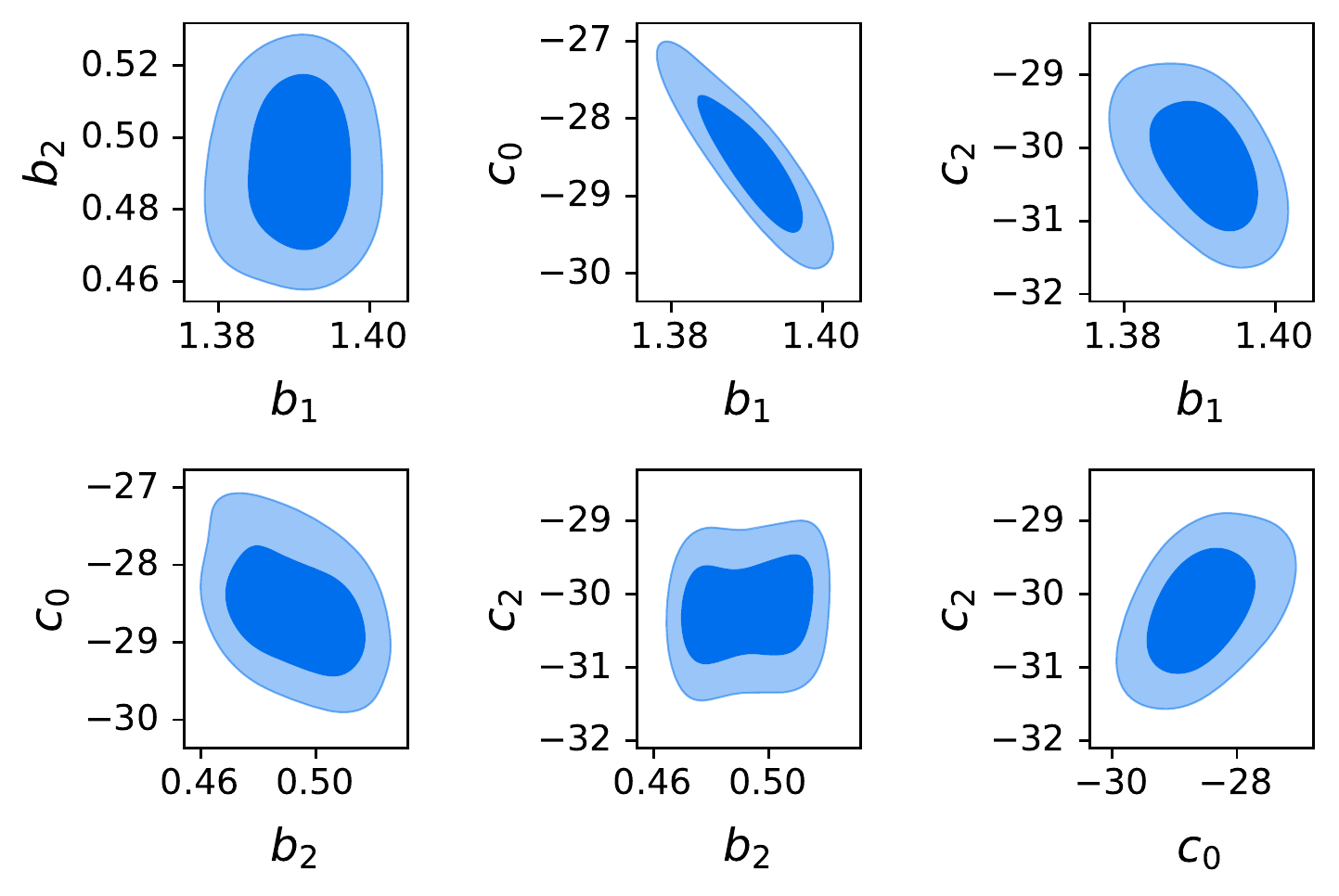} \\
\caption{\label{fig:CLs} Contour regions for bias and EFT parameters in the cosmology F5, 
\texttt{halos1} at $z=0.5$, corresponding to results in table \ref{tabla:BiasParams}. 
}
\end{figure}

\end{section}

\begin{section}{Connection to other works \label{smooth_K}}

In this section we would like to make connection to smoothing kernel templates widely used in the literature. 

By taking derivatives to the relation $\langle e^{i\ve J \cdot \ve A} \rangle = \exp \left[ \langle e^{i\ve J \cdot \ve A} \rangle_c \right]$ 
between cumulants and moments one arrives to \cite{Scoccimarro:2004tg,Vlah:2018ygt}:
\begin{align} \label{expcumM}
 &1+\mathcal{M}(\vk,\vx) =  \exp \left[\sum_{n=2}^\infty \frac{i^n}{n!} k_{i_1}\cdots k_{i_n}\langle \Delta u_{i_1}\cdots \Delta u_{i_n}\rangle_c \right] 
 \Big\{  1 + \langle (\delta_1 + \delta_2) e^{i\vk\cdot\Delta \ve u} \rangle_c  \nonumber\\
 &\qquad +  \langle \delta_1 e^{i\vk\cdot\Delta \ve u} \rangle_c   \langle \delta_2 e^{i\vk\cdot\Delta \ve u} \rangle_c   + \langle \delta_1 \delta_2 e^{i\vk\cdot\Delta \ve u} \rangle_c       \Big\}.
\end{align}
 The exponential prefactor can be recast as
\begin{align} \label{FoGexp}
 e^{- (k \mu f_0 \sigma_v)^2 } \exp\!\Big[\! -\frac{1}{2} \big( k_ik_j \langle \Delta u_{i} \Delta u_{j}\rangle_c -2  (k \mu f_0 \sigma_v)^2\big) 
 - \frac{i}{6}k_ik_jk_k \langle \Delta u_{i} \Delta u_{j} \Delta u_{k}\rangle_c +\cdots \Big],
\end{align}
such that in the second exponential we removed the zero lag correlator $\langle u_i(0)u_j(0) \rangle = 2 \sigma^2_v f_0^2 \hat{n}_i \hat{n}_j$.
We further expand the cumulants in the braces of the RHS of eq.~(\ref{expcumM}), 
\begin{align}
&\langle (\delta_1 + \delta_2) e^{i\vk\cdot\Delta \ve u} \rangle_c =  ik_i\langle (\delta_1 + \delta_2) \Delta u_i \rangle - \frac{1}{2}k_ik_j \langle (\delta_1 + \delta_2) \Delta u_i \Delta u_j \rangle, 
                           \\
& \langle \delta_1 e^{i\vk\cdot\Delta \ve u} \rangle_c   \langle \delta_2 e^{i\vk\cdot\Delta \ve u} \rangle_c = -k_ik_j  \langle \delta_1 \Delta u_i \rangle \langle  \delta_2 \Delta u_i \rangle,  \\
& \langle \delta_1 \delta_2 e^{i\vk\cdot\Delta \ve u} \rangle_c = \langle \delta_1 \delta_2 \rangle +ik_i\langle \delta_1 \delta_2 \Delta u_i \rangle,   
\end{align}
valid at 1-loop for primordial Gaussian fields. 
The idea behind the smoothing kernel is to expand the second exponential in the RHS of eq.~(\ref{FoGexp}) and the terms $e^{i\vk\cdot\Delta \ve u}$ inside the cumulants
of the terms within the curly brackets of eq.~(\ref{expcumM}). Lengthy but straightforward algebra leads to 
\begin{align}
 & 1+\mathcal{M}(\vk,\vx) = e^{-k^2\mu^2 f_0^2 \sigma_v^2} 
           \Bigg[ 1 + \sum_{n=0}^4 \frac{i^n}{n!}k_{i_1}\cdots k_{i_n}\Xi^{n}_{i_1\dots i_n}(\vx)  \nonumber\\
 &\quad + (k\mu f_0 \sigma_v )^2\Big[\langle \delta_1 \delta_2 \rangle  +  ik_i\langle (\delta_1 + \delta_2) \Delta u_i \rangle -\frac{1}{2}k_ik_j \langle \Delta u_i \Delta u_j \rangle\Big] \Bigg],
\end{align}
such that the inverse Fourier transform of eq.~(\ref{RSDPS}) can be performed analytically to give
\begin{align} \label{SKPs}
 P_s^\text{SK}(\vk) &= e^{-k^2\mu^2 f_0^2 \sigma_v^2} \Big[P_s^{K,\text{NL}}(k,\mu) + A(k,\mu) + B(k,\mu) + C(k,\mu) \Big].
\end{align} 
Note that by expanding the exponential we recover the moment expansion approach PS  [eq.~\eqref{PSME}].
This model has been used in several works, showing to fit reasonably well to simulations,  see e.g.  \cite{Jalilvand:2019brk}, when $\sigma_v^2$ is considered as a free parameter. 
This approach has the disadvantage of breaking Galilean invariance  
that occurs when splitting $\langle \Delta u_{i}\Delta u_{j}\rangle_c= 2\langle u_{i}(0) u_{j}(0)\rangle_c - 2\langle u_{i}(\vx) u_{j}(0)\rangle_c$, and expanding the second term out of the exponential. 
The consequence of this is the IR divergence we have seen in $C(k,\mu)$, that in the ME expression [eq.~\eqref{PSME}] is cured by the term $- (k \mu f_0 \sigma_v)^2 P_s^K(k,\mu)$, which is missing in $P_s^\text{SK}(\vk)$.

Other smoothing kernel models exist in the literature, but all of them suffer similar problems.  
The most popular is perhaps the TNS model \cite{Taruya:2010mx}, which seems cannot be deduced within the scheme we adopt, but we can obtain it by simply omitting the function $C(k,\mu)$,
\begin{align} \label{TNSPs}
 P_s^\text{TNS}(k,\mu) &= e^{-k^2\mu^2 f_0^2 \sigma_v^2} \Big[P_s^{K,\text{NL}}(k,\mu) +  A(k,\mu) + B(k,\mu) \Big].
\end{align} 
This approach does not have the IR divergence, but as we discussed, the $B(k,\mu)$ has terms UV sensitive which should be canceled by the $C(k,\mu)$ function.  

Another popular phenomenological model was given by Scoccimarro \cite{Scoccimarro:2004tg}, consisting in neglecting the corrections $A(k,\mu)$ and  $D(k,\mu)$,
\begin{equation} \label{PSSc04}
 P_s^\text{Sc04}(k,\mu) =  e^{-k^2 \mu^2 f_0^2 \sigma^2_v }  P_s^{K,\text{NL}}(k,\mu).
\end{equation}
hence this model is free UV and IR divergencies, but is incomplete in the sense that not all 1-loop corrections are considered.

\end{section}

\begin{section}{Summary and Conclusions}

The interplay of high accuracy PT and simulations is an important tool to test both approaches at linear and quasilinear scales. The physical insights from PT and EFT techniques provide the ability to comprehend features in N-point statistics that are valuable to identify effects from different origins, particularly with regards to alternative cosmologies. In the present work we have developed an RSD theory for generalized kernels, beyond EdS, which can be applied to $\Lambda$CDM with or without massive neutrinos, MG models, and dark energy sourced by a nontrivial scale dependent Poisson equation. To show its applicability, we focused on the $\Lambda$CDM and  HS $f(R)$ models.

The approach developed here is valid for models that comply with eqs.~(\ref{PoissonEq}), (\ref{FEcontpre}), and (\ref{FEEulerpre}). The source of the Poisson eq.~(\ref{PoissonEq}) defines the alternative DE/gravity model to consider. The SPT kernels are then computed by substituting eqs.~(\ref{deltanK}) and (\ref{thetanK}) into eqs.~(\ref{FScontEq}) and (\ref{FSEulerEq}) to solve them iteratively. In this way we found the kernels $F_n$ and $G_n$ necessary for the computations to 1-loop that are given in the main text for general $A(k,t)$ and $S$ functions, or in appendix \ref{app:fRHS} for the specific case of HS. However, we remark that the approach we follow thereafter is independent of the specific kernels. With these at hand, one can readily compute the PS for dark matter perturbations using the expressions at the end of section \ref{sect:basic_model}. We then followed ref.~\cite{McDonald:2009dh} to construct the renormalized bias expansion that incorporates new pieces that reflects the complexity of generalized cosmologies, mainly encoded in the kernels themselves. With the biasing theory included, PS for tracers are given in section \ref{PS_1dim}.      

In section \ref{RSD_PT_GC} we developed the main results of this work, that is, a formalism to compute the redshift-space PS for generalized cosmologies. We employed the density-weighted velocity field moments expansion approach \cite{Scoccimarro:2004tg}, based in a formalism that was recently employed for standard cosmology \cite{Vlah:2018ygt,Chen:2020fxs}. We expand the moments in terms of the $I^\tm_{n}(k)$ functions, eq.~(\ref{Pmdef}), that had proven to be useful in \cite{Jalilvand:2019brk,Taruya:2010mx}, and here, for our RSD modeling. We computed the different momenta in section \ref{momenta_comp}. We then expanded the momenta to arrive to the ME approach PS, eq.~(\ref{PSME}), that is an extension of the known expression to generalized cosmologies. The different pieces of the latter expression are all necessary to have a correct large- and small-scale behavior, as proven in section \ref{large-small-behav}. To complete the PS expansion we also computed the EFT counterterms that are necessary to account for backreactions from small-scale physics and also for the damping of the PS along the line-of-sight direction in the particular case of RSD. We also adopted an IR-resummation prescription to correctly model the smearing of BAO due to non-perturbative large scale bulk flows, that also avoids unphysical large oscillations for small-scale modes.           

We finally compared our formalism against simulations, finding very good agreement between our EFT, IR-resummed theory and the state-of-the-art \verb|ELEPHANT|, MG $N$-body suite of simulations. We made comparisons using two ranges of halo masses at redshifts $z=0.5$ and $z=1$. Within error-bars, we find that the matching is good up to scales $k\sim 0.27 \, \text{Mpc}^{-1} h$ for the monopole, and slightly smaller $k$ for the quadrupole and hexadecapole. These comparisons demonstrate that our RSD modeling is overall successful at recovering the anisotropic redshift-space halo PS across a broad range of halo masses, cosmological redshifts and deviations from the base $\Lambda$CDM case. They also suggest, we argue, that our model is a promising tool to construct theoretical templates, such as \cite{Giblin:2019iit,Ramachandra:2020lue,Alam:2020jdv}, to estimate cosmological parameters using real data obtained from surveys.  However, to do this, numerical obstacles may arise because the involved 2-dimensional integrations are constructed from general kernels that are solutions to systems of lengthy differential equations. Such solutions should be obtained at each bin of the 2-dimensional quadratures since these do not have analytical expressions for arbitrary internal momenta, making thus the computations considerably slower than in the corresponding $\Lambda$CDM counterpart, where it is customary to use the analytical EdS kernels. Furthermore, the methods to accelerate these integrals by means of FFTlog methods \cite{Schmittfull:2016jsw,McEwen:2016fjn,Chudaykin:2020aoj} do not apply here, and one has to rely on brute force, which becomes computationally challenging if one needs to span a broad range of parameters in a realistic Monte Carlo Markov Chain analysis.  It is important to consider such analysis in future works, since  ultimately it will tell us if gravity can be tested using theoretical PT templates and large scales cosmological data from upcoming surveys such as DESI and LSST.  That is,  without a detailed parameter estimation it is not trivial to conclude to what extent the MG linear PS and kernels are degenerated with biasing, EFT counterterms, and stochasticity. A first step towards this direction was recently taken in \cite{Liu:2021weo}, through a Fisher analysis. On the other hand, from several works (e.g \cite{Aviles:2017aor,Koyama:2009me}), it is known that the MG screenings are necessary to fit the real space matter power spectrum modeling with simulations. These screenings are non-linear and hence the use of MG kernels $F_n$, instead of EdS, is necessary. However, for tracers in redshift-space the situation is different and it is possible that the MG effects become degenerated with the free parameters at the quasi-linear scales reached by PT/EFT.

To our knowledge, this work presents the most accurate to-date theoretical PT work in the redshift-space PS for MG, being the only one that accounts for beyond linear local biasing, incorporating also EFT counterterms and IR-resummation. Though our theory is based on existing standard cosmology works, mainly \cite{Scoccimarro:2004tg,McDonald:2009dh,Vlah:2018ygt,Chen:2020fxs,Ivanov:2018gjr,Ivanov:2019pdj}, 
we organize the theory differently such that, in our opinion, it is easier to compare between different RSD approaches; and further, 
we find expressions for all the modeling contributions that are valid for kernels beyond EdS.

\end{section}

\acknowledgments

We thank Oliver Philcox, Shi-Fan Chen, Kazuya Koyama, Gustavo Niz, Matteo Cataneo, Benjamin Bose, and Shun Saito for useful discussions and suggestions.
AA, JLCC, and MARM~acknowledge support by CONACyT project 283151. AA also acknowledges partial support to CONACyT Ciencia de Frontera grant No.~102958. GV acknowledges financial support by NSF grant  AST-1813694. BL is supported by the European Research Council (ERC) through Starting Grant ERC-StG716532-PUNCA, and by the STFC through grants No. ST/T000244/1 and ST/P000541/1. The work of RB is supported by DoE grant DE-SC0011838, NASA ATP grant 80NSSC18K0695, NASA ROSES grant 12-EUCLID12-0004 and funding related to the Roman High Latitude Survey Science Investigation Team. 
The simulations described in this work used the DiRAC Data Centric system at Durham University, operated by the Institute for Computational Cosmology on behalf of the STFC DiRAC HPC Facility (\url{www.dirac.ac.uk}). This equipment was funded by BIS National E-infrastructure capital grant ST/K00042X/1, STFC capital grants ST/H008519/1, ST/K00087X/1, STFC DiRAC Operations grant ST/K003267/1 and Durham University. DiRAC is part of the National E-Infrastructure.

\appendix

\begin{section}{$I^m_n$ functions}\label{app:ImnFunctions}

For a rotational scalar function $S(\vk,\vp)=S(k,p,x)$, with $x\equiv \hat{\vk}\cdot\hat{\vp}$, we use the relation (see also Appendix D of \cite{Philcox:2020srd})
\begin{align} \label{angularint}
 \int \Dk{p} (\hat{\vp} \cdot \vhn )^n S(\vk,\vp)  &= 
 \sum_{m=0}^n (\hat{\vk}\cdot\vhn)^m \int \Dk{p} G_{nm}(x)  S(k,p,x), 
\end{align}
where 
\begin{align}\label{Gnm}
G_{nm}(x) &=   \sum_{\ell=0}^n \frac{(1+(-1)^{\ell+n}) (2 \ell+1)}{2(1+\ell+n)}  \binom{\ell}{m} \binom{2 \ell}{\ell}  \binom{\frac{\ell+m-1}{2}}{\ell} \nonumber\\
&\times 
{}_3F_2(\frac{1-\ell}{2},-\frac{\ell}{2},\frac{1}{2} (-1-\ell-n);\frac{1}{2}-\ell,\frac{1}{2} (1-\ell-n); 1 )
\mP_\ell(x)
\end{align}
where ${}_3F_2(\ve a;\ve b;z)$ is the generalized hypergeometric function of the kind ($p=3$, $q=2$) evaluated at $z=1$ \cite{DLMF-16}, and $\mP_\ell(x)$ is the Legendre polynomial of degree $\ell$.

We demonstrate the above expression: starting from \eqref{angularint} one arrives to
\begin{equation}
\sum_{m=0}^n \mu^m G_{nm}(x) = \sum_{\ell=0}^n A_{\ell}^n \mP_\ell(\mu)\mP_\ell(x)    
\end{equation}
where $\mP_\ell$ are the Legendre polynomials and $A_{\ell}^n$ is obtained from
\begin{equation}
    y^n = \sum_{\ell=0}^n A_{\ell}^n \mP_\ell(y), 
\end{equation}
hence
\begin{align}
 A_{\ell}^n &=  \frac{(1+(-1)^{\ell+n}) (2 \ell+1)}{2^{\ell+1}(1+\ell+n)}  \binom{2 \ell}{\ell}
   {}_3F_2(\frac{1-\ell}{2},-\frac{\ell}{2},\frac{1}{2} (-1-\ell-n);\frac{1}{2}-\ell,\frac{1}{2} (1-\ell-n); 1 ).
\end{align}
We further use $\mP_\ell(\mu) = \sum_{m=0}^{\ell} B^\ell_m \mu^m$, with
\begin{equation}
  B^\ell_m = 2^{\ell} \binom{\ell}{m} \binom{\frac{\ell+m-1}{2}}{\ell}.  
\end{equation}
Now, since $A_{\ell}^n=0$ for $\ell>n$ and $ B^\ell_m = 0$ for $m>\ell$, we arrive to
\begin{equation}
G_{nm}(x) = \sum_{\ell=0}^n A_{\ell}^n  B^\ell_m \mP_\ell(x),    
\end{equation}
which is eq.~\eqref{Gnm}. An equivalent formula for $G_{nm}$ was recently found in \cite{Philcox:2020srd}.

We can derive $G_{nn}=  \mP_n(x)$, and $G_{nm}=0$ for $n<m$ or if $n+m$ is an odd integer.  
With this formula one obtains all the $G_{nm}(x)$ functions listed in TNS paper \cite{Taruya:2010mx}, with the exception of $G_{55}(x) = \mP_5(x)$ that in \cite{Taruya:2010mx} contains a typo. Note also that the indices are inverted $G_{nm}(x)^\text{Here} = G_{mn}(x)^\text{TNS paper}$.

From eq.~\eqref{angularint} we can write
\begin{align} \label{pdotnG}
 \int \Dk{p} (\vp \cdot \vhn)^n S(\vk,\vp)  &= 
 \sum_{m=0}^n \mu^m \int \Dk{p} p^n G_{nm}(\hat{\vk}\cdot\hat{\vp})  S(\vk,\vp) \nonumber\\
 &= \frac{k^3}{4\pi^2} \sum_{m=0}^n \mu^m  \int_0^\infty dr \, r^2 \int_{-1}^1 dx (kr)^n G_{nm}(x)  S(k,r,x), 
\end{align}
with $x=\hat{\vk}\cdot\hat{\vp}$ and $r= p/k$.

\begin{subsection}{A function}

Function $A(k,\mu)$, defined in eq.~(\ref{defA_mu_k}), can be written as
\begin{equation}
 A(k,\mu) = \mu^2 \big[ f_0 I^{1,udd}_1(k)  + f_0^2 I^{2,uud}_{1}(k)\big] +  \mu^4 \big[f_0^2 I^{2,uud}_{2}(k) + f_0^3 I^{3,uuu}_{2}(k) \big] + \mu^6 f_0^3 I^{3,uuu}_{3}(k),
\end{equation}
with the use of eq.~\eqref{angularint}, where
\begin{align}\label{I1udd1ATracers}
 I^{1,udd}_1(k) &= \frac{k^3}{4\pi^2} \int_0^\infty dr \int_{-1}^1 dx \,
 \Big\{ A_{11}(\vk,\vp)P_L(k) + \tilde{A}_{11}(\vk,\vp)P_L(k r)  \Big\} \frac{P_L(|\vk-\vp|)}{(1+r^2-2 r x)^2} \nonumber\\
 &\quad +  \frac{k^3}{4\pi^2} \int_0^\infty dr \int_{-1}^1 dx \,  a_{11}(\vk,\vp) P_L(k) P_L(k r),
\end{align}
\begin{align}\label{I2uud12ATracers}
 I^{2,uud}_{n  }(k) &=   \frac{k^3}{4\pi^2} \int_0^\infty dr \int_{-1}^1 dx \,
 \Big\{ A_{n2}(\vk,\vp)P_L(k) + \tilde{A}_{n2}(\vk,\vp)P_L(k r)  \Big\} \frac{P_L(|\vk-\vp|)}{(1+r^2-2 r x)^2} \nonumber\\
 &\quad +  b_1  \frac{k^3}{4\pi^2} \int_0^\infty dr \int_{-1}^1 dx \,  a_{n2}(\vk,\vp) P_L(k) P_L(k r),    \qquad (n=1,2)
\end{align}
\begin{align}\label{I3uud23ATracers}
 I^{3,uuu}_{n  }(k) &=   \frac{k^3}{4\pi^2} \int_0^\infty dr \int_{-1}^1 dx \,
 \Big\{ A_{n3}(\vk,\vp)P_L(k) + \tilde{A}_{n3}(\vk,\vp)P_L(k r)  \Big\} \frac{P_L(|\vk-\vp|)}{(1+r^2-2 r x)^2} \nonumber\\
 &\quad +  b_1  \frac{k^3}{4\pi^2} \int_0^\infty dr \int_{-1}^1 dx \,  a_{n3}(\vk,\vp) P_L(k) P_L(k r),    \qquad (n=2,3)
\end{align}
with $A_{ab}$ and $a_{ab}$ functions
\begin{align}
\frac{A_{11}(\vk,\vp)}{(1+r^2-2 r x)^2} &= 2 b_1^2 r x G_2(-\vk,\vk-\vp) \nonumber\\
                                        &+ b_1^2 \frac{2r^2(1-rx)}{1+r^2-2 r x}G_1(|\vk-\vp|) \left\{  F_2(-\vk,\vk-\vp) + \frac{b_2}{2b_1 } + \frac{b_s}{b_1} S(-\vk,\vk-\vp) \right\},  \\
\frac{\tilde{A}_{11}(\vk,\vp)}{(1+r^2-2 r x)^2} &= 2 b_1^2 \left[ G_1(p) r x +  G_1(|\vk-\vp|) \frac{r^2(1-rx)}{1+r^2-2 r x}   \right]  \nonumber\\
                                                &\quad \times   \left\{  F_2(\vp,\vk-\vp) + \frac{b_2}{2b_1 } + \frac{b_s}{b_1} S(\vp,\vk-\vp) \right\}, \\
        a_{11}(\vk,\vp) &=         2 b_1^2 G_2(-\vk,-\vp) \frac{r^2(1-r x)}{1+r^2-2 r x}  \nonumber\\
                        &\quad  + 2 b_1^2 r x G_1(p)  \left\{   F_2(-\vk,\vp) +\frac{b_2}{2b_1 } + \frac{b_s}{b_1} S(-\vk,\vp) \right\}, 
\end{align}

\begin{align}
\frac{A_{12}(\vk,\vp)}{(1+r^2-2 r x)^2} &=      - b_1 \frac{r^2(1-x^2)}{1+r^2-2 r x} G_2(-\vk,\vk-\vp)G_1(|\vk-\vp|), \\                                    
 \frac{\tilde{A}_{12}(\vk,\vp)}{(1+r^2-2 r x)^2} &= -b_1\frac{r^2(1-x^2)}{1+r^2-2 r x} G_1(p)G_1(|\vk-\vp|) \Big\{  F_2(\vp,\vk-\vp) + \frac{b_2}{2b_1} + \frac{b_s}{b_1} S(\vp,\vk-\vp) \Big\} \\
        a_{12}(\vk,\vp) &= -b_1 \frac{r^2(1-x^2)}{1+r^2-2 r x} G_2(\vp,-\vk)G_1(p),  
\end{align}
\begin{align}
& \frac{A_{22}(k,x,r)}{(1+r^2-2rx)^2} = b_1\left[ \frac{r^2(1-3x^2)+2xr  }{1+r^2-2rx}G_1(|\vk-\vp|) + 2xr G_1(k) \right] G_2(-\vk,\vk-\vp) \nonumber\\
 &\qquad\qquad + b_1 \frac{2r^2(1-r x)}{1+r^2-2rx} G_1(|\vk-\vp|)G_1(k) \Big\{ F_2(-\vk,\vk-\vp) + \frac{b_2}{2b_1 } + \frac{b_s}{b_1 } S(\vk-\vp,-\vk) \Big\}, \nonumber\\
 & \frac{\tilde{A}_{22}(k,x,r)}{(1+r^2-2rx)^2} = b_1\left[   \frac{2r^2(1-r x)}{1+r^2-2rx} G_1(|\vk-\vp|) + 2xr G_1(p) \right] G_2(\vp,\vk-\vp) \nonumber\\
 &\qquad\qquad +  b_1\frac{r^2(1-3x^2)+2xr }{1+r^2-2rx} G_1(|\vk-\vp|)G_1(p) \Big\{F_2(\vp,\vk-\vp) +  \frac{b_2}{2b_1 } + \frac{b_s}{b_1 } S(\vp,\vk -\vp)  \Big\}, \nonumber\\
& a_{22}(k,x,r) = b_1\left[ \frac{r^2(1-3x^2)+2xr  }{1+r^2-2rx}G_1(p) +  \frac{2r^2(1-r x)}{1+r^2-2rx}  G_1(k) \right] G_2(-\vk,\vp) \nonumber\\
 &\qquad\qquad +     2 b_1 xr G_1(p)G_1(k) \Big\{ F_2(-\vk,\vp) +  \frac{b_2}{2b_1 } + \frac{b_s}{b_1 } S(\vp,-\vk) \Big\}.
\end{align}

\begin{align}
 \frac{A_{23}(k,x,r)}{(1+r^2-2rx)^2}         &= \frac{r^2 (x^2-1)}{1+r^2-2rx} G_2(-\vk,\vk-\vp)G_1(k)G_1(|\vk-\vp|) , \\
 \frac{\tilde{A}_{23}(k,x,r)}{(1+r^2-2rx)^2} &= \frac{r^2 (x^2-1)}{1+r^2-2rx} G_2(\vp,\vk-\vp)G_1(p)G_1(|\vk-\vp|) , \\
 a_{23}(k,x,r)                               &= \frac{r^2 (x^2-1)}{1+r^2-2rx} G_2(-\vk,\vp)G_1(k)G_1(p).
\end{align}

\begin{align}
 \frac{A_{33}(k,x,r)}{(1+r^2-2rx)^2}         &= \frac{r^2(1-3x^2)+2 r x}{1+r^2- 2 rx} G_2(-\vk,\vk-\vp)G_1(k)G_1(|\vk-\vp|) , \\
 \frac{\tilde{A}_{33}(k,x,r)}{(1+r^2-2rx)^2} &= \frac{r^2(1-3x^2)+2 r x}{1+r^2- 2 rx} G_2(\vp,\vk-\vp)G_1(p)G_1(|\vk-\vp|) , \\
 a_{33}(k,x,r)                               &= \frac{r^2(1-3x^2)+2 r x}{1+r^2- 2 rx} G_2(-\vk,\vp)G_1(k)G_1(p).
\end{align}

All these functions $A_{ab}$ reduce to the corresponding in TNS paper for the unbiased case ($b_1=1$, $b_2=b_s=0$) and with EdS kernels $G_1$, $G_2$, and $F_2$.

\end{subsection}

\begin{subsection}{D function} \label{App:D_function}
The $D(k,\mu)$ function in eq. (\ref{D_k_mu_def}) is 

\begin{align}
 D(k,\mu) &= \mu^2 \big[f_0^2 I^{2,uudd}_1(k) + f_0^3 I^{3,uuud}_1(k) +  f_0^4 I^{4,uuuu}_1(k) \big] \nonumber\\ 
 &\quad    + \mu^4 \big[f_0^2 I^{2,uudd}_2(k) + f_0^3 I^{3,uuud}_2(k) +  f_0^4 I^{4,uuuu}_2(k) \big]  \nonumber\\ 
 &\quad    + \mu^6 \big[ f_0^3 I^{3,uuud}_3(k) +  f_0^4 I^{4,uuuu}_3(k) \big] +  \mu^8 f_0^4 I^{4,uuuu}_4(k), 
\end{align}
with
\begin{align} 
I^{2,uudd}_{n}(k) &= I^{2,uudd}_{n\,B}(k) + I^{2,uudd}_{n\, C}(k)    - \delta_{n1} k^2 \sigma^2_v P_{\delta\delta}(k),   \label{I2nuudd}        & \\
I^{3,uuud}_{n}(k) &= I^{3,uuud}_{n\,B}(k) + I^{3,uuud}_{n\, C}(k)    - \delta_{n2} 2 k^2 \sigma^2_v P_{\delta\theta}(k)   ,   \label{I3nuuud}   & \\
I^{4,uuuu}_{n}(k) &= I^{4,uuuu}_{n\,B}(k) + I^{4,uuuu}_{n\, C}(k)    - \delta_{n3} k^2 \sigma^2_v P_{\theta\theta}(k),      \label{I4nuuuu}
\end{align}
according to the splitting $D= B + C - (k \sigma_v f_0  \mu)^2 P^K_s$.

\begin{align}
 I^{2,uudd}_{n\,B}(k) &= \frac{k^3}{4\pi^2}   \int_0^\infty dr \int_{-1}^1 dx \, B^n_{11}(r,x) \frac{P_{\delta\theta}(kr)P_{\delta\theta}(|\vk-\vp|)}{1+r^2-2rx}, 
\end{align}
with
\begin{align}
  B^1_{11}(r,x)&= \frac{r^2}{2}(x^2-1), \qquad B^2_{11}(r,x) = \frac{r}{2}(r - 3rx^2 + 2x).
\end{align}

\begin{align}
 I^{2,uudd}_{1\,C}(k) &= \frac{k^3}{4\pi^2}   \int_0^\infty dr  \int_{-1}^1 dx \,\frac{1}{4}(1-x^2) \Bigg\{  
           P_{\delta\delta}(k\sqrt{1+r^2-2rx})P_{\theta\theta}(kr) \nonumber\\
           &\qquad\qquad\qquad\qquad + r^4 \frac{P_{\delta\delta}(kr) P_{\theta\theta}(k\sqrt{1+r^2-2rx})}{(1+r^2-2rx)^2}  \Bigg\}, \\
 I^{2,uudd}_{2\,C}(k) &= \frac{k^3}{4\pi^2}   \int_0^\infty dr  \int_{-1}^1 dx \, \Bigg\{  
          \frac{1}{4}(3x^2-1) P_{\delta\delta}(k\sqrt{1+r^2-2rx})P_{\theta\theta}(kr) \nonumber\\
           &\qquad + \frac{r^2}{4} \big[2 - 4 r x + r^2 (3 x^2 - 1)\big] \frac{P_{\delta\delta}(kr) P_{\theta\theta}(k\sqrt{1+r^2-2rx})}{(1+r^2-2rx)^2} \Bigg\}.
\end{align}
We note these expressions have no counterpart in $B$, because the product here is $P_{\delta\delta}P_{\theta\theta}$, while the products for functions
$B$ are $P_{\delta\theta}P_{\delta\theta}$, $P_{\delta\theta}P_{\theta\theta}$,  and $P_{\theta\theta}P_{\theta\theta}$. Therefore,
functions $C^{1}_{11}$ or $C^{2}_{11}$, with the same meaning as those relative to $B$, do not exist.

Functions with $\tm=3$ are
\begin{align}
 I^{3,uuud}_{n\, C}(k) &=         -\frac{k^3}{4\pi^2}   \int_0^\infty dr  \int_{-1}^1 dx \, C^{n}_{21}(k,r,x) \frac{P_{\theta\theta}(k\sqrt{1+r^2-2rx})P_{\delta\theta}(kr)}{(1+r^2-2rx)^2} \nonumber\\
                       &\quad     -\frac{k^3}{4\pi^2}   \int_0^\infty dr  \int_{-1}^1 dx \, C^{n}_{12}(k,r,x) \frac{P_{\delta\theta}(k\sqrt{1+r^2-2rx})P_{\theta\theta}(kr)}{(1+r^2-2rx)}, \\
 I^{3,uuud}_{n\, B}(k) &=         -\frac{k^3}{4\pi^2}   \int_0^\infty dr  \int_{-1}^1 dx \, B^{n}_{21}(k,r,x) \frac{P_{\theta\theta}(k\sqrt{1+r^2-2rx})P_{\delta\theta}(kr)}{(1+r^2-2rx)^2} \nonumber\\
                       &\quad     -\frac{k^3}{4\pi^2}   \int_0^\infty dr  \int_{-1}^1 dx \, B^{n}_{12}(k,r,x) \frac{P_{\delta\theta}(k\sqrt{1+r^2-2rx})P_{\theta\theta}(kr)}{(1+r^2-2rx)},
\end{align}
with
\begin{align}
 C^{1}_{21}(r,x) &=   -\frac{3r^4}{8}(x^2-1)^2  = -B^{1}_{21}(r,x),\\
 C^{1}_{12}(r,x) &=   -\frac{3r^2}{8}(x^2-1)^2  = -B^{1}_{21}(r,x),
\end{align}
showing that indeed,  $I^{3,uuud}_{1\,D}(k) = I^{3,uuud}_{1\, B}(k) + I^{3,uuud}_{1\, C}(k) =0$, such that there is not $f_0^3 \mu^2$ term in $D(k,\mu)$ function. The rest of the functions are
\begin{align}
 C^{2}_{21}(r,x) &=   -\frac{r^2}{4}  (1 - x^2) (2 - 3 r^2 - 12 r x + 15 r^2 x^2),  \\
 C^{2}_{12}(r,x) &=   -\frac{1}{4}  (1 - x^2) (2 - 3 r^2 - 12 r x + 15 r^2 x^2),  \\
 B^{2}_{21}(r,x) &=   \frac{3}{4} r^2 (-1 + x^2) (-2 + r^2 + 6 r x - 5 r^2 x^2), \\
 B^{2}_{12}(r,x) &=   -\frac{3}{4} r (-1 + x^2) (-r - 2 x + 5 r x^2),
\end{align}
and
\begin{align}
 C^{3}_{21}(r,x) &=  -\frac{r^2}{8} (-4 + 3 r^2 + 24 r x + 12 x^2 - 30 r^2 x^2 - 40 r x^3 +  35 r^2 x^4),   \\
 C^{3}_{12}(r,x) &=  -\frac{1}{8} (-4 + 3 r^2 + 24 r x + 12 x^2 - 30 r^2 x^2 - 40 r x^3 +  35 r^2 x^4), \\
 B^{3}_{21}(r,x)  &= \frac{r}{8}\Big\{-8 x +  r \big[  -12 + 36 x^2 + 12 r x (3 - 5 x^2)  + r^2 (3 - 30 x^2 + 35 x^4)   \big]    \Big\},  \\
 B^{3}_{12}(r,x) &= \frac{r}{8} \big[ 4 x (3 - 5 x^2) + r (3 - 30 x^2 + 35 x^4)  \big].
\end{align}

Functions with $\tm=4$ are
\begin{align}
 I^{4,uuuu}_{n\, B}(k) &= \frac{k^3}{4\pi^2}   \int_0^\infty dr  \int_{-1}^1 dx B^{n}_{22}(r,x)
 \frac{P_{\theta\theta}(kr) P_{\theta\theta}(k\sqrt{1+r^2-2rx})}{(1+r^2-2rx)^2},  \\
 I^{4,uuuu}_{n\, C}(k) &= \frac{k^3}{4\pi^2}   \int_0^\infty dr  \int_{-1}^1 dx C^{n}_{22}(r,x)
 \frac{P_{\theta\theta}(kr) P_{\theta\theta}(k\sqrt{1+r^2-2rx})}{(1+r^2-2rx)^2},
\end{align}
with
\begin{align}
C^1_{22} &= -\frac{5r^4}{16} (-1 + x^2)^3 = -B^1_{22},\\
C^2_{22} &= \frac{3r^2}{16}(-1 + x^2)^2 (7 - 5 r^2 - 30 r x + 35 r^2 x^2), \\
C^3_{22} &= 
-\frac{1}{16} (-1 + x^2) \Big[4 +  3 r \Big\{-16 x +  r \big[ -14 + 70 x^2 + 20 r x (3 - 7 x^2) \nonumber\\ 
  &\quad \qquad + 
         5 r^2 (1 - 14 x^2 + 21 x^4) \big] \Big\} \Big], \\
C^4_{22} &=\frac{1}{16}  \Big[ -4 + 12 x^2 + 16 r x (3 - 5 x^2) + 7 r^2 (3 - 30 x^2 + 35 x^4) \nonumber\\ 
  &\quad \qquad - 6 r^3 x (15 - 70 x^2 + 63 x^4) + r^4 \big\{ -5 + 21 x^2 (5 - 15 x^2 + 11 x^4) \big\} \Big],
\end{align}
and
\begin{align}
B^2_{22} &= -\frac{3r^2}{16}(-1 + x^2)^2 (6 - 5 r^2 - 30 r x + 35 r^2 x^2), \\
B^3_{22} &= \frac{3 r}{16}(x^2-1) \Big[-8 x + 
   r \big\{-12 + 60 x^2 + 20 r x (3 - 7 x^2) + 
      5 r^2 (1 - 14 x^2 + 21 x^4) \big\} \Big],\\
B^4_{22} &=   \frac{r}{16} \Big[ 8 x (-3 + 5 x^2) - 6 r (3 - 30 x^2 + 35 x^4) +  6 r^2 x (15 - 70 x^2 + 63 x^4) \nonumber\\
  &\quad \qquad + r^3 \big\{(5 - 21 x^2 (5 - 15 x^2 + 11 x^4) \big\}  \Big].
\end{align}
The relation $C^1_{22} + B^1_{22}=0$ implies that,  $I^{4}_{1}(k) =0$, such that there is not $f_0^4 \mu^2$ term in $D(k,\mu)$ function.

We stress out that all our functions $B^n_{ab}$ are identical to those presented in TNS paper, this is because all are constructed out of products of correlators of linear fields. The only
difference is that a factor $f(k)/f_0$ is introduced each time a velocity field $\theta$ is present. The biasing of these functions is implicit in the power spectra  $P_{\delta\delta}$
and $P_{\theta\theta}$ and cross-power spectrum $P_{\delta\theta}$. Meaning that 
\begin{equation}
 B(k,\mu;f_0) \, \longrightarrow \, b_1^4 B(k,\mu,f_0/b_1), \qquad 
 C(k,\mu;f_0) \, \longrightarrow \, b_1^4 C(k,\mu,f_0/b_1). 
\end{equation}
As we mentioned in section \ref{sect:BiasExp}, we restrict the inclusion of curvature bias to $\mathcal{O}(P_L)$ terms, so it does not appear in the above expressions. 

\end{subsection}

\end{section}

\begin{section}{On the f(R) Hu-Sawicki model}\label{app:fRHS} 
Throughout this work we apply our results to the Hu-Sawicki $f(R)$ model \cite{Hu:2007nk}, with $n=1$ and  amplitude $f_{R0} = -10^{-6},-10^{-5} -10^{-4} $, called F6, F5 and F4, respectively. We stress that results for other gravity models are straightforward to develop, following our formalism. Models defined in the Einstein frame can be put also in our frame by using field redefinitions; see \cite{Aviles:2018qotF}. In this appendix we summarize the main aspects needed from the HS models.   

A general $f(R)$ action is written as 
\begin{equation} \label{f_R_action}
S= \int d^4 x \sqrt{-g} \left( R + f(R) \right) +  \int d^4 x \sqrt{-g} {\cal L}_m.
\end{equation}
Variations with respect to the metric of this action lead to the field equations
\begin{equation}\label{fRFE}
 G_{\mu\nu} + f_R R_{\mu\nu} - \nabla_\mu \nabla_\nu f_R - \left( \frac{f}{2} - \square f_R \right)g_{\mu\nu} = 8 \pi G T_{\mu\nu},   
\end{equation}
where $f_R \equiv \frac{df(R)}{dR}$, that represents the scalar field degree of freedom of the theory. 
By taking the trace to this equation one obtains
\begin{equation} \label{TeqfR}
3 \square f_R = R(1-f_R) + 2 f - 8 \pi G \rho,
\end{equation}
where we use a dark matter fluid with $T^{\mu}_\mu = -\rho$.

Considering a perturbed Friedmann-Robertson-Walker line element 
\begin{equation}
ds^2 = - (1+2\Phi) dt^2 + a(t)^2 (1-2\Psi) d\vx^2 ,
\end{equation}
the fluid perturbation $\Delta \rho = \bar{\rho} \delta$ and the associated scalar field perturbation  
$ \delta f_{R} = f_R - \bar{f_R}$, $R= \bar{R} + \delta R$, where the bar indicates background quantities and $\bar{R} \equiv R(\bar{f}_R)$, the perturbative field equations in Fourier space are \cite{Koyama:2009me,Aviles:2017aor}: 
\begin{eqnarray}
- \frac{k^2}{a^2} \Phi &=& 4 \pi G \bar{\rho} \delta + \frac{1}{2} \frac{k^2}{a^2} \delta f_{R},
\label{PoissonEq_fR} \\
3 \frac{k^2}{a^2} \delta f_{R} 
&=&  8 \pi G \bar{\rho} \delta - M_1(k) \delta f_{R}  - {\delta \cal I}(\delta f_{R}), 
\label{KG_fR}
\\
  \Psi - \Phi &=&  \delta f_{R},  
\end{eqnarray}
where $m \equiv (M_1 /3)^{1/2}$ represents the mass of scalar field. On scales larger than $m^{-1}$, the scalar field does not propagate and one recovers the  $\Lambda$CDM model. On the other hand, on small scales ($k \rightarrow \infty$, 
$ \mu \rightarrow 4/3$, see eq.~\ref{mu_Eq}), so gravity is enhanced by one third, $G_{\rm eff} \rightarrow (4/3)G$. 

Substituting eq.~(\ref{KG_fR}) in  (\ref{PoissonEq_fR}), one arrives to eq.~(\ref{PoissonEq}) identifying the functions: 
\begin{eqnarray}
\mu(k,t) &=& 1 + \frac{k^2/a^2}{3 k^2/a^2 + M_1(k)} ,
\label{mu_Eq} \\
S(\vk) &=& -\frac{1}{2} \, \frac{k^2/a^2}{3 k^2/a^2 + M_1(k)}  {\delta \cal I},
\label{S_Eq}
\end{eqnarray}
${\delta \cal I}$ represents nonlinear interactions which can be expanded as: 
\begin{align}\label{selfIntexp}
\delta \mathcal{I}(\delta f_{R})
&= \frac{1}{2} \int \frac{d^3 k_1 d^3 k_2}
{(2 \pi)^3} \delta_D(\vk -\vk_{12}) M_2(\vk_1, \vk_2)
\delta f_{R}(\vk_1) \delta f_{R}(\vk_2) \nonumber\\
&\quad + \frac{1}{6}
\int \frac{d^3 k_1 d^3 k_2 d^3 k_3}{(2 \pi)^6}
\delta_D(\vk - \vk_{123}) M_3(\vk_1, \vk_2, \vk_3)
\delta f_{R}(\vk_1) \delta f_{R}(\vk_2) \delta f_{R}(\vk_3) + \cdots,
\end{align}
where the functions $M_i$  are in general scale and time dependent and are determined by the $f(R)$ model.

Giving eq.~(\ref{mu_Eq}), we can now determine $A(k,t)$ using eq.~(\ref{Ak_Eq}), as 
\begin{equation} \label{AktHS}
A(k,t) = 4 \pi G \bar{\rho} \, \mu(k,t) = 
4 \pi G \bar{\rho} \left( 1 + \frac{k^2/a^2}{3 k^2/a^2 + M_1(k)} \right),  
\end{equation}
that enters in eq.~(\ref{1stOrderEq}) to determine the growth function to first order $D_{+}(k,t)$.  It only remains to determine the functions $M_i$ to have all the needed functions to be able to integrate the system. These functions are given for $f(R)$ gravity by 

\begin{equation} \label{potexp}
 \delta R = \sum_i \frac{1}{n!} M_n (\delta f_R)^n, \qquad M_n \equiv \frac{d^n R(f_R)}{d f_{R}^n} \Bigg|_{f_R = \bar{f}_R}. 
\end{equation}

For the  $n=1$ HS model can be expressed as: 
\begin{equation}\label{HSfR}
 f(R)= - \frac{c_1 R}{c_2 R/M^2 + 1}.
\end{equation}

 In order to have an effective $\Lambda$CDM model at background level, the energy scale is chosen to be $M^2= H_0^2 \Omega_{m0}$ and $c_1/c_2 = 6 \Omega_\Lambda / \Omega_{m0}$. 

The functions $M_1$, $M_2$ and $M_3$ are:
\begin{align}
M_1(a) = \frac{3}{2}  \frac{H_0^2}{|f_{R0}|} \frac{(\Omega_{m0} a^{-3} + 4 \Omega_\Lambda)^3}{(\Omega_{m0}  + 4 \Omega_\Lambda)^2}, \label{M1}
\end{align}
\begin{align}
M_2(a) = \frac{9}{4}  \frac{H_0^2}{|f_{R0}|^2} \frac{(\Omega_{m0} a^{-3} + 4 \Omega_\Lambda)^5}{(\Omega_{m0}  + 4 \Omega_\Lambda)^4}, 
\end{align}
\begin{align}
M_3(a)= \frac{45}{8}  \frac{H_0^2}{|f_{R0}|^3} \frac{(\Omega_{m0} a^{-3} + 4 \Omega_\Lambda)^7}{(\Omega_{m0}  + 4 \Omega_\Lambda)^6}. 
\end{align}
These functions depend only on the background evolution since they are the coefficients of the expansion of a scalar field potential 
about its background value, see eq.~(\ref{potexp}).

The second order source entering eq.~\eqref{DAeveq} becomes
\begin{equation}
\mathcal{S}^{(2)}(\vk_1,\vk_2) = -\left(\frac{2A_0}{3} \right)^2 \frac{ M_2(\vk_1,\vk_2) k^2/a^2}{6 \Pi(k)\Pi(k_1)\Pi(k_2)},     
\end{equation} \label{defPi}
where $\Pi(k) \equiv \frac{1}{3 a^2}(3 k^2 + M_1 a^2)$. 
\end{section}

\begin{section}{SPT generalized kernels}\label{app:Kernels}

In SPT the expansion is performed directly to the overdensity and velocity fields, $\delta = \delta^{(1)} + \delta^{(2)}+\cdots$ and
$\theta = \theta^{(1)} + \theta^{(2)}+\cdots$, which we have written as a Taylor Fourier expansion in eqs.~\eqref{deltanK} and \eqref{thetanK}.
The $F_n$ and $G_n$ kernels are usually obtained by solving iteratively continuity and Euler equations, eqs.~(\ref{FScontEq}) and (\ref{FSEulerEq}).
In this notation, the linear growth functions $D_+$ are 
kept attached to the linear fields because they are scale dependent and cannot be pulled out of the integral;
and the $G_n$ kernels carry the linear growth rate $f(\vk,t)$. 
Our notation coincides with 
that of ref.~\cite{Taruya:2016jdt} except for a minus sign in $G_n$, but differs from the most used notations that factorize the $f$ factors.

Instead of solving for $F_n$ and $G_n$ directly, we obtain them from known LPT kernels, using the mappings developed in \cite{Aviles:2018saf}. In LPT one follows the trajectories $\vx$ of 
particles initially located at a Lagrangian position $\vq$,
\begin{equation}\label{LcToEc}
x^i(\vq,t) = q^i + \Psi^i(\vq,t),    
\end{equation}
with the Lagrangian displacement vector field $\Psi$ which in PT is written as
\begin{align}
 \Psi_i(\vp) &=   i \sum_{m=1}^{\infty} \frac{1}{m!} \underset{\vp_{1\cdots m} = \vp}{\int}  L_{i}^{(m)}(\vp_1,\dots,\vp_m;t)
 \delta_L(\vp_1) \cdots \delta_L(\vp_{n}), 
\end{align}
with $L_{i}^{(m)}$ the LPT kernels.
For $F_n$ we have the relation \cite{Aviles:2018saf}
\begin{equation}
F_n(\vk_1,\dots,\vk_n) = \sum_{\ell=1}^n \sum_{m_1+\cdots+m_\ell=n} \frac{k_{i_1}\cdots k_{i_\ell}}{\ell! m_1!\cdots m_\ell!} 
L_{i_i}^{(m_1)}(\vk_{1}, \dots,\vk_{m_1}) \cdots L_{i_\ell}^{(m_\ell)}(\vk_{m_{\ell-1}}\cdots, \vk_{m_\ell}).    
\end{equation}
Once symmetrized, for $F_2$ and $G_2$ we obtain, see also \cite{Matsubara:2011ck,Rampf:2012xa}, 
\begin{align}
 F_2(\vk_1,\vk_2) &=  \frac{1}{2} \Big[ k_iL^{(2)}_i(\vk_1,\vk_2) + k_i k_j L_i^{(1)}(\vk_1)L_j^{(1)}(\vk_2) \Big], \label{LPTtoF2}\\
 F_3(\vk_1,\vk_2,\vk_3) &= \frac{1}{3!}\Big[ k_i L^{(3)s}_i(\vk_1,\vk_2,\vk_3)  
 + k_ik_j (L^{(2)}_i(\vk_1,\vk_2)L^{(1)}_j(\vk_3) + \text{cyclic})\nonumber\\
 &\qquad + k_ik_jk_kL^{(1)}_i(\vk_1)L^{(1)}_j(\vk_2)L^{(1)}_k(\vk_3) \Big]. \label{LPTtoF3}
\end{align}

Computing the $G_n$ kernels is a little more messy than the $F_n$, but still straightforward.
The velocity field is given by $v_i(\vx,t) = d\vx/d\tau = a \dot{\Psi}_i(\vq,t)$.
We use 
\begin{equation}
 \frac{\partial\,}{\partial q^i} = \frac{\partial x^j}{\partial q^i}\frac{\partial\,}{ \partial x^j} = J_{ji} \frac{\partial\,}{ \partial x^j}, 
\end{equation}
with $J_{ij}=\partial x^i/\partial q^j$ the Jacobian matrix of the coordinate transformation \eqref{LcToEc} and $J$ its determinant, 
to get $\nabla_{\vx i} = (J^{-1})_{ji} \nabla_{ j}$ with $(J^{-1})_{ji} = (2J)^{-1}\epsilon_{ikp}\epsilon_{jqr}J_{kq}J_{pr}$
or
\begin{equation}
 J(J^{-1})_{ji} =  \delta_{ij} + (\delta_{ij}\delta_{ab} - \delta_{ia}\delta_{jb})\Psi_{a,b} + \frac{1}{2}\epsilon_{ikp}\epsilon_{jqr} \Psi_{k,q} \Psi_{p,r}.
\end{equation}
Hence
 \begin{equation}
 J \frac{1}{a} \nabla_{\vx\,i}v_i = \dot{\Psi}_{i,i} + \Psi_{j,j}\dot{\Psi}_{i,i} - \Psi_{i,j}\dot{\Psi}_{i,j} 
 + \frac{1}{2}\epsilon_{ikp}\epsilon_{jqr} \Psi_{k,q} \Psi_{p,r}\dot{\Psi}_{i,j}.
 \end{equation}
The Fourier transform of the velocity divergence yields 
\begin{align} \label{tfpsid}
-  H f \theta(\vk) &= \int d^3 x e^{-i\vk \cdot \vx} \frac{1}{a} \nabla_{\vx\,i}v_i 
               =  \int d^3 q e^{-i\vk \cdot \vq} e^{-i\vk\cdot \Psi(\vq,t)} J(\vq,t) \frac{1}{a} \nabla_{\vx\,i}v_i \nonumber\\
              &=  \int d^3 q e^{-i\vk \cdot \vq} \sum_{\ell=0}^{\infty} \frac{1}{\ell!}(-ik_a \Psi_a)^{\ell}  \dot{\Psi}_{i,i}    \nonumber\\
              &\quad +  (\delta_{ij}\delta_{ab} - \delta_{ia}\delta_{jb}) \int d^3 q e^{-i\vk \cdot \vq} \Psi_{a,b}\dot{\Psi}_{i,j} \sum_{\ell=0}^{\infty}\frac{1}{\ell!} (-ik_k \Psi_k)^{\ell}   \nonumber\\
              &\quad +  \frac{1}{2}\epsilon_{ikp}\epsilon_{jqr}  \int d^3 q e^{-i\vk \cdot \vq}   \Psi_{k,q} \Psi_{p,r}\dot{\Psi}_{i,j} \sum_{\ell=0}^{\infty} \frac{1}{\ell!}(-ik_s \Psi_s)^{\ell}.  
\end{align}
On the other hand by taking the derivative of the displacement field
\begin{align}
 \dot{\Psi}_i(\vp) &=   i \sum_{m=1}^{\infty} \frac{1}{m!} \underset{\vp_{1\cdots m} = \vp}{\int}  L_{i}^{'(m)}(\vp_1,\dots,\vp_m)
 \delta_L(\vp_1) \cdots \delta_L(\vp_{n}), 
\end{align}
with
\begin{equation}
 L_{i}^{'(m)}(\vp_1,\dots,\vp_m) =  \dot{L}_{i}^{(m)}(\vp_1,\dots,\vp_m) + H L_{i}^{(m)}(\vp_1,\dots,\vp_m)(f(\vp_1) + \cdots + f(\vp_m)).
\end{equation}
 
\begin{itemize}
    \item To second order, using eq.~\eqref{tfpsid},
\begin{align} \label{2orderExp}
 -f H \theta^{(2)}(\vk) 
              &=  \int d^3 q e^{-i\vk \cdot \vq}  \dot{\Psi}^{(2)}_{i,i}(\vq,t) 
                 - i k_a \int d^3 q e^{-i\vk \cdot \vq} \Psi^{(1)}_a(\vq,t)  \dot{\Psi}^{(1)}_{i,i}(\vq,t)\nonumber\\
              &\quad +  (\delta_{ij}\delta_{ab} - \delta_{ia}\delta_{jb}) \int d^3 q e^{-i\vk \cdot \vq} \Psi^{(1)}_{a,b}(\vq,t)\dot{\Psi}^{(1)}_{i,j}(\vq,t) \\
              &\equiv I_A + I_B + I_C,
\end{align}
with
\begin{align} \label{2orderIa}
I_A(\vk) &= \int d^3 q e^{-i\vk \cdot \vq}  \nabla_i \dot{\Psi}^{(2)}_{i}(\vq,t) =   \int d^3 q e^{-i\vk \cdot \vq}  \nabla_i \int \Dk{p} e^{i\vp \cdot \vq }\dot{\Psi}^{(2)}_{i}(\vp,t) \nonumber \\
   &= \int \Dk{p} i \vp_i \dot{\Psi}^{(2)}_{i}(\vp,t)     \int d^3q e^{i(\vp -\vk) \cdot \vq } = i \vk_i  \dot{\Psi}^{(2)}_{i}(\vk,t)  \nonumber \\
  &= - \ikk  \frac{1}{2} (\vk_1 + \vk_2)_i L^{'(2)}_i(\vk_1,\vk_2)\delta_1\delta_2,
\end{align}
\begin{align} \label{2orderIb}
I_B(\vk) &=- \ikk  (\vk_1 + \vk_2)_a  k_2^i  L^{(1)}_a(\vk_1)  L^{'(1)}_i(\vk_2)\delta_1\delta_2 \\
  &= - \ikk  \frac{1}{2} (\vk_1 + \vk_2)_a \left[ k_2^i  L^{(1)}_a(\vk_1)  L^{'(1)}_i(\vk_2) + k_1^i L^{(1)}_a(\vk_2)   L^{'(1)}_i(\vk_1) \right] \delta_1\delta_2.
\end{align}
Now, using
\begin{align}
 \int d^3 q e^{-i\vk \cdot \vq} \Psi^{(1)}_{a,b}(\vq)\dot{\Psi}^{(1)}_{i,j}(\vq) =  \ikk k_1^b k_2^j L_a^{(1)}(\vk_1) L_i^{'(1)}(\vk_2) \delta_1 \delta_2,
\end{align}
the last term is
\begin{align} \label{2orderIc}
I_C(\vk) 
  &=  (\delta_{ij}\delta_{ab} - \delta_{ia}\delta_{jb}) \int d^3 q e^{-i\vk \cdot \vq} \Psi^{(1)}_{a,b}(\vq,t)\dot{\Psi}^{(1)}_{i,j}(\vq,t) \nonumber\\
  &=  \ikk \Big[ k_1^a L_a^{(1)}(\vk_1)k_2^i L_i^{'(1)}(\vk_2) - \vk_1\cdot\vk_2 L_a^{(1)}(\vk_1)L_a^{'(1)}(\vk_2) \Big] \delta_1 \delta_2 \\
  &=  - \ikk \frac{1}{2} \Big[\vk_1\cdot\vk_2 \big(L_a^{(1)}(\vk_1)L_a^{'(1)}(\vk_2) + L_a^{(1)}(\vk_2)L_a^{'(1)}(\vk_1) \big)  \nonumber\\
  &\qquad\qquad - k_1^a k_2^i L_a^{(1)}(\vk_1) L_i^{'(1)}(\vk_2) - k_2^a k_1^i L_a^{(1)}(\vk_2) L_i^{'(1)}(\vk_1) \Big] \delta_1 \delta_2.
\end{align}
Summing the three contributions and identifying to \eqref{thetanK}, we obtain 
\begin{align}
2 f H G_2(\vk_1,\vk_2) &=  (\vk_1 + \vk_2)_i L^{'(2)}_i(\vk_1,\vk_2) \nonumber\\ 
&+ (\vk_1 + \vk_2)_a \left[ k_2^i  L^{(1)}_a(\vk_1)  L^{'(1)}_i(\vk_2) + k_1^i L^{(1)}_a(\vk_2)   L^{'(1)}_i(\vk_1) \right] \nonumber\\
&-  k_1^a k_2^i L_a^{(1)}(\vk_1) L_i^{'(1)}(\vk_2) - k_2^a k_1^i L_a^{(1)}(\vk_2) L_i^{'(1)}(\vk_1) \nonumber\\
&+ \vk_1\cdot\vk_2 \big(L_a^{(1)}(\vk_1)L_a^{'(1)}(\vk_2) + L_a^{(1)}(\vk_2)L_a^{'(1)}(\vk_1) \big) \label{G2fromLPTpre}
\end{align} 

\item To third order, using eq.~\eqref{tfpsid},
\begin{align}
 - f H \theta^{(3)}(\vk) 
              &=  \int d^3 q e^{-i\vk \cdot \vq}  \dot{\Psi}^{(3)}_{i,i}    \nonumber\\
              &+  \int d^3 q e^{-i\vk \cdot \vq} (-ik_a \Psi^{(1)}_a)  \dot{\Psi}^{(2)}_{i,i}    \nonumber\\
              &+  \int d^3 q e^{-i\vk \cdot \vq} (-ik_a \Psi^{(2)}_a)  \dot{\Psi}^{(1)}_{i,i}    \nonumber\\
              &+  \frac{1}{2}\int d^3 q e^{-i\vk \cdot \vq}  (-ik_a \Psi^{(1)}_a)(-ik_b \Psi^{(1)}_b)  \dot{\Psi}^{(1)}_{i,i}    \nonumber\\
              &+  (\delta_{ij}\delta_{ab} - \delta_{ia}\delta_{jb}) \int d^3 q e^{-i\vk \cdot \vq} \Psi_{a,b}^{(1)}\dot{\Psi}^{(2)}_{i,j}    \nonumber\\
              &+  (\delta_{ij}\delta_{ab} - \delta_{ia}\delta_{jb}) \int d^3 q e^{-i\vk \cdot \vq} \Psi_{a,b}^{(2)}\dot{\Psi}_{i,j}^{(1)}   \nonumber\\
              &+  (\delta_{ij}\delta_{ab} - \delta_{ia}\delta_{jb}) \int d^3 q e^{-i\vk \cdot \vq} \Psi_{a,b}^{(1)}\dot{\Psi}^{(1)}_{i,j}(-ik_k \Psi_k^{(1)})   \nonumber\\
              &+  \frac{1}{2}\epsilon_{ikp}\epsilon_{jqr}  \int d^3 q e^{-i\vk \cdot \vq}   \Psi^{(1)}_{k,q} \Psi^{(1)}_{p,r}\dot{\Psi}^{(1)}_{i,j}. 
\end{align} 
The algebra is considerably larger than for the second order case, but it follows the same basic principles from which we arrive to
\begin{align}
& G_3(\vk_1,\vk_2,\vk_3) = \nonumber\\
&   \frac{1}{2} C_3 \Gamma^{f}_3(\vk_1,\vk_2,\vk_3)  \nonumber \\
&+  \frac{1}{3} C_2 \Gamma^{f}_2(\vk_2,\vk_3) \frac{\vk \cdot \vk_1}{k_1^2} + \text{(2 cyclic perm)}   \nonumber\\
& + \frac{1}{6} C_2 \Gamma_2(\vk_2,\vk_3) \frac{\vk \cdot \vk_{23}}{k_{23}^2} \frac{f_1}{f_0} + \text{(2 cyclic perm)}     \nonumber\\
& + \frac{1}{6} \frac{(\vk \cdot \vk_1)(\vk \cdot \vk_2)}{k_1^2 k_2^2} \frac{f_3}{f_0} + \text{(2 cyclic perm)}    \nonumber\\
& -  \frac{1}{3} C_2 \Gamma_2^f(\vk_2,\vk_3)\left( 1- \frac{(\vk_1\cdot \vk_{23})^2}{k_1^2 k_{23}^2}  \right)  + \text{(2 cyclic perm)} \nonumber\\
& - \frac{1}{6} C_2 \Gamma_2(\vk_2,\vk_3) \left( 1- \frac{(\vk_1\cdot \vk_{23})^2}{k_1^2 k_{23}^2}  \right) \frac{f_1}{f_0}   + \text{(2 cyclic perm)}            \nonumber\\
& - \frac{1}{3} \frac{\vk \cdot \vk_1}{k_1^2}  \left( 1- \frac{(\vk_2\cdot \vk_{3})^2}{k_2^2 k_{3}^2}  \right)\frac{f_2+f_3}{2 f_0} + \text{(2 cyclic perm)} \nonumber\\
& + 3  \Bigg[ 1 -  \frac{(\vk_1 \cdot \vk_2)^2}{k_1^2 k_2^2}   -  \frac{(\vk_2 \cdot \vk_3)^2}{k_2^2k_3^2} 
-  \frac{(\vk_3\cdot\vk_1)^2}{k_3^2 k_1^2} \nonumber\\
&\quad \qquad +  2 \frac{(\vk_1\cdot\vk_2)(\vk_2\cdot\vk_3)(\vk_3\cdot\vk_1)}{k_1^2k_2^2k_3^2} \Bigg]\frac{f_1+f_2+f_3}{3 f_0}. \label{G3fromSPTpre}
\end{align}
We have used the scalar kernels $C_n \Gamma_n$ for the transverse piece of the Lagrangian displacement, $\vk \cdot \Psi$, defined as
\begin{align}
C_n \Gamma_n(\vk_1,\dots,\vk_n) &= k^{i}_{1\cdots n} L^{(n)}_i(\vk_1,\dots,\vk_n), \\
C_n \Gamma_n^f(\vk_1,\dots,\vk_n) &= \frac{1}{n H f_0} k^{i}_{1\cdots n} L^{'(n)}_i(\vk_1,\dots,\vk_n),
\end{align}
with $C_n$ a set of numbers, from which the first two are fixed to $C_1=1$ and $C_2=3/7$ for simplicity.
Since for 1-loop, 2-point statistics, the transverse component of Lagrangian displacements project out, one can use the scalar $\Gamma$ instead of vector $\ve L$ kernels without loss of generality.

\end{itemize}

\bigskip

We have related the second and third order SPT kernels to the LPT kernels and their derivatives through eqs.~\eqref{LPTtoF2}, \eqref{LPTtoF3}, \eqref{G2fromLPTpre} and \eqref{G3fromSPTpre}. We take the
LPT kernels from ref.~\cite{Aviles:2017aor}. For $F_2$ and $G_2$ we arrive at the expressions of eqs.~(\ref{F2_kernel}) and (\ref{G2_kernel}).

For the third order kernels $F_3,G_3(\vk_1,\vk_2,\vk_3)$ we are interested in double-squeezed configurations, on which $\vk_1=\vk$ and $\vk_3=-\vk_2=\vp$. 
To write the final expressions for these kernels we consider 
the first order scalar kernels, that are $\Gamma_1(\vk)=1$ and $\Gamma_1^f(\vk)=f(k)/f_0$. The second order scalar kernels are
\begin{align} \label{Gamma2}
\Gamma_2(\vp_1,\vp_2) &= \left[\mA(\vp_1,\vp_2) - \mB(\vp_1,\vp_2) \frac{(\vp_1 \cdot \vp_2)^2}{p_1^2 p_2^2}\right],\\
\Gamma^f_2(\vp_1,\vp_2) 
     &= \Gamma_2(\vp_1,\vp_2) \frac{f(p_1) + f(p_2)}{2 f_0} 
     +  \frac{1}{2f_0 H}\left[\dot{\mA}(\vp_1,\vp_2) - \dot{\mB}(\vp_1,\vp_2) \frac{(\vp_1 \cdot \vp_2)^2}{p_1^2 p_2^2}\right], 
\end{align}
where $\mA,\mB=\mA,\mB(\vp_1,\vp_2)$ are the second order growth functions  defined by eqs.~(\ref{AandBdef}). The third order kernels are
\begin{align}
C_3 \Gamma_3[\vp_1,\vp_2,\vp_3] &= \frac{D_+^{(3)s}(\vp_1,\vp_2,\vp_3)}{D_+(\vp_1)D_+(\vp_2)D_+(\vp_3)},  \\
C_3 \Gamma^f_3[\vp_1,\vp_2,\vp_3] &= \frac{1}{3f_0} \frac{\frac{d\,}{d \ln a} D_+^{(3)s}(\vp_1,\vp_2,\vp_3)}{D_+(\vp_1)D_+(\vp_2)D_+(\vp_3)}.
\end{align}
with the (symmetric) third order growth function $D_+^{(3)s}$ as given in \cite{Aviles:2017aor}.\footnote{After correcting a trivial typo, as in the expressions of appendix of \cite{Aviles:2018qotF}, or eq.~(A5) of \cite{Valogiannis:2019xed}.} Actually, we will not use the value of $C_3$ at all, so 
we can let it free. But by defining $C_2=3/7$ we make the notation simpler.

The third order kernels are 
\begin{align} \label{F3_kernel}
& F_3(\vk,-\vp,\vp) = 
   \frac{1}{6} C_3 \Gamma_3(\vk,-\vp,\vp)  
+  \frac{1}{3} C_2 \frac{\vk\cdot(\vk-\vp) \vk \cdot \vp}{p^2|\vk-\vp|^2} \Gamma_2(\vk,-\vp) - \frac{1}{6} \frac{(\vk \cdot \vp)^2}{p^4}   
\end{align}

\begin{align} \label{G3_kernel}
& G_3(\vk,-\vp,\vp) = 
   \frac{1}{2} C_3 \Gamma^{f}_3(\vk,-\vp,\vp)  
    +  \frac{2}{3} C_2 \frac{\vk \cdot \vp}{p^2} \Gamma^{f}_2(\vk,-\vp)  
    + \frac{1}{3} C_2 \frac{f(p)}{f_0}  \Gamma_2(\vk,-\vp) \frac{\vk \cdot (\vk-\vp)}{|\vk-\vp|^2}    \nonumber\\
&\quad - \frac{1}{6} \frac{(\vk \cdot \vp)^2}{p^4} \frac{f(k)}{f_0}     
       -  \frac{1}{3} C_2  \left[ 2 \Gamma_2^f(\vk,-\vp) + \Gamma(\vk,-\vp) \frac{f(p)}{f_0} \right]\left[1 -\frac{(\vp \cdot (\vk-\vp))^2}{p^2|\vk-\vp|^2}  \right],
\end{align}
which are valid upon integration with $d^3p$.
It is worth noticing that the second term in the second line yields zero in $\Lambda$CDM because  $\int_{-1}^1 dx\, x(1-x^2) = 0$. But in general  cosmologies do not because of the $x$ dependence of functions $\A$ and $\B$. 
Notoriously, for DGP or cubic Galileons, $\B$ is only time dependent and the dependence on $x$ of $\A$ contains only zero and second Legendre Polynomials, such that
$x(\A - \B x^2)$ is odd in $x$ and hence that term is also zero.

We notice that the approximation of static kernels, usually taken in  $\Lambda$CDM and exact for EdS, corresponds to
\begin{equation}
 \Gamma_n^f \simeq  \Gamma_n,  \qquad \text{($\Lambda$CDM)}.
\end{equation}

\end{section}

 \bibliographystyle{JHEP}  
 \bibliography{bib.bib}

\end{document}